\UseRawInputEncoding
\documentclass[twoside,twocolumn,9pt]{article}
\usepackage{extsizes}
\usepackage[super,sort&compress,comma]{natbib} 
\usepackage[version=3]{mhchem}
\usepackage[left=1.5cm, right=1.5cm, top=1.785cm, bottom=2.0cm]{geometry}
\usepackage{balance}
\usepackage{mathptmx}
\usepackage{sectsty}
\usepackage{xurl}
\usepackage{graphicx} 
\usepackage{stfloats}
 \usepackage{booktabs}
\usepackage{fancyvrb}
\usepackage{fvextra}
\usepackage{lastpage}
\usepackage[format=plain,justification=justified,singlelinecheck=false,font={stretch=1.125,small,sf},labelfont=bf,labelsep=space]{caption}
\usepackage{float}
\usepackage{fancyhdr}
\usepackage{fnpos}
\usepackage[english]{babel}
\addto{\captionsenglish}{%
  
} 
\usepackage[normalem]{ulem}
\usepackage{array}
\usepackage{droidsans}
\usepackage{charter}
\usepackage[T1]{fontenc}
\usepackage[usenames,dvipsnames]{xcolor}
\usepackage{setspace}
\usepackage[compact]{titlesec}
\usepackage{hyperref}

\usepackage{epstopdf}

\definecolor{cream}{RGB}{222,217,201}

\begin{document}

\pagestyle{fancy}
\thispagestyle{plain}
\fancypagestyle{plain}{
\renewcommand{\headrulewidth}{0pt}
}

\makeFNbottom
\makeatletter
\renewcommand\LARGE{\@setfontsize\LARGE{15pt}{17}}
\renewcommand\Large{\@setfontsize\Large{12pt}{14}}
\renewcommand\large{\@setfontsize\large{10pt}{12}}
\renewcommand\footnotesize{\@setfontsize\footnotesize{7pt}{10}}
\makeatother

\renewcommand{\thefootnote}{\fnsymbol{footnote}}
\renewcommand\footnoterule{\vspace*{1pt}%
\color{cream}\hrule width 3.5in height 0.4pt \color{black}\vspace*{5pt}} 
\setcounter{secnumdepth}{5}

\renewcommand{\dbltopfraction}{0.95}      
\renewcommand{\textfraction}{0.05}        
\renewcommand{\dblfloatpagefraction}{0.9} 

\makeatletter 
\renewcommand\@biblabel[1]{#1}            
\renewcommand\@makefntext[1]%
{\noindent\makebox[0pt][r]{\@thefnmark\,}#1}
\makeatother 
\renewcommand{\figurename}{\small{Fig.}~}
\sectionfont{\sffamily\Large}
\subsectionfont{\normalsize}
\subsubsectionfont{\bf}
\setstretch{1.125} 
\setlength{\skip\footins}{0.8cm}
\setlength{\footnotesep}{0.25cm}
\setlength{\jot}{10pt}
\titlespacing*{\section}{0pt}{4pt}{4pt}
\titlespacing*{\subsection}{0pt}{15pt}{1pt}

\fancyfoot{}
\fancyfoot[LO,RE]{\vspace{-7.1pt}\includegraphics[height=9pt]{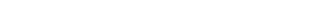}}
\fancyfoot[CO]{\vspace{-7.1pt}\hspace{13.2cm}\includegraphics{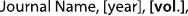}}
\fancyfoot[CE]{\vspace{-7.2pt}\hspace{-14.2cm}\includegraphics{head_foot/RF}}
\fancyfoot[RO]{\footnotesize{\sffamily{1--\pageref{LastPage} ~\textbar  \hspace{2pt}\thepage}}}
\fancyfoot[LE]{\footnotesize{\sffamily{\thepage~\textbar\hspace{3.45cm} 1--\pageref{LastPage}}}}
\fancyhead{}
\renewcommand{\headrulewidth}{0pt} 
\renewcommand{\footrulewidth}{0pt}
\setlength{\arrayrulewidth}{1pt}
\setlength{\columnsep}{6.5mm}
\setlength\bibsep{1pt}

\makeatletter 
\newlength{\figrulesep} 
\setlength{\figrulesep}{0.5\textfloatsep} 

\newcommand{\topfigrule}{\vspace*{-1pt}%
\noindent{\color{cream}\rule[-\figrulesep]{\columnwidth}{1.5pt}} }

\newcommand{\botfigrule}{\vspace*{-2pt}%
\noindent{\color{cream}\rule[\figrulesep]{\columnwidth}{1.5pt}} }

\newcommand{\dblfigrule}{\vspace*{-1pt}%
\noindent{\color{cream}\rule[-\figrulesep]{\textwidth}{1.5pt}} }

\makeatother

\twocolumn[
  \begin{@twocolumnfalse}
{\includegraphics[height=30pt]{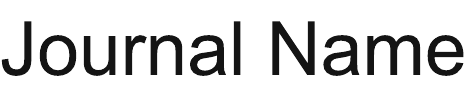}\hfill\raisebox{0pt}[0pt][0pt]{\includegraphics[height=55pt]{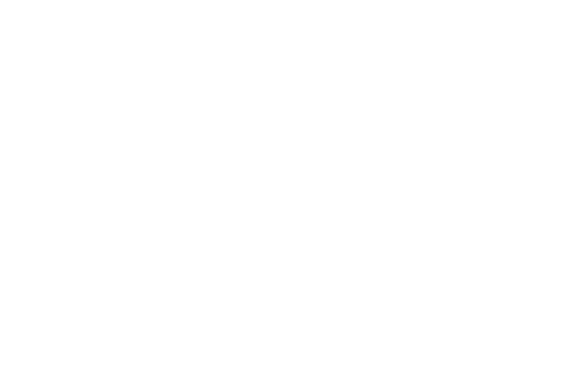}}\\[1ex]
\includegraphics[width=18.5cm]{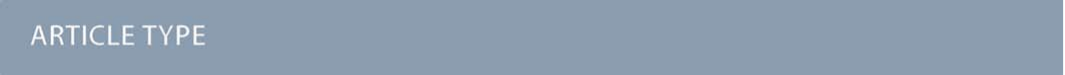}}\par
\vspace{1em}
\sffamily
\begin{tabular}{m{4.5cm} p{13.5cm} }

\includegraphics{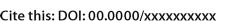} & \noindent\LARGE{\textbf{LeMat-Synth: a multi-modal toolbox to curate broad
synthesis procedure databases from scientific literature$^\dag$}} \\
\vspace{0.3cm} & \vspace{0.3cm} \\

 & \noindent\large{
 Magdalena Lederbauer,\textit{$^{a,b,*}$} 
 Siddharth Betala,\textit{$^{b}$} 
 Valerie Gentzke,\textit{$^{c}$} 
 Anamaria Leonescu,\textit{$^{c}$} 
 Amine Sehaba,\textit{$^{d,e}$} 
 Faris Flaifil,\textit{$^{g}$} 
 Ayush Jain,\textit{$^{h}$} 
 Alfonso Amayuelas,\textit{$^{i}$} 
 Nikhil Yelamarthy,\textit{$^{c}$} 
 Xiyao Li,\textit{$^{b}$} 
 Gr\'{e}goire Germain,\textit{$^{b}$} 
 Stefano Ribes,\textit{$^{j}$} 
 Stefan P. Schmid,\textit{$^{k}$}
 Alexandre Nozadze,\textit{$^{k}$} 
 Anna Kelmanson,\textit{$^{l}$} 
 Sudheesh Kumar Ethirajan,\textit{$^{m}$} 
 Mohd Zaki,\textit{$^{n}$} 
 Elton Pan,\textit{$^{a}$} 
 Georgia Channing,\textit{$^{o}$} 
 Connor W. Coley,\textit{$^{a}$} 
 Philippe Schwaller,\textit{$^{k}$} 
 Roc\'{\i}o Mercado,\textit{$^{i}$} 
 Alexandre Duval,\textit{$^{b}$} 
 Mathilde L. D. Franckel,\textit{$^{p,b,*}$} 
 Samuel P. Gleason\textit{$^{b,*}$}
 } 
 \\

\includegraphics{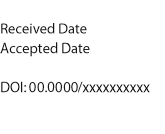} & \noindent\normalsize{Wide access to advanced experimental methods in materials science has given rise to an abundance of procedural knowledge, which is scattered across decades of scientific literature and recorded in unstructured formats that are challenging to analyze systematically.
In this work, we present \texttt{LeMat-Synth Parser}, a modular, open-source, and multi-modal extraction toolbox that utilizes large language models (LLMs) and vision language models (VLMs) to automatically structure synthesis protocols and performance metrics extracted from both text and figures of publications.
Applying \texttt{LeMat-Synth Parser} to 81K open-access publications, we curate \texttt{LeMat-Synth}, an extensive dataset of 58K synthesis procedures and to our knowledge the largest and most diverse structured inorganic materials synthesis dataset to date, covering 35 synthesis methods and 16 material classes based on a domain-specific ontology.
We validate extraction quality against annotations by domain experts and a scalable LLM-as-a-judge framework, and benchmark a suite of models to identify optimal configurations and characterize cross-model biases.
To demonstrate the extensibility of \texttt{LeMat-Synth Parser}, we apply it to two distinct domains. First, we link synthesis protocols and catalyst identity to thermocatalytic performance across a corpus of ammonia-decomposition publications. Second, we cross-validate text- and figure-reported critical transition temperatures across 1,384 superconductivity papers, then use the validated pipeline to recover $T_c$ for every composition in a sample series.
We release \texttt{LeMat-Synth Parser} and the \texttt{LeMat-Synth} dataset openly on \href{https://github.com/LeMaterial/lematerial-llm-synthesis}{GitHub} and \href{https://huggingface.co/datasets/LeMaterial/LeMat-Synth}{Hugging Face}.

} \\

\end{tabular}

 \end{@twocolumnfalse} \vspace{0.6cm}

  ]

\renewcommand*\rmdefault{bch}\normalfont\upshape
\rmfamily
\section*{}
\vspace{-1cm}

\footnotetext{
\textit{
\textsuperscript{a}~Massachusetts Institute of Technology, Cambridge, MA, USA,
\textsuperscript{b}~Entalpic, Paris, France,
\textsuperscript{c}~Independent Researcher,
\textsuperscript{d}~URM MAP, Lyon, France,
\textsuperscript{e}~LIRIS, Lyon, France,
\textsuperscript{f}~Université Paris-Saclay
\textsuperscript{g}~Conservatoire National des Arts et Métiers (CNAM), France,
\textsuperscript{h}~Georgia Tech, Atlanta, Georgia, USA,
\textsuperscript{i}~University of California Santa Barbara, Santa Barbara, USA,
\textsuperscript{j}~Chalmers University of Technology and University of Gothenburg, Gothenburg, Sweden,
\textsuperscript{k}~ETH Zurich, Zurich, Switzerland,
\textsuperscript{l}~EPFL, Lausanne, Switzerland,
\textsuperscript{m}~University of California, Davis, CA, USA,
\textsuperscript{n}~Johns Hopkins University, Baltimore, MD, USA,
\textsuperscript{o}~Hugging Face, New York City, NY, USA,
\textsuperscript{p}~Imperial College London, London, UK,
\textsuperscript{*}~E-mail: magled@mit.edu, m.franckel25@imperial.ac.uk, samuel.gleason@entalpic.ai%
}}

\footnotetext{\dag~Supplementary Information available: extended methods (data acquisition and PDF post-processing, ontology, figure segmentation and digitization), the full human-expert vs.\ LLM-as-a-judge agreement analysis with worked disagreement examples and uncertainty quantification, VLM figure-extraction benchmarks, per-category dataset statistics, linking failure-mode analyses for both case studies, and all system prompts and configurations.}




\section{Introduction}
Materials discovery underpins advances in energy conversion~\cite{zhang2013nanomaterials}, energy storage~\cite{liu2010advanced}, catalysis~\cite{vogt2022concept}, and many other technologies.~\cite{butler2018machine}
Associated challenges have driven decades of experimental and computational materials research, resulting in an extensive body of literature.
Initiatives such as the Materials Genome Initiative~\cite{MaterialsGenomeInitiative}, Harvard Clean Energy Project~\cite{HarvardCleanEnergyProject}, and Materials Project~\cite{MaterialsProject} exemplify the increasing emphasis on computational approaches to materials discovery.
However, these efforts primarily rely on \textit{in silico} property prediction and are constrained by the limited availability of structured experimental synthesis and characterization data in the literature.

\begin{figure*}[tp]
\centering
\includegraphics[width=\textwidth]{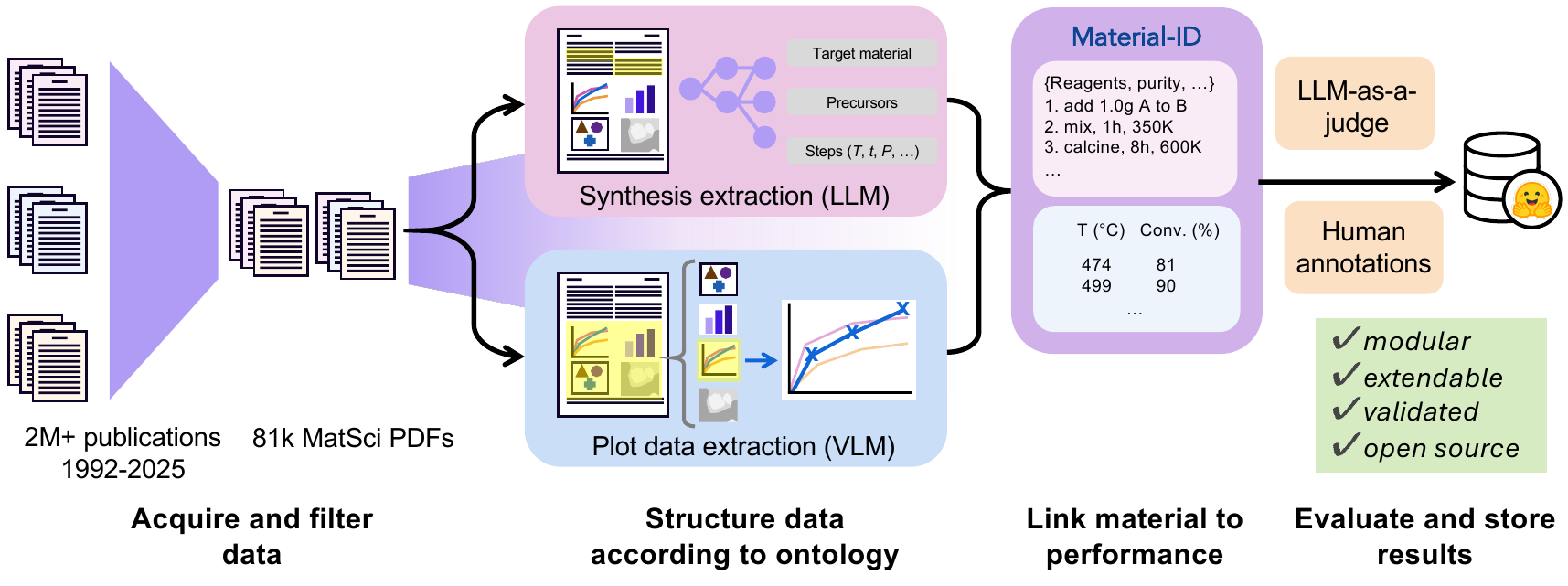}
\caption{Overview of the pipeline used to curate \texttt{LeMat-Synth} by linking synthesis procedures with performance data. We collect data from a corpus $>$2M open-access papers from arXiv\cite{arxiv}, ChemRxiv\cite{chemrxiv}, and Semantic Scholar\cite{semanticscholar,kim2025towards}, which is filtered to 81K materials science publications. For papers containing synthesis procedures, both textual and visual content are extracted. Materials and synthesis protocols are parsed using a structured LLM-based extraction pipeline. Figures are segmented, classified, and digitized using computer vision models and VLMs, producing structured representations of plotted data. To link performance measurements to synthesis protocols, plot series are matched to identified materials using a language model that reads series labels, figure captions, surrounding text, and axis metadata. Plots are pre-filtered to retain performance-relevant data (\textit{e.g.}, conversion vs. temperature) while excluding characterization figures. Linking quality is subsequently assessed using an LLM-as-judge framework that scores candidate mappings and identifies predefined failure modes, such as material-name mismatches or precursor–product confusion. The resulting records are validated and assembled into a standardized, extensible database of synthesis procedures and associated performance metrics.
}\label{fig:overview}
\end{figure*}

In organic chemistry, structured reaction data has long been available at scale, largely through decades of expert curation in commercial databases such as CAS SciFinder and Reaxys~\cite{CAS_SciFinder, Reaxys}, and more recently through community deposition in open repositories such as the Open Reaction Database~\cite{kearnes2021open}.
Automated extraction has played a narrower role~\cite{Vaucher2019NatureCom}, succeeding mainly in the patent space, which is more standardized and legally required to be complete, enabling the mining of millions of reactions~\cite{lowe2012extraction}. These structured resources have in turn powered machine learning (ML) approaches for reaction prediction and synthesis planning.~\cite{coley2017prediction, segler2018planning} For inorganic materials synthesis, no comparable curation infrastructure exists, and although materials patents and literature are plentiful, they lack the standardization and canonical machine-readable structure representations that made large-scale reaction mining from organic chemistry patents tractable.~\cite{sun2025critical, karpovich2023interpretable}
With the growing relevance of ML in materials discovery, the absence of large and high-quality materials synthesis datasets is a major bottleneck.
Addressing this knowledge gap requires scalable automated methods for extracting, standardizing, and structuring synthesis protocols from the existing literature.

In pursuit of this goal, early systems such as ChemicalTagger~\cite{ChemicalTagger}, often used in combination with LeadMine~\cite{LeadMine}, ChemDataExtractor~\cite{ChemDataExtractor}, and OSRA~\cite{OSRA}, demonstrated the feasibility of extracting structured chemical information from scientific literature.
Subsequent work extended these ideas to materials science through rule-based and classical natural language processing methods for extracting synthesis information and constructing machine-readable databases~\cite{Olivetti2020APR,kim2017materials,wang2022ulsa,kononova2019text,mahbub2020text}. 
More recently, ML and LLM-based approaches have further improved the robustness of extraction, enabling datasets such as ZeoSyn and the automated extraction of organic synthesis procedures~\cite{pan2024zeosyn,ai2024extracting}.

Despite these advances, recent benchmark studies indicate that extracting synthesis protocols remains challenging when publications describe multiple materials or complex workflows.~\cite{hira2024reconstructing,alampara2025probing}
Composition-centric extraction approaches often fail to accurately associate processing conditions with the relevant material instances, even when utilizing advanced LLMs.
This highlights the need for material-level ontologies and standardized evaluation benchmarks for experimental information extraction~\cite{litxbenchbenchmark}.

Recently, LLMs and VLMs have enabled more sophisticated extraction of materials synthesis data.
These developments include automated synthesis parameter extraction from scientific text~\cite{lee2024text, Walker2023Extracting, Lee2025Data} and multi-modal parsing of materials literature using LLM-based agents and knowledge representations~\cite{da2024automated}.
\citet{prein2025language} further demonstrated the use of LLM-generated synthesis protocols to augment experimental synthesis data, while \citet{pan2026synreason} used reinforcement learning with experimental feedback to improve synthesis reasoning in LLMs. See \citet{schilling2025text} for a review of text mining and foundation models in chemistry and materials science.

Because many current pipelines remain text-centric, their capacity to capture information distributed across figures, tables, supplementary materials, and experimental schematics has been limited.
Extracted outputs frequently require extensive post-processing prior to downstream analysis or database integration~\cite{kim2025towards,hira2024reconstructing}.
Despite the capabilities of modern foundation models, challenges such as hallucination, factual consistency, reproducibility, and domain adaptation continue to impede reliable materials knowledge extraction~\cite{miret2025enabling}.
Closing this gap requires extraction pipelines that treat figures and tables as primary sources of data rather than as a supplement to the text. It also requires developing a quantitative understanding of how such pipelines fail (e.g., hallucination, omission, and misattribution rates).

Most prior work operates on relatively small literature corpora, typically comprising only a few hundred to a few thousand publications~\cite{kim2017materials,kononova2019text,mahbub2020text,pan2024zeosyn,kim2025towards}, yielding datasets with limited scope.
Such datasets are insufficient to support robust, generalizable synthesis prediction models for a broad class of materials.
Moreover, existing studies typically evaluate a single LLM or extraction pipeline rather than systematically comparing multiple LLMs and VLMs across extraction and evaluation tasks~\cite{lee2024text,da2024automated,prein2025language,litxbenchbenchmark}.
Differences in prompting strategies, output schemas, reasoning capabilities, and multi-modal performance further complicate reproducibility and hinder fair benchmarking across studies~\cite{miret2025enabling,schilling2025text,kim2025towards}.
These limitations define a clear need for scalable and standardized extraction pipelines that generate high-quality, structured synthesis datasets from large scientific corpora and enable rigorous comparison of modern LLMs and VLMs.

Here, we present \texttt{LeMat-Synth Parser}, a modular extraction toolbox that integrates LLMs and VLMs to curate literature from diverse materials science domains, and extracts synthesis protocols and digitizes performance data from text and figures at scale.
\texttt{LeMat-Synth Parser} is supported by a domain-specific ontology covering 35 synthesis methods and 16 material classes, which maps unstructured synthesis steps and experimental conditions into a standardized format.
Demonstrated here on two domains, the underlying approach is applicable to a wide range of inorganic materials systems whose literature exhibits comparable reporting conventions.
We conducted a thermocatalysis study that links synthesis protocols to catalytic performance for ammonia decomposition, validated on 23 manually-annotated papers and mapping which metal-support combinations and synthesis routes achieve high conversion and which remain untested, and a superconductivity study across 1,384 papers, using text-reported $T_c$ values to validate a figure-extraction pipeline that we then apply at full corpus scale to recover $T_c$ for every composition in a sample series.
To support data-driven materials research, we release \texttt{LeMat-Synth}, a structured dataset comprising 58K synthesis procedures (30K in a curated high-quality subset) extracted from 81K open-access materials science papers.
To our knowledge, this is among the broadest structured synthesis datasets in terms of synthesis methods and material classes covered classes\cite{kim2017materials,kononova2019text,pan2024zeosyn}.
We assess extraction quality through an evaluation pipeline that combines annotations by human domain experts with LLM-based scores~\cite{kim2025towards}.
We additionally compare four LLMs and three VLMs across extraction and evaluation roles, letting practitioners select the model best suited to their own extraction task and offering guidance on cross-model evaluation bias.
Collectively, these contributions establish a scalable foundation for transforming materials literature into structured, machine-readable knowledge that facilitates data-driven synthesis planning and accelerates materials discovery.

\section{Methods}

\begin{figure}[h]
    \centering
    \includegraphics[width=\linewidth]{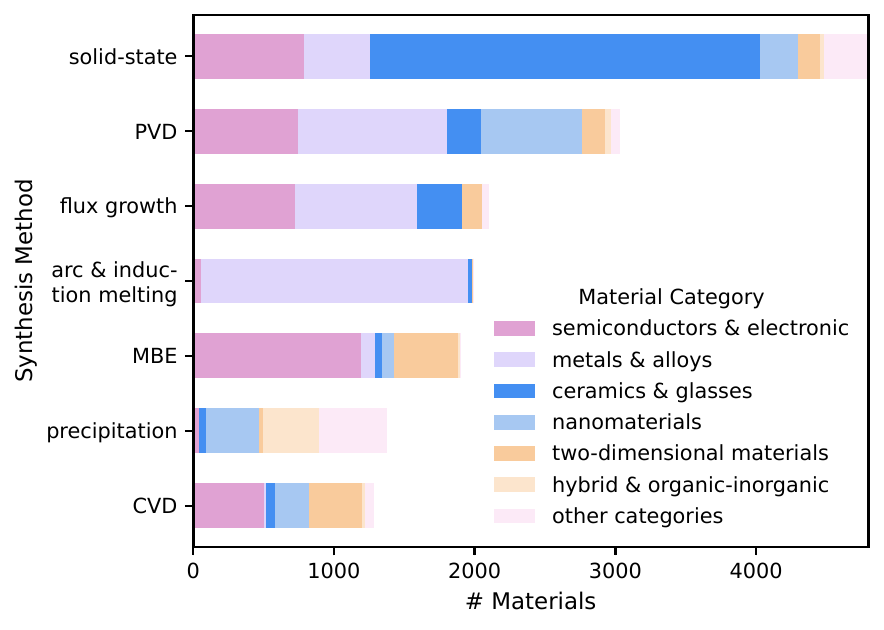}
      \caption{The diversity of synthesis methods covered by \texttt{LeMat-Synth}, shown for the 30K-procedure high-quality subset. For the seven most common synthesis methods (of 35 total, ``other'' excluded), each bar is colored by the distribution of material categories synthesized with that method. Full per-category and per-method breakdowns, source-stacked variants, and coverage of the material-class $\times$ synthesis-method space are provided in SI Section S3.1. Abbreviations: PVD = Physical vapor deposition, CVD = chemical vapor deposition, MBE = molecular beam epitaxy.}
  \label{fig:dataset-stats}
\end{figure}

The complete workflow, illustrated in Fig.~\ref{fig:overview}, processes full-text publications to generate structured synthesis protocols and digitized performance data. We provide further implementation details and technical specifications in SI Section S2. Fig.~\ref{fig:dataset-stats} summarizes the dataset, which we characterize further in SI Section S3.1.

\texttt{LeMat-Synth Parser} is organized around four interchangeable components that together define a domain-specific instantiation, as detailed in the SI (Fig. S1). A corpus loader specifies how to fetch and parse publications from a given source. A domain ontology, a typed Pydantic schema defining the structured output for the target domain, can inherit from the \texttt{GeneralSynthesisOntology}. A \texttt{PlotFilterConfig} dataclass specifies which figure types are relevant via axis labels and keywords. Finally, an evaluation rubric, the LLM-as-a-judge prompt, and scoring criteria are defined in a YAML configuration. Changing these four components without modifying the underlying codebase fully re-instantiates the pipeline for a new domain, which we demonstrate for two scientifically distinct domains in Section~\ref{sec:casestudies}.

\subsection{Data curation}
We aggregated a corpus of materials science publications from three open-access sources: (1) the arXiv \cite{arxiv} preprint server, from which we extracted 2M papers, including 381K from the condensed matter (cond-mat) category, covering years 1992--2025, (2) the ChemRxiv \cite{chemrxiv} preprint server, from which we extracted all 30K papers, and included over 2,900 chemistry papers from selected categories in materials science, and (3) a curated corpus of materials science papers sourced via the Semantic Scholar API \cite{semanticscholar} (Open Materials Guide, OMG24~\cite{kim2025towards}) consisting of $>$17K papers reporting synthesis procedures. For arXiv and ChemRxiv, an LLM classifier (Mistral-Small-3.1-24B) reads each paper's full text and returns whether it describes a synthesis procedure, while we use the OMG24 corpus as pre-filtered by its authors. From this initial corpus, we identified 81K publications containing explicit synthesis procedures. We extracted the full-text content and figures from each relevant work, followed by standardized post-processing to ensure consistent formatting across sources. We describe detailed preprocessing steps and filtering criteria in SI Section S3.2.

\subsection{Extraction of synthesis protocols}
To systematically structure synthesis protocols, we first define an ontology that represents each synthesis as a sequence of discrete operations. Each operation specifies an action (\textit{e.g.}, heating, mixing, annealing), precursors (including material names, quantities, and purities), and experimental conditions (temperature, pressure, duration of different steps). We implement this ontology as a typed \texttt{Pydantic} schema\cite{pydantic} and the \texttt{DSPy} framework \cite{khattab2023dspy} to ensure consistency and facilitate downstream analysis. We fix the base ontology (the core operation, precursor, and condition fields) to preserve a common schema across domains. Users extend it for specialized domains by adding fields, types, and options at the domain-specific subclass level (see SI Section S2, step 3). This design trades materials-space coverage for a common schema. Because domain ontologies must inherit from a shared base, the framework does not attempt to natively represent every materials class (\textit{e.g.}, formulations, composite geometries, or multi-component systems require additional domain-specific fields, and device-level assemblies fall outside its scope entirely).
For each paper, the framework identifies target materials, extracts all corresponding unique synthesis procedures, and generates structured JSON outputs containing the relevant information.

\subsection{Data extraction from figures}

Scientific publications frequently present quantitative results through charts and graphs that complement textual descriptions~\cite{midway2020principles,hira2024reconstructing}. To capture this information, our pipeline extracts numerical data from plots that commonly display synthesis-dependent properties such as electrical conductivity or catalytic performance. Our processing workflow consists of two stages. First, we segment multi-panel figures  into individual subplots and classify each as containing either quantitative or qualitative content, using \texttt{Florence-2} \cite{xiao2024florence}, which we fine-tuned for joint subplot segmentation and classification using low-rank adaptation (LoRA)~\cite{hu2021lora}. The model performs subplot detection and quantitative/qualitative classification in a single forward pass, enabling qualitative subplots to be filtered out efficiently. Second, we digitize the identified charts using a VLM, extracting datapoints such as $(x, y)$ coordinate pairs along with metadata including axis labels, units, titles, and series identifiers. We select the specific VLM per case study based on manual annotation and agreement with human-curated ground truth. For both the thermocatalysis and superconductivity case studies, Qwen3.5-397B-A17B\cite{Qwen3_5_2026} performed best. For details, see SI Section S5.

\subsection{Performance-to-synthesis linking}
To connect extracted plot data with the corresponding synthesis protocols, we developed an automated linking step. For each plot, an LLM (\texttt{SeriesMaterialLinker}) matches the extracted series names (\textit{e.g.}, legend entries) to the list of identified materials, considering the series labels, figure context, and axis metadata. The model returns structured mappings with confidence scores, which we validate against the known material and series lists to assess hallucinated associations. Prior to linking, we filter plots by axis characteristics to retain only performance-relevant data (\textit{e.g.}, conversion vs. temperature for catalysis), excluding spectra, images or illustrations of workflows. We assess linking quality with an LLM-as-judge that scores each mapping across four criteria and flags nine predefined failure modes: name mismatches, one-to-many synthesis, many-to-one figure, sample-code failure, precursor–product confusion, characterization confusion, dual-axis error, false negatives, and false positives. This enables scalable, traceable association of synthesis procedures with their resulting material properties and performance data.

\subsection{Validation by human domain experts}

A team of eleven human domain experts with MSc and PhD training in materials science and chemistry provided the ground truth annotations for a subset of extracted synthesis procedures. We developed an evaluation protocol that combines these expert ground truth annotations with a scalable LLM-as-a-judge framework. For this evaluation step, the experts annotated 80 synthesis procedures from 36 papers (sampled 1:1:1 across the three data sources), with the experts collectively discussing potential edge cases and ambiguous instances.
Expert annotations assess seven criteria on a 1--5 scoring scale: structural completeness, material extraction, process steps, equipment identification, condition extraction, semantic accuracy, and format compliance; we outline these criteria in the SI Section S4.

\section{Results and discussion}

\subsection{Extraction of synthesis procedures}\label{sec:results-extraction}

\begin{figure*}[t]
    \centering
    \includegraphics[width=1\linewidth]{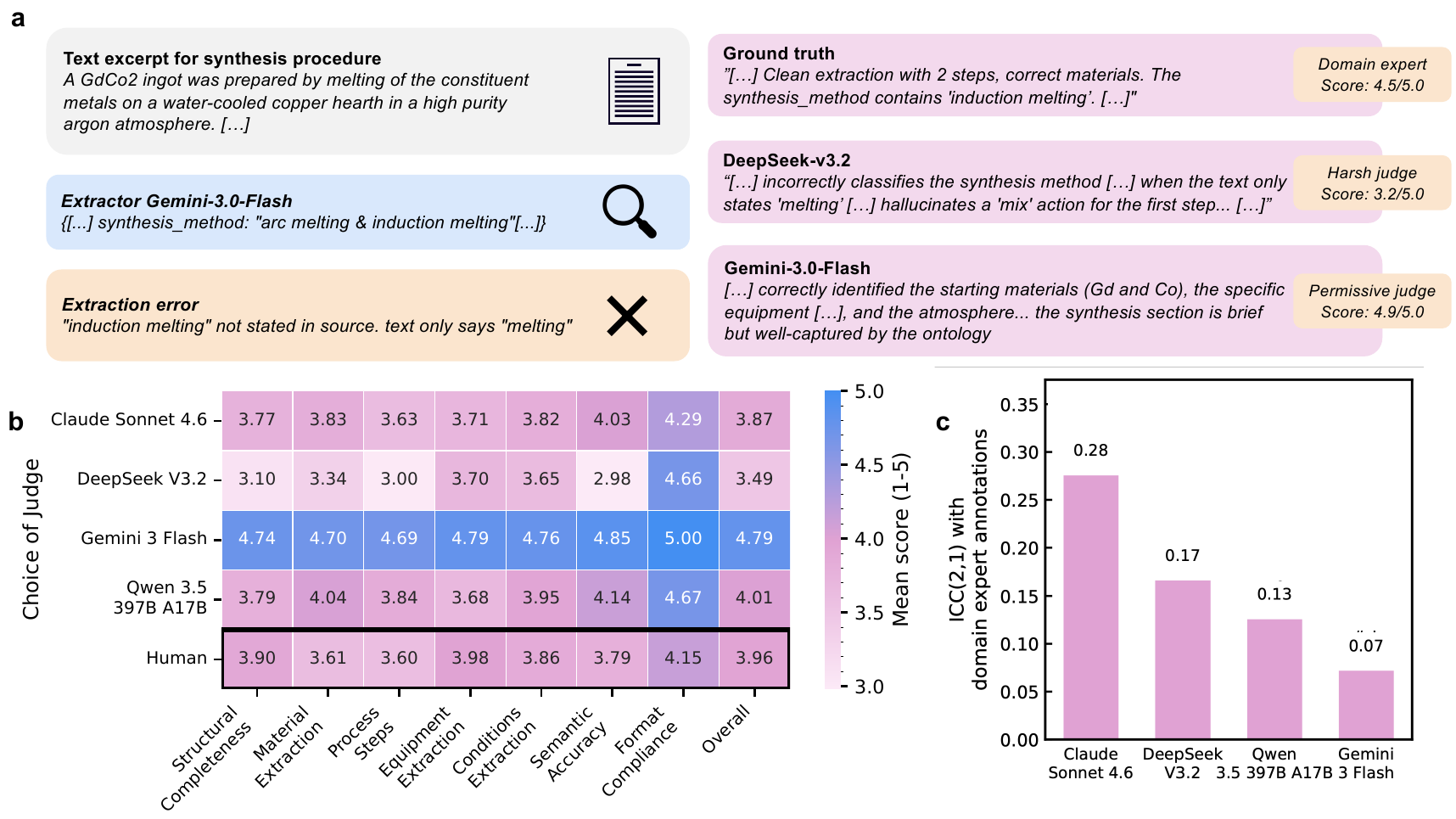}
    \caption{\textbf{Overview of the judge and extractor choice for the LLM-as-a-judge evaluation framework.} \textbf{(a)} Concrete example of judge disagreement on \ce{GdCo2} extracted by Gemini-3-Flash. The human annotator accepted the extraction (synthesis method and starting materials both match ground truth), yet the five independent graders assign overall scores spanning 1.7 points (3.2 to 4.9), with Gemini-3-Flash the most permissive and DeepSeek-V3.2 the harshest. Additional examples are provided in SI Section S4. Gemini-3-Flash is the highest scorer in all six disagreement examples checked and DeepSeek-V3.2 the lowest in five of six examples, indicating systematic differences in judge calibration. \textbf{(b)} Mean rubric score per dimension for four candidate LLM judges and domain experts. The judges disagree with each other more than they disagree with the human expert. Gemini-3-Flash scores highest on every criterion, DeepSeek-V3.2 scores below the human on all of them, and Claude-Sonnet-4.6 is close to the human overall. Per-criterion mean differences between each judge and the human are given in SI Table S6.
    \textbf{(c)} Agreement between LLM judges and humans by ICC(2,1), a two-way random-effects intraclass correlation that measures how closely a judge's absolute scores track the human's, correcting for each judge's own scale and offset. Claude-Sonnet-4.6 ranks highest (0.28), then DeepSeek-V3.2 (0.17), Qwen3.5-397B-A17B (0.13), and Gemini-3-Flash lowest (0.07). With a paper-level bootstrap standard error of 0.11--0.15 (SI Table S10) these estimates overlap substantially. The ranking is corroborated by the extractor-ranking agreement in SI Table S11.}

    \label{fig:synthesis-extraction-scores}
\end{figure*}

We benchmark four LLMs (Claude-Sonnet-4.6\cite{ClaudeSonnet46SystemCard2026}, DeepSeek-V3.2\cite{DeepSeekV32_2025}, Gemini-3-Flash\cite{Gemini3FlashModelCard2025}, Qwen3.5-397B-A17B\cite{Qwen3_5_2026}) across synthesis extraction and evaluation roles. For the extraction task, we evaluate models as both extractor (generating structured outputs) and judge (scoring extraction quality, see Fig.~\ref{fig:synthesis-extraction-scores}), yielding a $4\times4$ extractor–judge matrix.
Human experts scored 80 synthesis procedures as ground truth. Because we score each procedure once per extractor and not every extractor produces a scoreable, name-matchable output for every procedure, the extractor--judge matrix is restricted to the $n=59$ (procedure, extractor) pairs for which a human score and all four judges' scores are jointly available.
We select the judge model as the one with the highest bias-corrected agreement with human scores (ICC(2,1), leave-one-out excluding each judge's own self-scored cells) across all extractors. Consequentially, we choose the extractor model as the one achieving the highest mean score under the selected judge. We find that Claude-Sonnet-4.6 is the best-agreeing judge (ICC(2,1)$_{\text{loo}}=0.28$, Spearman correlation $\rho=0.41$) and simultaneously the top-ranked extractor (mean score 3.67) according to human judgment. These ICC estimates carry a paper-level bootstrap SE of 0.11--0.15 (SI Table S10), however, making the ordering among judges indicative. Every candidate judge model ranks Claude-Sonnet-4.6 as the best extractor. Critically, we test for self-preference bias, namely whether judge models systematically inflate scores for outputs from models in the same family, and find that Gemini-3-Flash exhibits substantial self-preference (self-scored outputs average 1.28 points higher than its scores of peer models', a difference-in-differences of $+0.44$ against the human reference), while Claude-Sonnet-4.6 and Qwen3.5-397B-A17B show negligible self-preference and DeepSeek-V3.2 is, if anything, harsher on its own outputs (SI Table S11).

By these criteria, Claude-Sonnet-4.6 is the highest-quality choice for both extraction and judging. However, deploying it across the full 81K-paper corpus is cost-prohibitive at scale. We therefore deploy the open-weight models Qwen3.5-397B-A17B as the extractor and DeepSeek-V3.2 as the judge for the full-corpus run: DeepSeek-V3.2 is the second-best-agreeing judge in our benchmark (ICC(2,1)$_{\text{loo}}=0.17$), and Qwen3.5-397B-A17B is the third-best extractor by human judgment (mean score 3.25), at a substantially lower per-token inference cost. With this benchmark, we characterize and justify the deliberate quality--cost trade-off of using open weight models for extraction.


Distribution of extraction scores across synthesis methods shows overall good performance (mean overall scores 2.9--4.0 on a 5-point scale across methods with $n>100$ papers), with well-characterized protocols like flux growth (3.97) and hydrothermal synthesis (3.88) achieving the highest scores.

\subsection{Dataset}\label{sec:results-dataset}

Applying the deployed extraction configuration (Qwen3.5-397B-A17B extractor, DeepSeek-V3.2 judge) across the full 81K-paper corpus yields \texttt{LeMat-Synth}: 58K synthesis procedures spanning 35 synthesis methods and 16 material classes (SI Fig. S2). We additionally release a curated high-quality subset of 30K procedures, filtered by judge-assigned overall score above 3.25 and more than one extracted synthesis step (Fig.~\ref{fig:dataset-stats}). We give full per-category and per-source breakdowns in the SI Section S3.1.

\subsection{Extraction of plot data}
For figure digitization on selected corpora, we evaluate three of the same models as VLMs (Claude-Sonnet-4.6, Gemini-3-Flash, Qwen3.5-397B-A17B) on an annotated set of held-out figures (per case study: \textit{e.g.}, 23 thermocatalysis papers, 152 catalyst compositions, 157 conversion-vs-temperature curves ground truth, see SI Table S21). We score a series as a true positive (TP) when the pipeline correctly matches a predicted curve to its ground-truth catalyst/support combination, a false positive (FP) if the model reports a curve with no matching ground-truth series, and a false negative (FN) when it misses a ground-truth series entirely. We compute RMSE on matched (TP) series only, with point-wise distances normalized by the ground-truth axis range of each plot (min--max scaling per figure, applied jointly to $x$- and $y$-axes). All figures quoted here are for the LLM-matched strategy (SI Table S21 also reports a fuzzy string-matching baseline). Gemini-3-Flash achieves the lowest RMSE (0.048) but the lowest recall (recall 0.375, precision 0.760, F1 0.502, 95 false negatives), missing the most true series.  Claude-Sonnet-4.6 achieves the highest recall (recall 0.704, precision 0.669, F1 0.686), and Qwen3.5-397B-A17B achieves the best overall balance (precision 0.817, recall 0.586, F1 0.682) at a substantially lower per-call cost than Claude-Sonnet-4.6. As shown in SI Table S21, we use Qwen3.5-397B-A17B as the figure-digitization VLM for the thermocatalysis case study, consistent with its selection as the extractor LLM for text-based synthesis extraction (Section~\ref{sec:results-extraction}). We prioritize Qwen3.5-397B-A17B's cost-efficiency at a small, quantified precision/recall trade-off relative to Claude-Sonnet-4.6. Full benchmarking results are provided in SI Section S7.

\subsection{Modularity of toolbox}\label{sec:casestudies}
A central design goal of \texttt{LeMat-Synth Parser} is applicability to a wide range of inorganic materials systems. Therefore, applying our pipeline to a particular scientific subdomain requires only small configurational changes to the workflow (see SI Fig. S1). Users accomplish this tuning by supplying a corpus, a \texttt{PlotFilterConfig} that informs the figure-processing stage which quantitative plots to target, and, where the default schema is insufficient, an extension of the base ontology with the entities and properties that characterize the domain. To demonstrate this, we apply the workflow to two domains of high interest to materials science: thermocatalysis and superconductors. In each case, we distill the extracted corpus into a compact ``Map-of-Science'' overview that summarizes what the literature collectively reveals about the subdomain, and offer insight regarding how this knowledge can be put to practical use.

\begin{figure*}[tp]
    \centering
    \includegraphics[width=0.8\textwidth]{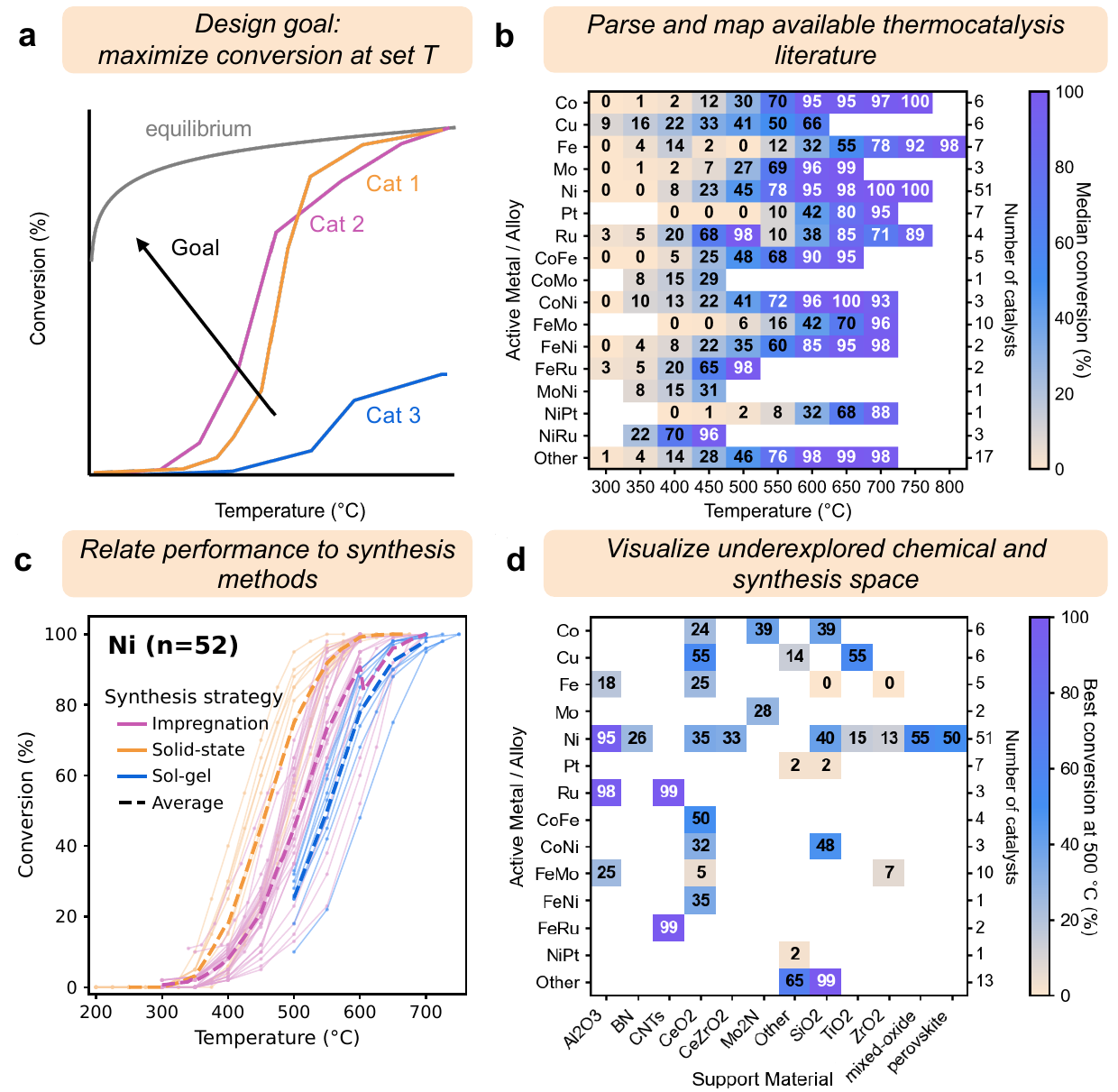}
        \caption{Overview of the thermocatalysis case study. (a) We formulate the design goal as identifying catalysts that maximize NH$_3$ conversion at a given operating temperature, moving performance toward the thermodynamic equilibrium curve\cite{Wen2025Ammonia}. (b) Conversion landscape as a function of active metal and reaction temperature, aggregated from all extracted conversion-vs-temperature curves. Color indicates median conversion (\%) at each temperature bin, revealing metal-dependent light-off behavior across the corpus. (c) Extracted conversion-vs-temperature curves for all Ni-based catalysts ($n=52$), colored by synthesis strategy (impregnation, sol-gel, solid-state). A single curve using co-precipitation as the synthesis method is omitted in this plot. Their corresponding averages over synthesis procedure are displayed in darker color, illustrating how the synthesis method relates to catalytic light-off performance for a single active metal. (d) Conversion landscape as a function of active metal and support/oxide at a fixed reference temperature (500\,\textdegree C), showing the \emph{best} (maximum) conversion recorded for each metal--support pair and exposing combinations of metal and support that are well- or under-represented in the literature. Counts in (d) reflect only curves with data at 500\,\textdegree C.}
    \label{fig:overview-thermocat}
\end{figure*}

\subsubsection{Case study: Thermocatalysis}\label{sec:casestudies-thermocat}
Heterogeneous catalysis is a domain in which the relationship between how a catalyst is made and how it performs is of substantial practical value, yet has proven difficult to map quantitatively across the literature. This is due to the wide diversity of synthesis routes, testing procedures, catalyst active phase, and support material, all of which have non-trivial impacts on catalyst performance \cite{Khan2022Recent}. Thermal catalysis performance, or activity, is conventionally reported as the percent conversion of an input molecule to a target product (or the yield of that product) as a function of temperature, and it is governed by a complex interplay of active metal, support, and synthesis procedure. The practical design objective, illustrated in Fig.~\ref{fig:overview-thermocat}a, is to reach high conversion at the lowest possible process temperature. For this case study, we focused on ammonia decomposition (also known as ammonia cracking), a thermocatalytic route to CO$_2$-free hydrogen production \cite{Khan2022Recent, Han2025Catalyst}. Ammonia cracking is a well-suited illustrative case for our pipeline because the field offers a sizable body of literature, yet remains relatively young, with much of the work conducted within the last 15 years~\cite{Wen2025Ammonia}.

To obtain the ground truth, we manually digitized every conversion-vs-temperature curve from 23 papers with WebPlotDigitizer~\cite{WebPlotDigitizer}, yielding 157 annotated curves spanning 152 distinct catalyst/support combinations~\cite{duan_mcm-412012,itoh_hydrogen_2002,noauthor_rare_2025,wu_ammonia_2020,chen_bimetallic_2020,do_turning_2024,fang_dispersed_2023,lucentini_catalytic_2019,maleki_coceo_2024,okura_ammonia_2018,podila_hydrogen_2016,teng_ru_2024,yi_plasma-assisted_2019,gao_plasma-assisted_2023,wu_engineering_2021,zhou_spatial_2021,dasireddy2021Fuel,meng_nh3_2024,podila_mgfe_2020,lorenzut_femo-based_2012,varisli_2012_cox,Srifa_2016_cox,chen_ru-based_2021}. We tie each curve to a specific catalyst and figure reference.
Mislinking of these performance curves to their synthesis records is rare. Across the Claude-Sonnet-4.6-scored papers, incorrect material-name matching (F1) and many-to-one figure mislinks (F3) each affect only a handful of papers (4 and 2 of the scored papers, respectively, SI Section S6), and one-to-many synthesis mislinks (F2) affects 2 of 26; the remaining six predefined failure modes (defined in SI Table S20) were not observed in this corpus.

Next, by aggregating the extracted data across the corpus, we generate a ``Map-of-Science'' for ammonia decomposition (Fig.~\ref{fig:overview-thermocat}b--d). Binning median conversion by active metal and reaction temperature (Fig.~\ref{fig:overview-thermocat}b) recovers several qualitative trends already established in the NH$_3$ decomposition literature~\cite{Wen2025Ammonia,Khan2022Recent}. Specifically, Ru catalysts and Ru-containing systems (\textit{e.g.}, NiRu) are the highest performing, showing nearly full conversion at the lowest temperatures ($\sim$500\,\textdegree C). Ni- and Co-based catalysts display intermediate performance, and Fe-based catalysts are notably less active. Ni in particular is heavily studied as a lower-cost replacement for Ru \cite{Su2023Review}, and much of the ammonia decomposition literature is devoted to improving the activity of Ni-based systems and to identifying other low-cost alternatives that outperform Ni.

Isolating a single active metal removes composition as a confounding variable and allows interrogation of the role of synthesis directly, although differences in support material are still present and influence performance. The corpus contains 53 Ni-based curves, of which 52 are plotted in Fig.~\ref{fig:overview-thermocat}c, where the single extracted co-precipitation curve is omitted for clarity, and 51 have data at exactly 500\,\textdegree C as shown in Fig.~\ref{fig:overview-thermocat}d. Performance ranges over several hundred \textdegree C and correlates strongly with preparation route. Solid-state synthesis tends to produce samples that activate at lower temperatures than their impregnation and sol-gel counterparts, though support identity and other unreported experimental conditions remain potential confounders. The relative underperformance of the sol-gel method on average is rather surprising, as the sol-gel method generally allows more fine control of the synthesis \cite{Rad2016Sol}. However, in ammonia cracking catalysis, the sol-gel method is typically used only for the synthesis of the support directly, rather than forming the entire supported catalyst \cite{Wang2019Ru, Chen2026High}. The catalyst is then placed on the sol-gel synthesized support using, often, the impregnation method. Importantly, this is still classified as impregnation synthesis by \texttt{LeMat-Synth Parser}. Therefore, the sol-gel synthesis examples shown in Fig.~\ref{fig:overview-thermocat}c reflect instances where the entire catalyst (active phase and support) were created using the sol-gel method. This is a relatively complicated synthesis route, and is not commonly employed in ammonia decomposition synthesis (there are only three publications in the curated corpus). Therefore, it is likely this field is not as mature as impregnation synthesis, and some of the success of the reported impregnation samples are actually involving supports made using sol-gel synthesis. 

The high performance of solid-state synthesis is more clear-cut. This can be understood from the fact that solid-state synthesis can produce smaller and dispersed particles on the support, with more nanoscale control of the active phase, than impregnation synthesis, on average \cite{yang2024mechanochemical, amrute2021mechanochemical}. This has been demonstrated directly in ammonia decomposition catalysis \cite{Lucentini2024Ni}. Specifically, solid-state synthesis can allow the user to skip the calcination step often required by the impregnation method, which can produce large aggregates, especially when high temperatures are needed for calcination \cite{Lucentini2024Ni}. Higher surface area to volume ratio and less aggregation are both commonly associated with higher catalytic performance, other things equal, leading to increased performance of solid-state synthesis. 


That such a spread persists at fixed active metal highlights the central premise of this work that a complicated interplay of synthesis and composition governs performance. This demonstrates our workflow's ability to elucidate synthesis-structure-property relationships from dozens of papers that would be impractical to assemble by hand.

Examining active metal versus support at a fixed reference temperature of 500\,\textdegree C (Fig.~\ref{fig:overview-thermocat}d) turns the same corpus into a map of what has, and has not, been studied. A handful of combinations dominate the reported literature and achieve high conversion (Ru/Al$_2$O$_3$, Ru/carbon nanotubes (CNTs), and Ni/Al$_2$O$_3$ all reach or exceed 95\% at 500\,\textdegree C), while most of the metal-support matrix is empty, exposing a large space of untested combinations. This directly serves the design objective of Fig.~\ref{fig:overview-thermocat}a by identifying the current best-performing systems and the regions most in need of further experimental data. Ni has been studied across many supports, consistent with its role as the leading low-cost candidate for replacing Ru, and the same view highlights supports that are well characterized for Ru but not yet represented for Ni, \textit{e.g.}, CNTs. Together with the synthesis-strategy breakdown above, this points to solid-state and impregnation routes as promising starting points for new Ni-based catalyst development.

\begin{figure*}[tp]
\centering
\includegraphics[width=\textwidth]{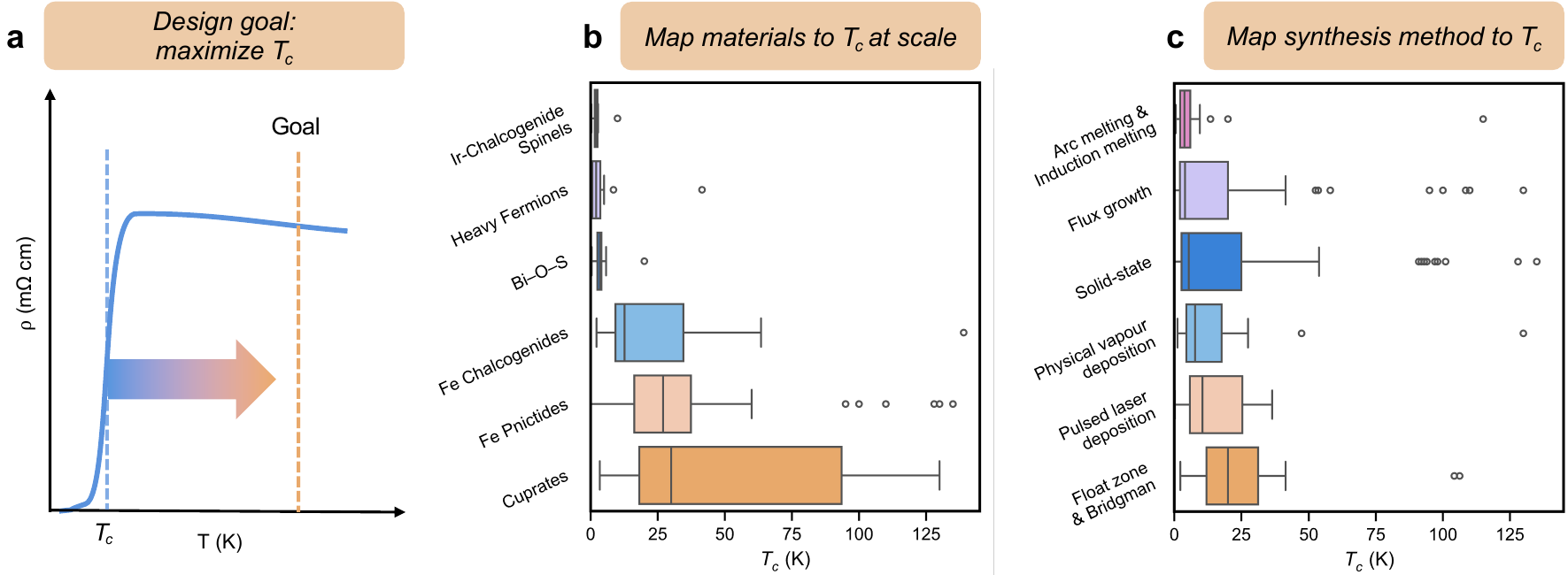}
\caption{\textbf{Overview of the superconductivity case study, extracting critical transition temperatures ($T_c$) from literature at scale.} \textbf{(a)} $T_c$ is defined as the temperature at which resistivity $\rho$ drops to zero during the superconducting transition. We formulate the design goal as identifying synthesis routes that maximize $T_c$. \textbf{(b)} Distribution of extracted $T_c$ for the six most frequent material families and \textbf{(c)} the six most frequent synthesis methods, computed on the filtered corpus with rows with an overall judge score greater than four (SI Section S9). Full plots covering all families and methods are provided in SI Fig. S14 and SI Fig. S13. The parity between human text-reported and VLM figure-extracted $T_c$ that validates this pipeline ($R^2=0.94$, MAE$=2.7$~K, $n=48$) is shown in SI Fig. S12.}\label{fig:overview-supercond}
\end{figure*}

\subsubsection{Case study: Superconductivity}

Superconductivity, the complete loss of electrical resistance below a critical temperature $T_c$ (Fig.~\ref{fig:overview-supercond}a), underpins technologies from MRI magnets to fusion confinement and quantum hardware, all of which are constrained by the cost of the cryogenic cooling they require. The design goal is to maximize $T_c$ at ambient pressure, beyond the 77 K boiling point of liquid N$_2$ and ultimately towards room temperature.
A century of discovery has been driven by empirical exploration, and $T_c$ has been demonstrated to be sensitive to composition, doping, and synthesis conditions~\cite{keimerQuantumMatterHightemperature2015}.
Existing $T_c$ compilations, however, are manually curated from the literature, and the field's reporting conventions introduce biases.
Authors typically discuss one or two materials in the text, usually the best-performing sample in a series, but the accompanying resistivity-vs-temperature plot often shows the entire series of measured compositions, many of whose $T_c$ values are never written out numerically anywhere in the paper.
A text-only pipeline would therefore recover a small, biased sample of the field's actual data.
The omitted points are disproportionately low-performing samples, and without them a downstream model cannot tell an unexplored region of composition space from one that was already tried and found unsuccessful, risking redundant exploration.
Combining both data sources lets us use the material each paper describes in text as a built-in ground truth for the figure-extraction pipeline, and then apply that validated pipeline to recover every other curve in the same plot that the text never reports. 
The omitted points are disproportionately low-performing samples, and without them a downstream model cannot tell an unexplored region of composition space from one that was already tried and found unsuccessful, risking redundant exploration.

We build this case study's corpus by starting from the full \texttt{LeMat-Synth-Papers} corpus, restricting to papers tagged as superconductor-related whose abstract mentions resistivity, then using an LLM to confirm that each remaining paper contains a resistance/resistivity-vs-temperature plot (excluding plots that sweep only magnetic field or pressure rather than composition).
This yields 1,384 papers. For each, a figure pipeline has a VLM read $T_{\text{onset}}$, $T_c$, and $T_{\text{zero}}$ off every $\rho(T)$ curve, scanning from high to low temperature and preferring a zoomed inset over the main panel when both are shown.
The main prompting challenge in the figure pipeline is telling a genuine superconducting transition (a sharp, near-vertical drop over a few K) apart from the gradual resistance decrease of ordinary metallic or Kondo-coherent behavior, which we address by explicitly warning against the latter in the extraction prompt (SI Section S10).

On a 40-paper manually annotated subset, the $T_c$ values our figure pipeline reads off the plots agree closely with the values the same papers report in text and a domain expert transcribed($R^2=0.94$, MAE$=2.7$~K, $n=48$ matched material rows, SI Fig. S12), confirming that the figure pipeline reads curves reliably where a text value exists for validation. We then apply the validated figure pipeline across the full corpus, giving material-resolved $T_c$ values, including for compositions and doping levels that never appear as numbers in the source text. Fig.~\ref{fig:overview-supercond}b and~c show the extracted $T_c$ distribution for the most frequent material families and synthesis methods in the dataset.
To keep this view to reliably extracted records, both panels are computed on the subset of papers whose judge \texttt{overall\_score} exceeds 4 ($n=543$ rows; SI Section S9). This scale gives broad coverage of the major superconductor families studied over the past two decades, including conventional low-$T_c$ superconductors, cuprates, and iron-based systems (see SI Fig. S15), while adding per-composition detail within each series, \textit{e.g.}, the full $T_c$-vs-$x$ trend across a Ca(Fe,Co)AsF-type substitution series, that individual papers only ever report for their single best composition.

Manual review of extractions flagged as implausibly high ($T_c \geq 150$~K, a regime essentially restricted to hydride superconductors under extreme pressure) surfaces the pipeline's main failure mode of misreading the composition axis of a $T_c$-vs-doping phase diagram as a temperature axis, or misreading a structural or charge-density-wave resistivity anomaly as a superconducting transition. This affects only 2.6\% of extracted $T_c$ rows (14 of 543 samples), of which all but one, a high-pressure hydride, are misreads (see SI Table S22).
Nonetheless, this failure mode is rare and flagged by its implausible threshold, and the resulting corpus reproduces the field's well-established historical progression toward higher-$T_c$ material families over time (SI Fig. S15), demonstrating the pipeline's utility for large-scale, literature-wide analysis of synthesis-property trends.

\subsection{Limitations}
We showcase the successful implementation of \texttt{LeMat-Synth Parser} for materials synthesis procedures and performance metric extraction, but highlight several limitations of the current workflow. The full-corpus \texttt{LeMat-Synth} run relies on the quality and coverage of open-access literature, which excludes closed-access publications and may bias domain coverage, though the thermocatalysis case study, drawn from a small expert-curated closed-access set, shows the pipeline is not restricted to open-access sources. Extractions are incomplete when synthesis procedures are described implicitly (\textit{i.e.}, multiple materials are synthesized the same way, but only one is given in the text) rather than as discrete, itemized steps.
The plot-filtering and linking stages do not currently account for confounding experimental variables (\textit{e.g.}, pressure, flow rate, gas composition) that affect reported performance independently of synthesis and composition. Future work could incorporate these as additional fields in the domain-specific \texttt{PlotFilterConfig} and ontology.
Chemical entity and material-identifier standardization across records is not yet universal, which complicates cross-paper aggregation; we expect this to be a substantially larger obstacle for materials domains where identity is not a single well-defined composition, such as composites, multi-component systems, or formulations with variable ratios. While we expect \texttt{LeMat-Synth Parser} to be applicable to a wide range of inorganic materials systems, it will not cover all of them equally well, and this identifier-matching gap is the clearest current limit on that coverage.

Finally, the pipeline verifies whether an extraction faithfully reflects the source document, not whether the reported procedure or measurement is in itself correct. This distinction matters for a corpus drawn largely from preprint servers, where results have not undergone peer review, where some subfields have a documented history of contested claims, and, while no direct analysis of LLM-generated studies has been performed in the materials domain, in fields like biology and public policy the growing volume of fully or partially AI-generated manuscripts raises the risk of extracting procedures that were never performed.\cite{suchak2025explosion, haider2024gpt}
However, because each record retains its provenance (source publication and date), users can restrict analyses to the subsets of the literature they trust, or utilize our pipeline on their own curated subset of papers for exploration of a specific materials science domain.


More broadly, while \texttt{LeMat-Synth Parser} excels at scale, extracting highly specific synthesis conditions and reliably tying them to individual performance measurements remains difficult for current frontier LLMs and VLMs. Nevertheless, the ever-growing scale of the open-access corpus means future work will need to address remaining gaps. We hope the modularity of our open-source toolbox makes it a practical foundation for scaling synthesis and performance extraction to new domains not yet explored via systematic data mining.

Beyond these limitations, we see \texttt{LeMat-Synth} as a foundation for downstream tasks that go beyond extraction itself.
The structured synthesis-property records could enable training and evaluation of models for synthesis planning, including precursor recommendation~\cite{he2023precursor} and joint precursor and reaction-condition generation~\cite{pan2026diffsyn}. Coupled with \texttt{LeMat-Synth Parser}'s ability to extract literature-wide synthesis-property trends at scale, this positions \texttt{LeMat-Synth} to support end-to-end materials discovery workflows that move from literature mining to the design of new synthesis routes.

\section{Conclusion}
We present both \texttt{LeMat-Synth Parser}, a multi-modal toolbox that combines LLMs and VLMs to extract structured synthesis procedures and performance data from scientific literature, and \texttt{LeMat-Synth}, a dataset containing 58K structured synthesis procedures (30K in a curated high-quality subset) extracted from 81K open-access materials science papers. \texttt{LeMat-Synth} covers 35 synthesis methods and 16 material classes. We validate \texttt{LeMat-Synth Parser} through domain expert annotations and a scalable LLM-as-a-judge evaluation, and benchmark it across four candidate LLMs and three VLMs to characterize extraction quality, judging reliability, and cross-model bias.

We demonstrate the toolbox's applicability across two scientifically distinct domains. In heterogeneous catalysis, we map synthesis-structure-performance relationships across a manually annotated corpus of 23 papers, where we manually validated the extractions, establishing quantitative connections between synthesis protocols and catalytic performance and exposing metal-support combinations that remain untested. In superconductivity, we validate text- and figure-extracted critical temperatures against expert-transcribed text values on a 40-paper subset ($R^2=0.94$, $n=48$), then apply that pipeline across 1,384 papers to recover $T_c$ values for compositions that the source papers never report numerically. Downstream predictive models will need this data in order to distinguish unexplored composition space from compositions already tried and found unsuccessful.

By releasing both \texttt{LeMat-Synth} and \texttt{LeMat-Synth Parser} openly, we provide extensible infrastructure for transforming unstructured literature into machine-readable synthesis knowledge, allowing the possibility to model synthesis-structure-property relationships and support AI-driven materials discovery. We release the code, the extracted synthesis procedures and performance metrics, and the evaluation tools openly to accelerate the progress of data extraction in materials science.
Future work includes standardization of chemical entity identifiers across synthesis records, expansion to closed-access literature, development of screening tools to flag AI-generated or fabricated manuscripts before they enter a synthesis corpus, and development of synthesis prediction models trained on \texttt{LeMat-Synth}.

Unstructured text has long bottlenecked materials discovery due to inaccessible structured synthesis procedures. By turning decades of synthesis literature into a structured, queryable resource, and by providing an open, modular toolbox for extending this approach to new domains, \texttt{LeMat-Synth} and \texttt{LeMat-Synth Parser} contribute to data-driven materials discovery workflows that learn from the entirety of the materials science literature. Via \texttt{LeMat-Synth} and \texttt{LeMat-Synth Parser}, we determine quantitative insights that were previously impractical to extract at scale, and demonstrate how these can guide the next generation of materials development. 

\section*{Author contributions}
\textbf{M.L.}: Conceptualization, Methodology, Software, Investigation, Visualization, Writing – original draft. \textbf{S.B.}: Data curation, Conceptualization, Methodology, Software, Validation, Formal analysis, Investigation. \textbf{V.G.}: Data curation, Formal analysis, Investigation, Methodology, Software, Validation, Visualization. \textbf{A.L.}: Data curation, Conceptualization, Validation, Supervision \textbf{A.S.}: Data curation, Software, Validation. \textbf{F.F.}: Software, Data curation, Formal analysis, Visualization. \textbf{A.J.}: Formal analysis, Investigation, Methodology. \textbf{A.A.}: Data curation, Investigation. \textbf{N.Y.}: Software, Investigation, Validation. \textbf{X.L.}: Conceptualization, Formal analysis, Methodology, Software, Visualization. \textbf{G.G.}: Software. \textbf{S.R.}: Software. \textbf{S.P.S.}: Validation, Data curation, Investigation, Formal analysis. \textbf{A.N.}: Data curation. \textbf{A.K.}: Data curation, Software. \textbf{S.K.E.}: Data curation, Investigation. \textbf{M.Z.}: Data curation, Investigation. \textbf{E.P.}: Conceptualization, Visualization. \textbf{G.C.}: Data curation, Software, Methodology. \textbf{C.W.C.}: Supervision, Writing -- review and editing. \textbf{P.S.}: Supervision, Writing -- review and editing. \textbf{R.M.}: Supervision, Writing -- review and editing. \textbf{A.D.}: Funding acquisition. \textbf{M.L.D.F.}: Supervision, Data curation, Validation, Visualization, Writing – original draft. \textbf{S.P.G.}: Conceptualization, Methodology, Supervision, Writing – original draft. 

All authors contributed to reviewing and editing of the manuscript.

\section*{Conflicts of interest}
The authors declare no conflicts of interest.


\section*{Data availability}
The \texttt{LeMat-Synth} dataset is available on Hugging Face at \url{https://huggingface.co/datasets/LeMaterial/LeMat-Synth}, and the underlying paper corpus, including raw text, at \url{https://huggingface.co/datasets/LeMaterial/LeMat-Synth-Papers}. Source licence information is retained as a per-paper field in this corpus, and papers were filtered accordingly prior to release (open access). The \texttt{LeMat-Synth Parser} source code is available on GitHub at \url{https://github.com/LeMaterial/lematerial-llm-synthesis} under an Apache-2.0 licence. The fine-tuned Florence-2 image-segmentation model is available at \url{https://huggingface.co/amayuelas/plot-visualization-florence-2-lora-32}. The manual annotations of synthesis procedures and plots, the image-segmentation fine-tuning dataset, and open data from the case study are deposited on Zenodo upon publication. Because the thermocatalysis case study draws on closed-access journal articles, the data is available from the authors on request.


\section*{Acknowledgements}

The authors thank 
Ilya Mazalov, Jeremy Goumaz, Matthew Hart, Vasily Safronov, Thorben Prein, N. M. Anoop Krishnan, Dai Haiwen, Defne Circi, Kedar Hippalgaonkar, Luis Pinto, Sam Blouir for insightful discussions within the LeMaterial community. The authors thank Maroua Sehaba for support with data annotation for finetuning our image segmentation model.



\bibliography{LeMat-Synth} 
\bibliographystyle{rsc} 
\end{document}


\renewcommand{\thefigure}{S\arabic{figure}}
\renewcommand{\thetable}{S\arabic{table}}
\renewcommand{\theequation}{S\arabic{equation}}
\renewcommand{\thepage}{S\arabic{page}}
\setcounter{figure}{0}
\setcounter{table}{0}
\setcounter{equation}{0}
\setcounter{page}{1} 

\begin{center}
\section*{Supplementary Materials for\\ \scititle}

Magdalena~Lederbauer$^{\ast}$,
Siddharth~Betala,
Valerie~Gentzke,
Anamaria~Leonescu,
Amine~Sehaba,
Faris~Flaifil,
Ayush~Jain,
Alfonso~Amayuelas,
Nikhil~Yelamarthy,
Xiyao~Li,
Gr{'e}goire~Germain,
Stefano~Ribes,
Stefan~P.~Schmid,
Alexandre~Nozadze,
Anna~Kelmanson,
Sudheesh~Kumar~Ethirajan,
Mohd~Zaki,
Elton~Pan,
Georgia~Channing,
Connor~W.~Coley,
Philippe~Schwaller,
Roc{\'i}o~Mercado,
Alexandre~Duval,
Mathilde~L.~D.~Franckel$^{\ast}$,
Samuel~P.~Gleason$^{\ast}$\\
\small$^\ast$Corresponding author. Email: magled@mit.edu, m.franckel25@imperial.ac.uk, samuel.gleason@entalpic.ai
\end{center}
\vspace{-2cm} 
\tableofcontents
\newpage 



\subsection{Framework design \& generalization}

\begin{figure}
    \centering
    \includegraphics[width=\linewidth]{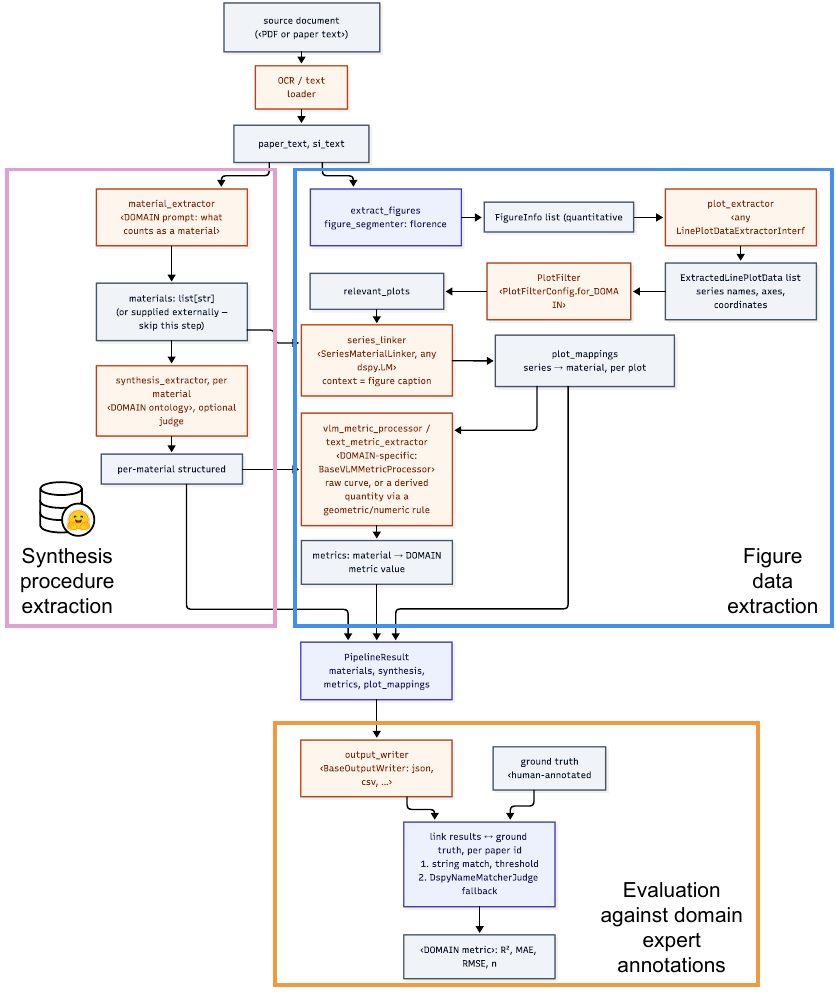}
    \caption{Overview of the pipeline implemented by \texttt{LeMat-Synth Parser}. The resulting database, \texttt{LeMat-Synth}, comprises extracted synthesis procedures from 81K open access papers. The synthesis procedures can be coupled to performance data from figures as illustrated in Sections \ref{app:casestudy-thermocat} and \ref{app:sec:supercond}.}
    \label{fig:app-pipeline}
\end{figure}

\texttt{LeMat-Synth Parser} is organized around four swappable components shown in Figure \ref{fig:app-pipeline}: a corpus loader (\texttt{PaperLoaderInterface}, specifying how publications are fetched and parsed from a given source), a domain ontology (\texttt{GeneralSynthesisOntology}, a typed Pydantic schema for the structured synthesis output), a \texttt{PlotFilterConfig} (specifying which figure axis labels/units indicate a domain-relevant plot), and an evaluation rubric (the LLM-as-a-judge prompt and scoring criteria, Section~\ref{app:subsec:prompts}). Re-instantiating the pipeline for a new domain means adapting these four components.

\paragraph{Adaptation steps}
\label{app:sec:generalization}
\begin{enumerate}
  \item \textbf{Define the corpus.} Choose a source (arXiv, ChemRxiv, a custom PDF folder, \ldots) and implement or reuse a \texttt{PaperLoaderInterface} for it; specify the keyword filter and/or LLM classifier prompt used to retain only relevant papers (Section~\ref{app:sec:data-acquisition}).
  \item \textbf{Configure \texttt{PlotFilterConfig}.} Specify which x/y-axis labels and units identify a domain-relevant plot. \texttt{PlotFilterConfig} ships four factory presets: \texttt{for\_catalysis()} (conversion/yield/activity vs.\ temperature, the default), \texttt{for\_superconductivity()} ($\rho$/R vs.\ temperature, with exclusion patterns for derivative and difference plots such as $d\rho/dT$ or $\rho-\rho_0$), \texttt{for\_electrochemistry()} (current/capacitance vs.\ potential), and \texttt{for\_coverage()} (adsorption/uptake vs.\ pressure, for porous-materials isotherms), and a fifth, \texttt{no\_filter()}, that disables axis filtering entirely. A new domain either selects one of these or constructs a custom \texttt{PlotFilterConfig} directly, e.g.\ for electrochemistry: \texttt{x\_axis\_labels=["potential", "voltage"]}, \texttt{x\_axis\_units=["V", "mV"]}, \texttt{y\_axis\_keywords=["current", "capacitance"]}.
  \item \textbf{Extend the ontology, if needed.} \texttt{GeneralSynthesisOntology} is designed to be subclassable with additional domain-specific fields. In practice, both case studies presented in this work use \texttt{GeneralSynthesisOntology} directly, without subclassing. Its existing fields (target compound, synthesis method, starting materials, process steps, equipment, conditions) were sufficient for both domains. Subclassing remains available for domains whose relevant synthesis parameters are not already covered (e.g.\ electrochemical deposition potential, applied field during growth).
  \item \textbf{Adapt the judge prompt.} Modify the evaluation-rubric YAML (Section~\ref{app:subsec:prompts}) to reflect domain-specific evaluation criteria, if the default seven-criterion rubric (structural completeness, material extraction, process steps, equipment, conditions, semantic accuracy, format compliance) does not adequately capture domain-specific correctness.
  \item \textbf{Validate.} Run the pipeline on a small manually annotated sample and compare LLM-judge scores against expert annotations (Section~\ref{app:sec:synth-extr-eval-llm-human}) before scaling to the full corpus.
\end{enumerate}

Corpus-specific keywords (Step 1), any new ontology fields and their allowed-value lists (Step 3, only if the default ontology is insufficient), and optionally an annotated validation sample (Step 5) are not automated by \texttt{LeMat-Synth Parser} and require domain expertise per new instantiation. Table~\ref{table:domain-recipe} summarizes the three domains instantiated in this work.

\begin{table}[h]
  \centering
 \caption{Comparison across the three domains instantiated in this work. The broad materials science run that produces the \texttt{LeMat-Synth} dataset is the default configuration. Thermocatalysis and superconductors are re-instantiations via \texttt{PlotFilterConfig} alone, without ontology subclassing or judge-rubric changes.}
  \label{table:domain-recipe}
  \resizebox{\linewidth}{!}{%
  \begin{tabular}{lccc}
    \toprule
    \textbf{Component} & \textbf{LeMat-Synth} & \textbf{Thermocatalysis} & \textbf{Superconductors} \\
    \midrule
    Corpus source & arXiv, ChemRxiv, OMG24 & Domain-expert papers & LeMat-Synth subset \\
    & & & (filtered by keywords and LLM) \\
    Number of papers & 80,940 & 23 & 1,384 \\
    Number of synthesis procedures & 58,345 & 157 & 2,794 \\
    Ontology class & \texttt{GeneralSynthesisOntology} & \texttt{GeneralSynthesisOntology} & \texttt{GeneralSynthesisOntology} \\
    \texttt{PlotFilterConfig} & \texttt{-} & \texttt{for\_catalysis()} & \texttt{for\_superconductivity()} \\
    \bottomrule
  \end{tabular}
  }
\end{table}

\subsection{Data}
\label{app:sec:data}

This section outlines details for the dataset (Section~\ref{app:sec:dataset-stats}) as well as the data curation (\ref{app:sec:data-acquisition}) presented in this work.

\subsubsection{Dataset statistics}\label{app:sec:dataset-stats}

We classify each material into a set of predetermined material categories and synthesis methods, as determined by the recommendation of domain experts on our team.

\paragraph{Material categories} With the goal of covering practically the entire space of material science synthesis, the following material categories were chosen by domain experts of our group and are employed in this work: metals \& alloys, ceramics \& glasses, polymers \& soft matter, composites, semiconductors \& electronic, nanomaterials, two-dimensional materials, framework \& porous materials, biomaterials \& biological, liquid materials, hybrid \& organic-inorganic, functional materials \& catalysts, energy \& sustainability, smart \& responsive materials, emerging \& quantum materials. Any category not covered in the list is assigned the label ``other''.

\paragraph{Synthesis methods} Similarly, the following synthesis methods were chosen by domain experts of our group and are employed in this work: PVD, CVD, arc discharge, ball milling, 
spray pyrolysis, 
electrospinning, 
sol-gel, 
hydrothermal, 
solvothermal, 
precipitation, 
coprecipitation, 
combustion, 
microwave-assisted, 
sonochemical, 
template-directed, 
solid-state, 
flux growth, 
float zone \& Bridgman, 
arc melting \& induction melting, 
spark plasma sintering, 
electrochemical deposition, 
chemical bath deposition, 
liquid-phase epitaxy, 
self-assembly, 
atomic layer deposition, 
molecular beam epitaxy, 
pulsed laser deposition, 
ion implantation, 
lithographic patterning, 
wet impregnation, 
incipient wetness impregnation, 
mechanical mixing, 
solution-based, 
mechanochemical.
Any category not covered in the list is assigned the label ``other''.



\begin{figure}
  \centering
  \begin{subfigure}[c]{0.4\linewidth}
    \includegraphics[angle=-90,width=\linewidth]{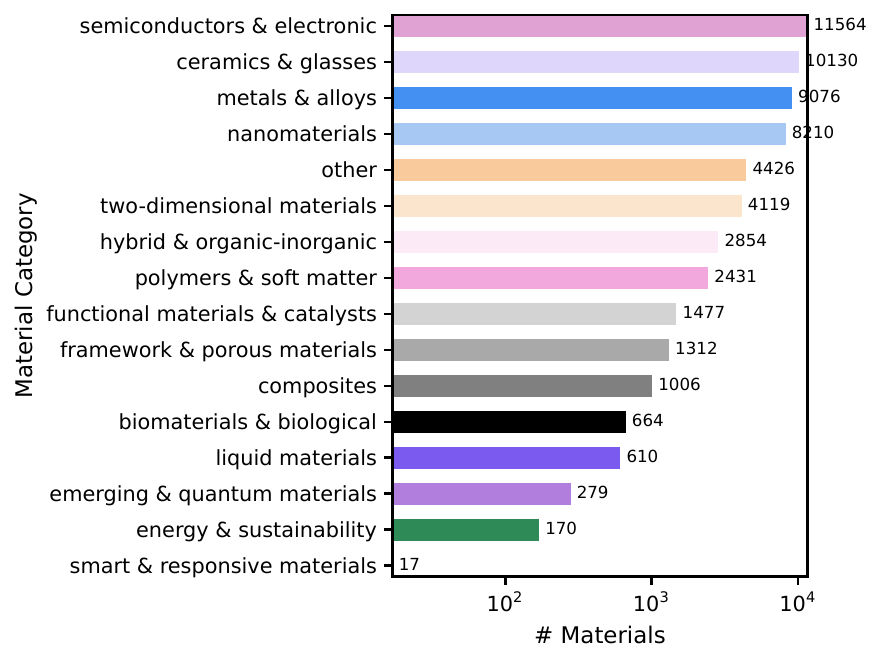}
    \caption{}
  \end{subfigure}\\
  \begin{subfigure}[c]{0.8\linewidth}
    \includegraphics[angle=-90,width=\linewidth]{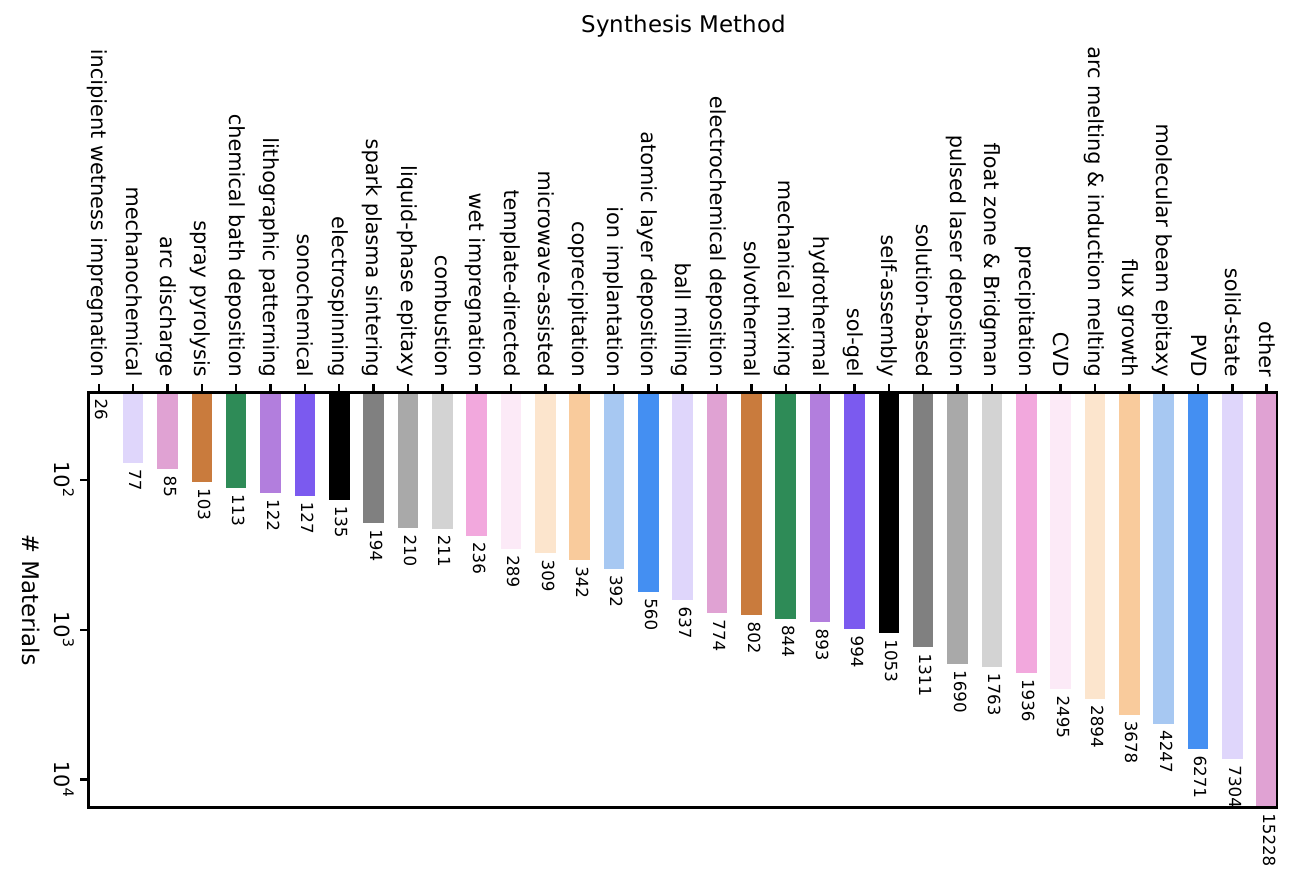}
    \caption{}
  \end{subfigure}
  \caption{Full breakdown of the \texttt{full} config (58.3K procedures). \textbf{(a)} All 16 material categories. \textbf{(b)} All 35 synthesis methods.}
  \label{fig:app-dataset-stats-overall}
\end{figure}
\begin{figure}
  \centering
  \begin{subfigure}[c]{0.4\linewidth}
    \includegraphics[angle=-90,width=\linewidth]{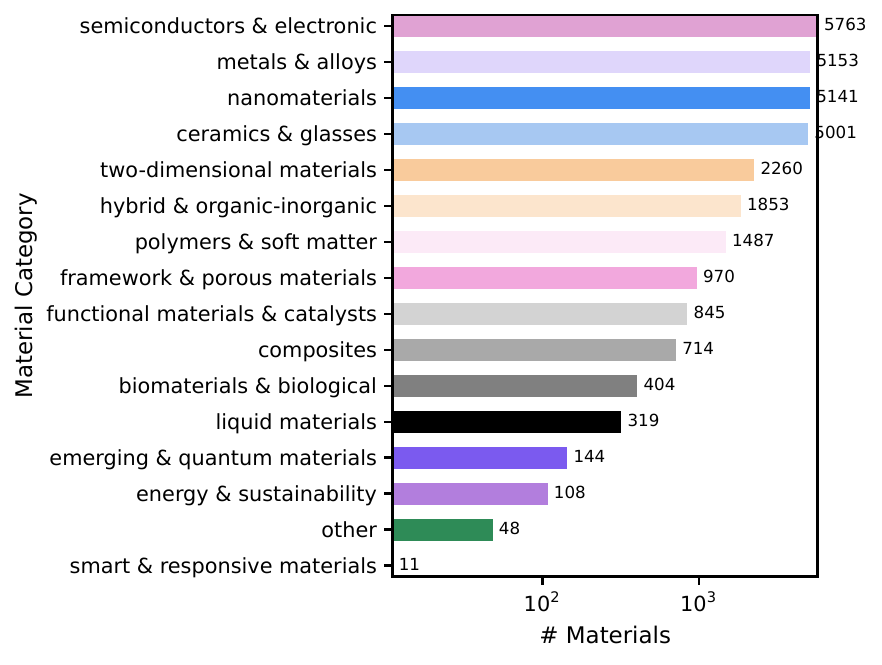}
    \caption{}
  \end{subfigure}\\
  \begin{subfigure}[c]{0.8\linewidth}
    \includegraphics[angle=-90,width=\linewidth]{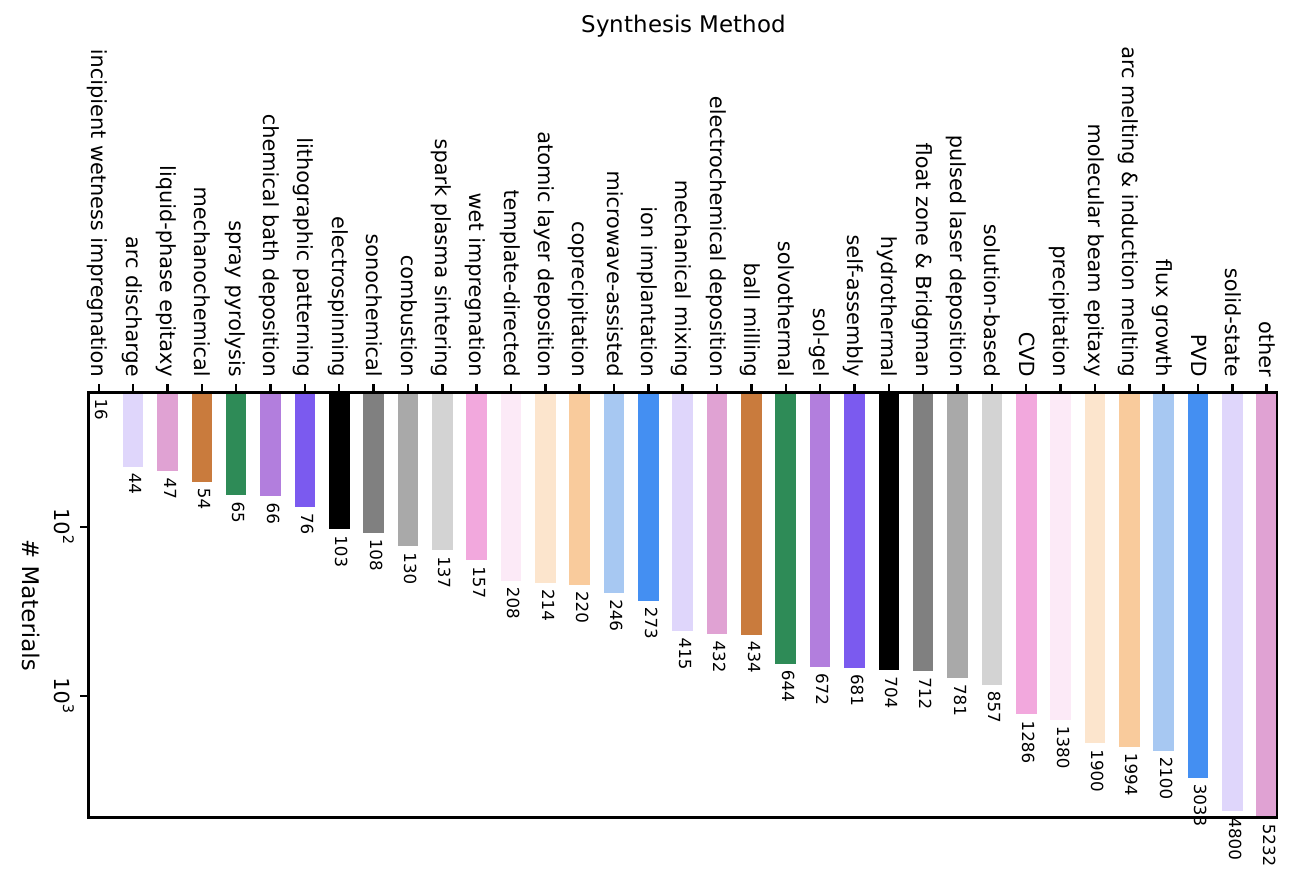}
    \caption{}
  \end{subfigure}
  \caption{Full breakdown of the \texttt{high\_score} config (30.2K procedures). \textbf{(a)} All 16 material categories. \textbf{(b)} All 35 synthesis methods.}
  \label{fig:app-dataset-stats-highscore}
\end{figure}
\begin{figure}
  \centering
  \includegraphics[width=0.7\linewidth]{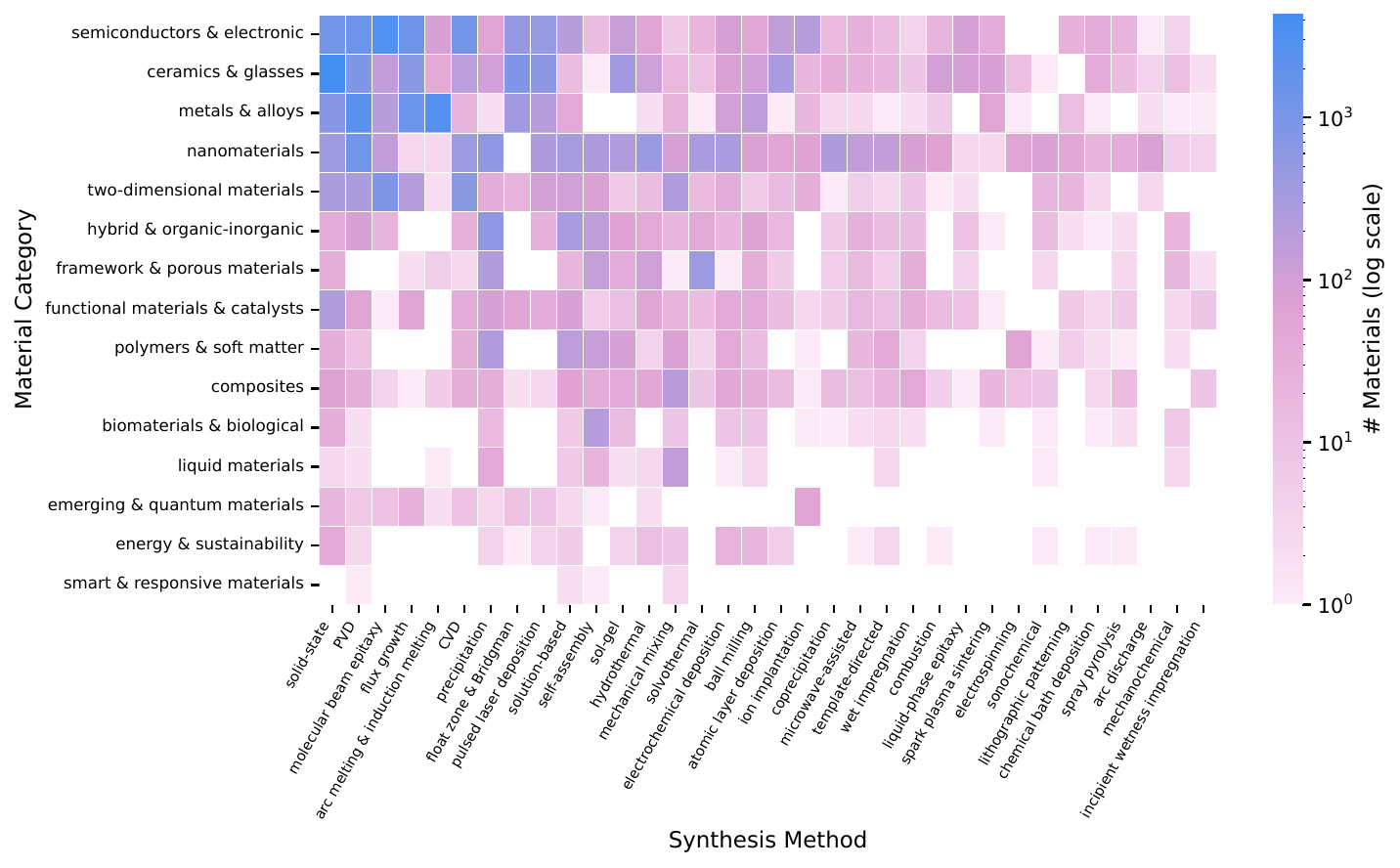}
  \caption{Coverage of the material-class $\times$ synthesis-method space for the \texttt{full} config (58.3K procedures). Each point is a (class, method) combination observed at least once in the data. Point size is proportional to procedure count.}
  \label{fig:app-dataset-diversity-overall}
\end{figure}
\begin{figure}
  \centering
  \includegraphics[width=0.7\linewidth]{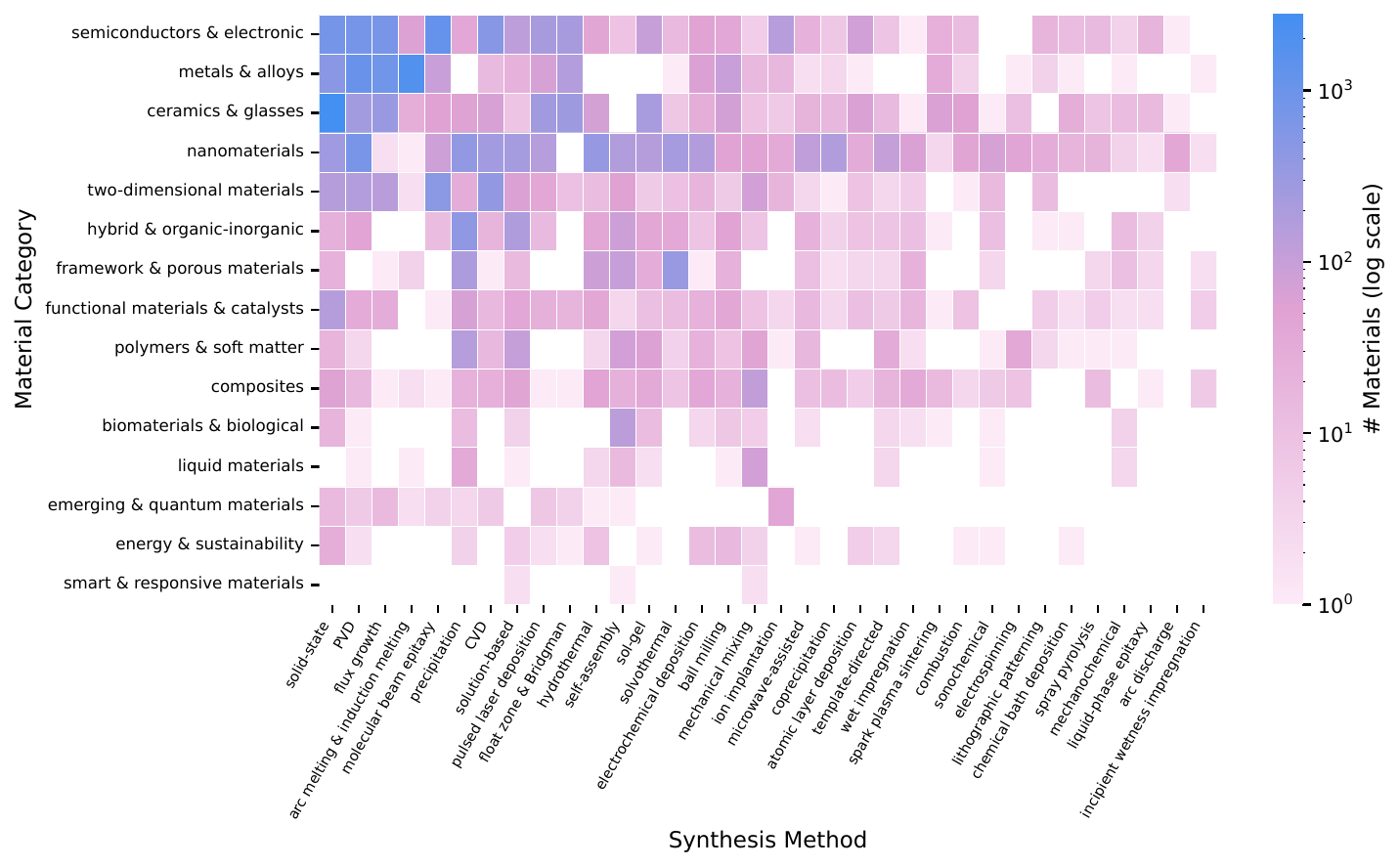}
  \caption{Coverage of the material-class $\times$ synthesis-method space for the \texttt{high\_score} config (30.2K procedures). Each point is a (class, method) combination observed at least once in the data. Point size is proportional to procedure count.}
  \label{fig:app-dataset-diversity-highscore}
\end{figure}
\begin{figure}
  \centering
  \includegraphics[width=0.6\linewidth]{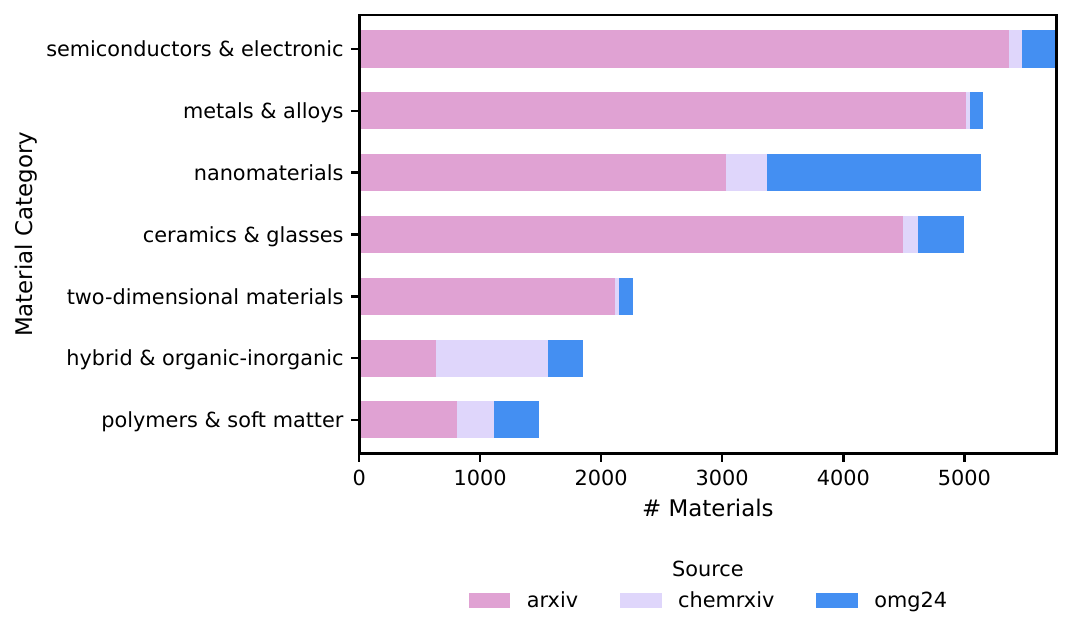}
  \caption{Composition of the \texttt{high\_score} config (30.2K procedures), stacked by source corpus / material category. Top-7 material categories, stacked by source corpus.}
  \label{fig:app-dataset-composition-highscore-category}
\end{figure}
\begin{figure}
  \centering
  
    \includegraphics[width=0.7\linewidth]{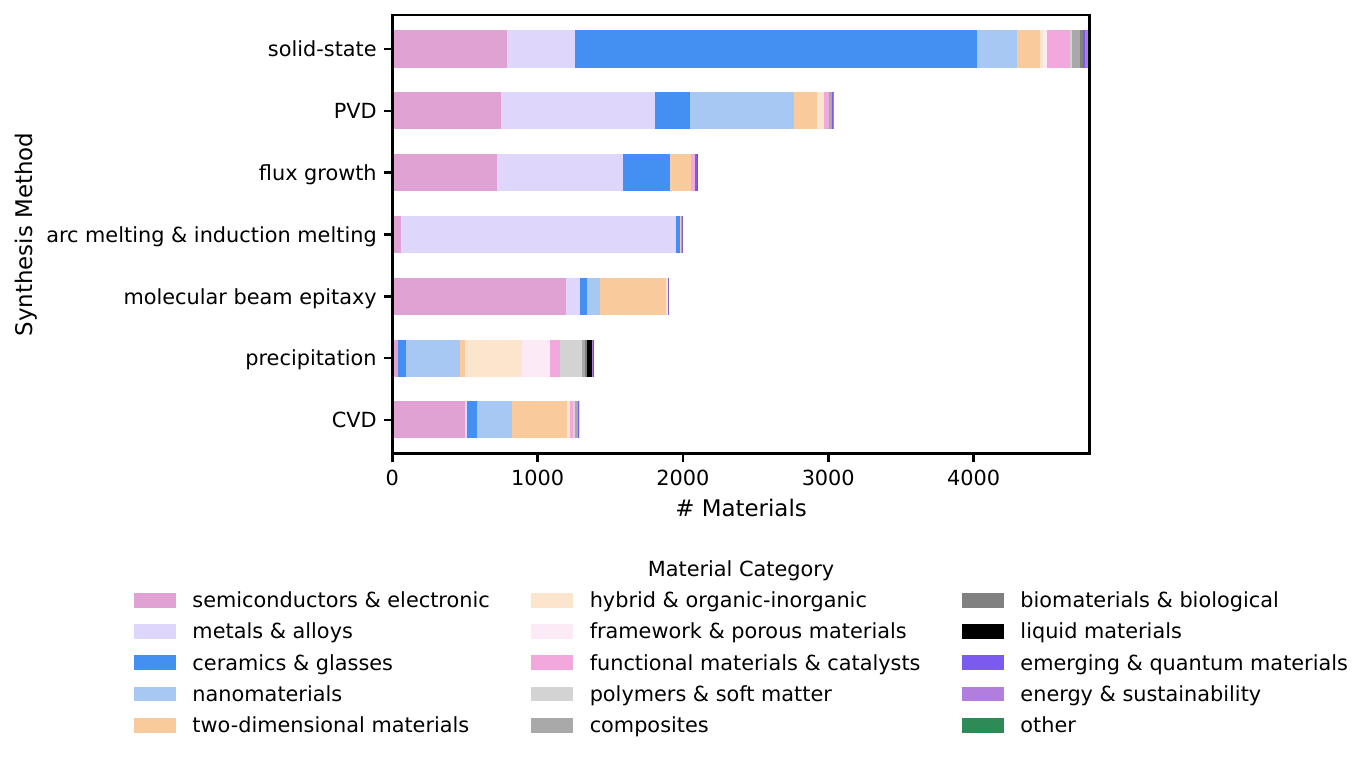}
  \caption{Composition of the \texttt{high\_score} config (30.2K procedures), stacked by source corpus / material category. Top-7 synthesis methods (``other'' excluded), stacked by material category.}
  \label{fig:app-dataset-composition-highscore}
\end{figure}
\begin{figure}
  \centering
  \begin{subfigure}{0.48\linewidth}
    \includegraphics[width=\linewidth]{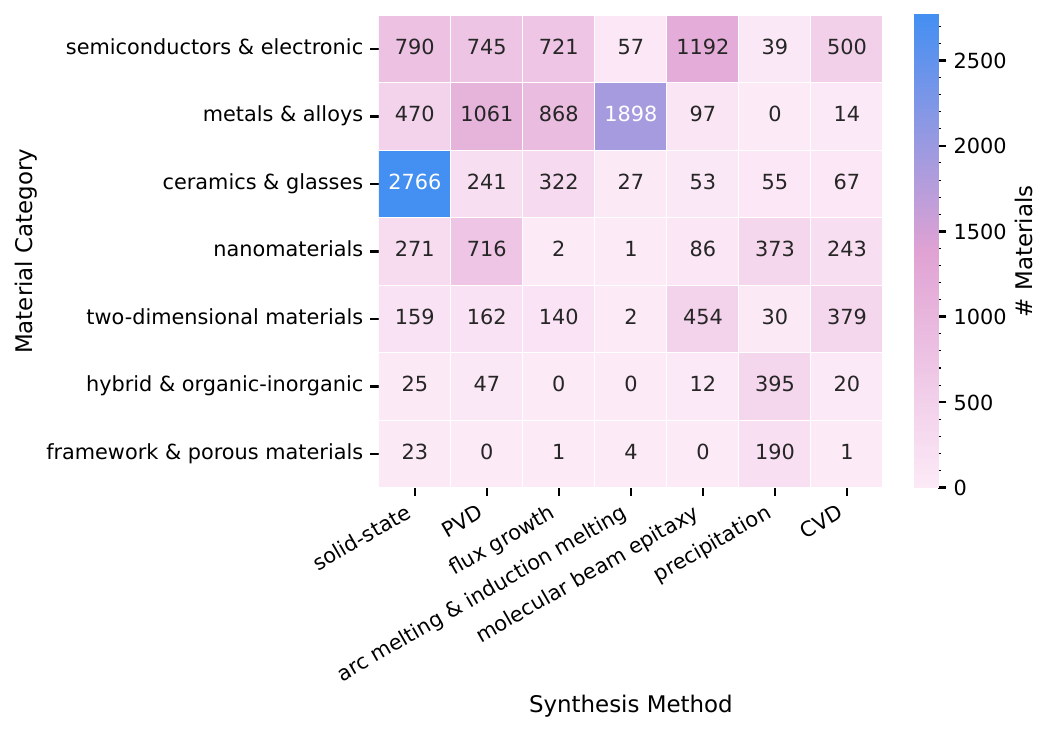}
    \caption{}
  \end{subfigure}
  \begin{subfigure}{0.48\linewidth}
    \includegraphics[width=\linewidth]{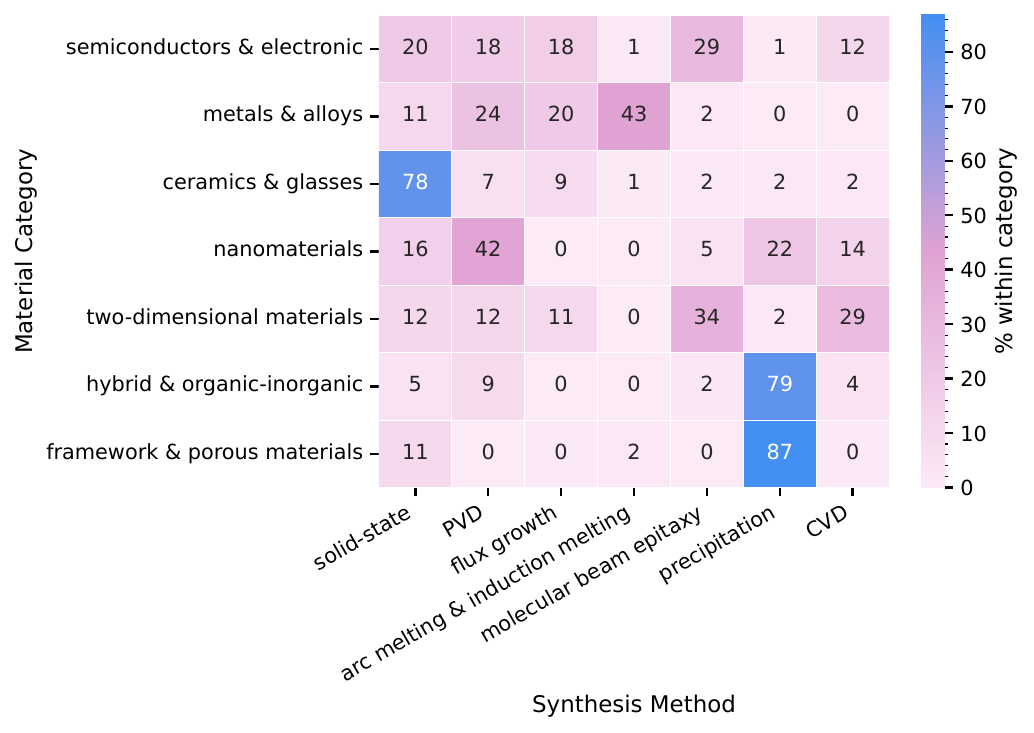}
    \caption{}
  \end{subfigure}
  \caption{Material-class $\times$ synthesis-method heatmaps (top-7 $\times$ top-7) for the \texttt{high\_score} config. \textbf{(a)} Raw procedure counts. \textbf{(b)} Row-normalized percentages.}
  \label{fig:app-dataset-heatmaps-highscore}
\end{figure}

The full \texttt{LeMat-Synth} dataset comprises 58.3K synthesis procedures across all three source corpora (arXiv, ChemRxiv, OMG24); we additionally release a stratified high-quality subset (\texttt{high\_score} config) of 30.2K procedures, restricted to those with a judge-assigned overall score $>3.25$ (out of 5) and more than one extracted synthesis step, for downstream users prioritizing precision over coverage. The score-only threshold used in an earlier release ($>4$) let through a cluster of degenerate zero- and one-step extractions that the judge nonetheless scored highly; adding the step-count criterion removes this failure mode entirely while roughly doubling the size of the subset. Figures~\ref{fig:app-dataset-stats-overall} and \ref{fig:app-dataset-diversity-overall} give the full category/method breakdown and material-class $\times$ synthesis-method coverage for the \texttt{full} config, while Figures~\ref{fig:app-dataset-stats-highscore}, \ref{fig:app-dataset-diversity-highscore}, Fig.~\ref{fig:app-dataset-composition-highscore-category}, \ref{fig:app-dataset-composition-highscore}, and Fig.~\ref{fig:app-dataset-heatmaps-highscore} give the same breakdown, coverage, top-7 composition, and heatmaps for the \texttt{high\_score} config.

\subsubsection{Data acquisition}
\label{app:sec:data-acquisition}

\paragraph{arXiv} From over two million articles on arXiv in total, we fetched 381,116 publications in the category \texttt{cond-mat} from 1992 to April 2025. We filtered down the corpus to 62,267 publications that contain synthesis procedures by parsing the PDF with \href{https://github.com/datalab-to/marker}{Marker} and calling Mistral-Small-3.1-24B-Instruct-2503~\cite{MistralSmall31_2025} on a cluster of 8xA100-PG509-200 with 40GB of memory each. The text from the PDF (if length larger the max tokens, chunk paper) is passed to the LLM to return whether it contains a synthesis procedure, the material name and category, see Section~\ref{app:subsec:prompts}.

\paragraph{ChemRxiv} From over 30,000 articles with the cutoff date of June 2025, we fetched 2,910 publications in the categories Solid State Chemistry, Solution Chemistry, Solvates, Spectroscopy (Inorg.), Structure, Supramolecular Chemistry (Inorg.), Supramolecular Chemistry (Org.), Surface, Surfactants, Thermal Conductors and Insulators, Thin Films, Wastes, Water Purification, with the ChemRxiv API. After filtering similarly to the arXiv data, we obtain 1,500 papers with synthesis procedures. If available, a supplementary file is appended to the main text.

\paragraph{Open Materials Guide 2024 (OMG24)} The data collection and curation from the Semantic Scholar API is described in \cite{kim2025towards}. It contains 17,667 synthesis procedures with ten different synthesis types from open access publications. We fetched the PDFs from the URLS provided in the  \href{https://huggingface.co/datasets/iknow-lab/open-materials-guide-2024}{published dataset}, downloaded them, and proceeded with parsing the text and images. As the papers in OMG24 are already pre-filtered to contain synthesis procedures, no filtering step is needed.

\paragraph{PDF post-processing}\label{app:par:pdf-processing}
To extract text and figures from PDFs obtained from the arXiv, we use \texttt{marker-pdf}, an open-source library, with \texttt{Gemini 2.0 flash} (\texttt{gemini-2.0-flash}\cite{Gemini2FlashModelCard2025}). We strip the images from the text, which is converted into Markdown format, and save the images separately, but such that they can be reinserted into the Markdown text. For the ChemRxiv and OMG24, we used \texttt{Mistral-OCR} (\texttt{mistral-ocr-latest}\cite{MistralOCR_2025}) to extract images and text in Markdown format. We empirically tested \texttt{Docling} \cite{deep2024docling}, an open source alternative to Mistral-OCR, and found Mistral-OCR to empirically perform better and infer results faster. For post-processing the text, we removed markdown image identifiers and the References section (= 50 lines after the heading References with regex).

Entries with an invalid synthesized-material name were subsequently filtered out to maintain data quality, applied uniformly across the \texttt{arxiv}, \texttt{chemrxiv}, and \texttt{omg24} splits of the \texttt{full} config. The largest single category is a generic, non-specific identifier ("Intermediate 1", "Compound B" etc., 26.02\% of all entries), followed by a number+letter pattern such as "8a" or "a23" (11.30\%), no valid synthesized material found (7.81\%), other invalid names (5.37\%), a single-character name (2.03\%), and names consisting only of numbers (0.01\%). The \texttt{high\_score} config is derived directly from this already name-filtered \texttt{full} config. Table~\ref{table:material-filtering} breaks down the removed entries by category, out of 122,972 total entries. This high dropout rate (52.6\% overall) highlights the need to standardize material identifiers to further make the database properly searchable and interoperable.

\begin{table}[h]
  \centering
  \caption{Breakdown of entries removed by the synthesized-material name filter, across all splits (\texttt{arxiv}, \texttt{chemrxiv}, \texttt{omg24}) of the \texttt{full} config, out of 122,972 total entries.}
  \label{table:material-filtering}
  \begin{tabular}{lrr}
    \toprule
    \textbf{Filter category} & \textbf{Removed} & \textbf{\%} \\
    \midrule
    No materials synthesized & 9,605 & 7.81\% \\
    Name is only numbers & 17 & 0.01\% \\
    Name is a single character & 2,500 & 2.03\% \\
    Number+letter / letter+number pattern (e.g.\ "8a", "a23") & 13,901 & 11.30\% \\
    Generic identifier (e.g.\ "Intermediate 1", "Compound B") & 31,994 & 26.02\% \\
    Other invalid name & 6,610 & 5.37\% \\
    \midrule
    \textbf{Total removed} & \textbf{64,627} & \textbf{52.55\%} \\
    \bottomrule
  \end{tabular}
\end{table}

\ref{table:filtering-per-split} reports the resulting per-split entry counts, before and after this filtering step, for the \texttt{full} config of LeMat-Synth. The \texttt{high\_score} config is derived directly from the (already name-filtered) \texttt{full} config, so it is not subject to a separate name-filtering step. Table~\ref{table:highscore-per-split} instead reports its construction from \texttt{full} via the score and step-count criteria described above.

\begin{table}[h]
  \centering
  \caption{Per-split entry counts before and after material-name filtering, for the \texttt{full} config of LeMat-Synth.}
  \label{table:filtering-per-split}
  \begin{tabular}{lrrr}
    \toprule
    \textbf{Split} & \textbf{Before} & \textbf{After} & \textbf{Removed} \\
    \midrule
    chemrxiv & 5,901 & 4,067 & 1,834 (31.1\%) \\
    omg24 & 10,625 & 6,202 & 4,423 (41.6\%) \\
    arxiv & 106,446 & 48,076 & 58,370 (54.8\%) \\
    \textbf{Total} & \textbf{122,972} & \textbf{58,345} & \textbf{64,627 (52.6\%)} \\
    \bottomrule
  \end{tabular}
\end{table}

\begin{table}[h]
  \centering
  \caption{Per-split entry counts for the \texttt{high\_score} config, filtered from \texttt{full} by judge-assigned overall score $>3.25$ and more than one extracted synthesis step.}
  \label{table:highscore-per-split}
  \begin{tabular}{lrrr}
    \toprule
    \textbf{Split} & \textbf{\texttt{full}} & \textbf{\texttt{high\_score}} & \textbf{Kept} \\
    \midrule
    chemrxiv & 4,067 & 2,747 & 67.5\% \\
    omg24 & 6,202 & 4,226 & 68.1\% \\
    arxiv & 48,076 & 23,248 & 48.4\% \\
    \textbf{Total} & \textbf{58,345} & \textbf{30,221} & \textbf{51.8\%} \\
    \bottomrule
  \end{tabular}
\end{table}
\subsection{Synthesis Extraction}
\label{app:sec:synthesis-extraction}

\paragraph{Ontology} \label{app:subsec:ontology}

\begin{figure}
    \centering
    \includegraphics[width=\linewidth]{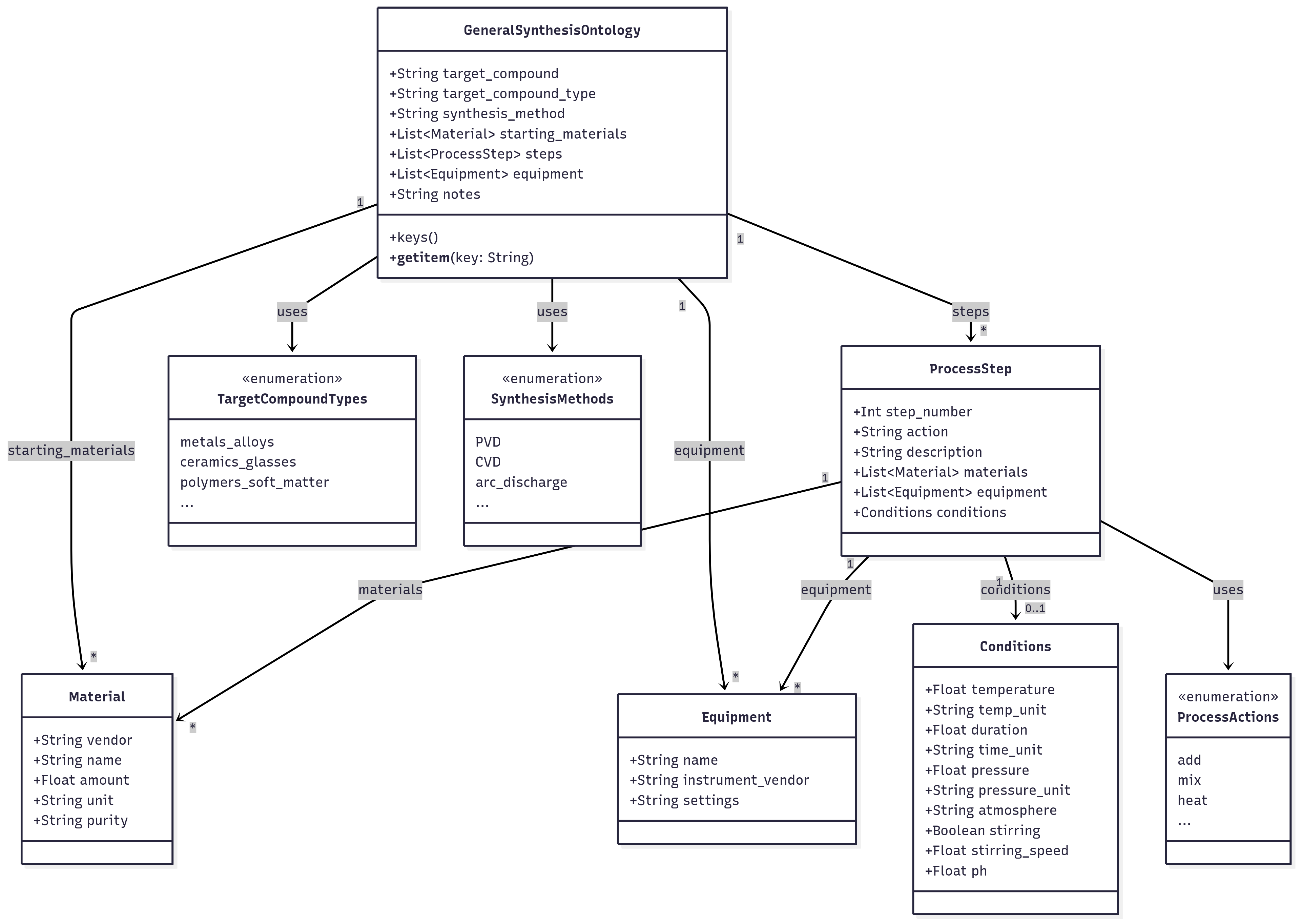}
    \caption{Visual representation of the hierarchical ontology for structuring synthesis procedures. The ontology organizes information from a global level (target compound, synthesis method) down to sequential process steps. Each step encapsulates detailed information about the specific actions, materials, equipment, and conditions involved, ensuring data consistency and machine-readability (Table~\ref{tab:synthesis-ontology}).}
    \label{fig:ontology}
\end{figure}

\ref{fig:ontology} and \ref{tab:synthesis-ontology} show the ontology developed in this work. We abstracted a \textit{broad} synthesis procedure as a sequence of steps with actions, conditions, equipment and an associated material, as well as starting materials. Note that in the library released in this work, the ontology can be adapted to custom cases, e.g. specialized syntheses for catalysts or polymers. The ontology can be adapted from the \texttt{GeneralSynthesisOntology} class.

\setlength{\LTcapwidth}{\linewidth}
\begin{longtable}{@{}p{3.5cm}p{3.5cm}p{8cm}@{}}
\caption{Detailed structure of the \texttt{GeneralSynthesisOntology} scheme for the standardized representation of a synthesis procedure. Note that the type (material category) and synthesis method are chosen from a pre-determined list of verbs. The General Synthesis Ontology contains the target compound, synthesis method, overall materials and equipment. The Process Steps object is sequential and contains ordered operations with specific actions, local materials, equipment, and conditions. Materials (Chemical identity, quantities, specifications, and vendor), Equipment (Instrumentation with settings and vendor information), Conditions (Environmental parameters: temperature, time, pressure, atmosphere, pH) are set.}
\label{tab:synthesis-ontology} \\
\toprule
\textbf{Component} & \textbf{Attributes} & \textbf{Description \& Examples} \\
\midrule
\endfirsthead

\caption[]{Detailed structure of the \texttt{GeneralSynthesisOntology} (continued)} \\
\toprule
\textbf{Component} & \textbf{Attributes} & \textbf{Description \& Examples} \\
\midrule
\endhead

\midrule
\multicolumn{3}{r}{\textit{Continued on next page}} \\
\endfoot

\bottomrule
\endlastfoot


\multirow{4}{3.5cm}{\textbf{Target Compound}} 
& compound & Chemical composition and description \\
& type & Material category: metals \& alloys, ceramics, nanomaterials, polymers, semiconductors, etc. \\
& synthesis method & Technique: sol-gel, hydrothermal, CVD, precipitation, electrodeposition, etc. \\
& notes & Additional observations or variations \\

\midrule

\multirow{5}{3.5cm}{\textbf{Material}} 
& name & Chemical name (e.g., Nickel Nitrate, Deionized Water) \\
& amount & Quantity used (numeric value) \\
& unit & Mass (g, mg), Volume (mL, L), Molar (mol, mmol), Concentration (M, mM), etc. \\
& vendor & Supplier information \\
& purity & Grade specification (99\%, ACS grade, etc.) \\

\midrule

\multirow{3}{3.5cm}{\textbf{Equipment}} 
& name & Instrument type (autoclave, tube furnace, magnetic stirrer) \\
& vendor & Manufacturer (Thermo Fisher, Agilent, Bruker, etc.) \\
& settings & Operating parameters (500 rpm, heating rate 5$^\circ$C/min) \\

\midrule

\multirow{6}{3.5cm}{\textbf{Conditions}} 
& temperature & Process temperature with units ($^\circ$C, K, $^\circ$F) \\
& duration & Time period with units (h, min, s, days) \\
& pressure & Applied pressure with units (atm, bar, Pa, torr) \\
& atmosphere & Gas environment (air, N$_2$, H$_2$, Ar, vacuum) \\
& stirring & Boolean and speed (rpm) \\
& pH & Solution acidity/basicity \\

\midrule

\multirow{6}{3.5cm}{\textbf{Process Step}} 
& step number & Sequential order in procedure \\
& action & Primary operation: add, mix, heat, cool, reflux, age, filter, wash, dry, etc. \\
& description & Detailed procedure text \\
& materials & List of materials used in this step \\
& equipment & List of equipment used in this step \\
& conditions & Environmental parameters for this step \\

\end{longtable}


\paragraph{Manual annotations} Eleven material scientists annotated a total of 36 papers, cross-checking each other's annotations, (\cite{Pruszynska-Karbownik2025_optical_BIC,Akbashev2023_SrIrO3,Gatner2005_GdCo2,Kapoor2020_GQDs,Petach2017_CTR,Tialiou2025_MetalloHydrogels,Lynn1996_Ferromagnetic_LaCaMnO,Yang2006_NormalState_BiCuO,Ren2016_Thermopower_CeAu2Si2,Benson2025_Fluorescence_Crystals,Chatterjee2025_Mechanochemical_Biomass,Wang2017_NobleMetalNanoparticles,Yang2022_CsBiI9Composite,Sharma2015_ImproperFerroelectricity,Labunov2013_FemtosecondLaserModification,Ibrahim2020_GadoliniaAddition,Li2018_FieldEmissionZnO,Zhang2019_MgB2CatalyzedGrowth,Han2016_KondoScattering,Li2023_DendriteFree_AlkaliMetal,Box2017_Indentation_FloatingElasticSheet,Goswami_Synthesis_Micelles_Magnetite_2023,Siemons_LaAlO3_SrTiO3_Interfaces_2023,Nasiri-Tabrizi_Thermal_Treatment_Fluorapatite_Titania_2023,Charmas2015_NiDopedCarbosils,
Martini_2019,Rota_2024,Singh_2024,sebastian2024enhancedhydrogenevolutionactivity,Fonseca_Guzman2023-zq,Wilson2024-zx,Tavan2023-th,Alvi2020-od,Biercuk_2006,bruning2006coexistenceopenclosedgallery,Hamawandi2020-hs} by inferring synthesis procedures from a sample picked at random among each of the following sources: arXiv, ChemRxiv, OMG24 (1 to 1 ratio, stratified sampling). The synthesis procedures were manually reviewed for correctness, completeness, and adherence to a pre-defined structured ontology. Note that this process ensured the relevant information was extracted as it was in the text, and didn’t aim to directly assess scientific accuracy. To the material scientists’ capacity, where relevant but ambiguous terms from the experimental workflows needed to be assessed, more than one annotator was consulted and a consensus was reached in order to maintain the consistency throughout the process.
Each validation assessed whether the LLM-extracted synthesis procedures were consistent with the original text. The annotators noted down any missing, incorrect or hallucinated content generated and attributed detailed scores for each procedure. A total of seven scoring criteria were used, ranging from 1 (poor) to 5 (excellent) in 0.5 increments:

\begin{itemize}
    \item \textbf{Structural completeness score}: Coverage of ontology-relevant information, including materials, synthesis steps, equipment, conditions, etc.
    \item \textbf{Material extraction score}: accuracy and completeness of the extracted materials, including names, quantities, units, and purities.
    \item \textbf{Process steps score}: correctness and organization of the procedural steps, including the sequence and classification of synthesis actions.
    \item \textbf{Equipment extraction score}: completeness and accuracy in identifying experimental apparatus, including vendor names and operational settings where available.
    \item \textbf{Conditions extraction score}: correctness of temperature, pressure, duration, and atmospheric conditions, along with unit consistency.
    \item \textbf{Semantic accuracy score}: the degree to which the structured extraction preserved the scientific meaning and contextual integrity of the original description.
    \item \textbf{Format compliance score}: adherence of the structured data to the ontology schema and data type requirements.
\end{itemize}

Finally, an \textbf{overall score} was computed as the mean of the individual criteria, with a final reasoning field summarizing strengths, weaknesses, and suggestions for improvement.

\paragraph{Domain expert/LLM-as-a-judge comparison}\label{app:sec:synth-extr-eval-llm-human}

\textbf{Sample sizes used in this section.} Several different $n$ appear below, because each analysis requires a different set of scores to be jointly present. They are nested as follows. Eleven experts annotated \textbf{80 synthesis procedures from 36 papers}. Of those papers, \textbf{34} yield at least one material that can be linked between the human annotation and some LLM extraction, and any single judge's agreement statistic rests on the \textbf{19--20 papers} for which that particular judge produced scores, which is the unit used for the paper-level bootstrap (Section~\ref{app:sec:bootstrap-uncertainty}). At the material level, the per-judge linked (human, LLM) score pairs number \textbf{$n=75$--$93$} depending on the judge (Table~\ref{table:human-llm-comparison}), or \textbf{$n=110$} when self-scored cells are retained instead of excluded leave-one-out (Table~\ref{table:loo-vs-all}). Separately, the $4\times4$ extractor--judge matrix reported in the main text requires a human score \emph{and} all four judges' scores on the same (procedure, extractor) pair, which is the stricter \textbf{$n=59$} set. Figure~3b and 3c in the main text report the per-judge quantities from Table~\ref{table:human-llm-comparison}, \textit{i.e.}\ the $n=75$--$93$ leave-one-out cell sets, not the $n=59$ matrix.

We compare four candidate judge LLMs (Claude-Sonnet-4.6\cite{ClaudeSonnet46SystemCard2026}, DeepSeek-V3.2\cite{DeepSeekV32_2025}, Gemini-3-Flash\cite{Gemini3FlashModelCard2025}, Qwen3.5-397B-A17B\cite{Qwen3_5_2026}) against domain-expert annotations. Agreement is summarized leave-one-out, \textit{i.e.}, excluding each judge's own self-scored extractions from its agreement computation, so that self-preference bias (\textit{vide infra}) cannot inflate a judge's apparent quality. Materials are linked between human and LLM annotations using \textbf{LLM-matched} alignment (string matching, augmented with an LLM name-matcher for materials the string matcher misses; see Section~\ref{app:sec:name-matcher-validation} below for validation and \ref{app:sec:string-vs-llm-match} for the string-matcher-only baseline this table would otherwise use, and how much statistical power LLM-matching recovers).

\begin{table}[t]
  \centering
    \caption{Agreement between each candidate judge LLM and human expert scores, leave-one-out (excluding self-scored cells), LLM-matched materials, ranked by ICC(2,1), which is the metric most robust to a judge's overall scale/offset and the criterion used for judge selection (main text). $\rho$: Spearman rank correlation with human \texttt{overall\_score}. $\kappa$: Cohen's kappa. \textbf{mean\_diff}: signed mean difference (positive = judge scores higher than human, \textit{i.e.}, more permissive). A raw mean-absolute-score-gap metric is omitted: it is confounded by a judge's overall scale/offset and would rank Gemini-3-Flash mid-table despite it having the lowest ICC(2,1) and $\kappa$ of the four judges. Note that Gemini-3-Flash is not uniformly worst: its Spearman $\rho$ of 0.343 is second-highest, reflecting that it ranks extractions reasonably well while calibrating their absolute scores poorly. Best and second-best ICC(2,1) are bolded and underlined, respectively.}
  \begin{tabular}{lcccccc}
    \toprule
    \textbf{Judge} & \textbf{ICC(2,1)}  & \textbf{ICC(3,1)} & $\rho$ & $\kappa$ & \textbf{mean\_diff} & \textbf{n} \\
    \midrule
    Claude-Sonnet-4.6   & \textbf{0.276} & 0.299 & 0.406 & 0.192 & $-0.295$ & 75 \\
    DeepSeek-V3.2       & \underline{0.166} & 0.195 & 0.196 & 0.212 & $-0.447$ & 87 \\
    Qwen3.5-397B-A17B   & 0.126 & 0.127 & 0.181 & 0.077 & $-0.132$ & 93 \\
    Gemini-3-Flash      & 0.072 & 0.121 & 0.343 & $-0.037$ & $+0.567$ & 75 \\
    \bottomrule
  \end{tabular}
  \label{table:human-llm-comparison}
\end{table}
Claude-Sonnet-4.6 achieves the highest absolute agreement with human scores (ICC(2,1) = 0.28) and the highest rank correlation ($\rho = 0.41$); we select it as the judge model on this basis (main text). All four judges' \texttt{mean\_diff} values show that agreement quality and scoring calibration are distinct: Gemini-3-Flash is systematically permissive (average $+0.57$ points above human), while DeepSeek-V3.2, Qwen3.5-397B-A17B, and Claude-Sonnet-4.6 are all systematically harsh ($-0.45$, $-0.13$, and $-0.30$ respectively). Discrepancies typically arise when literature descriptions are vague or incomplete. Experts may infer plausible synthesis details, whereas an overly strict judge penalizes under-specified inputs, and an overly permissive judge fails to flag genuine errors (see examples below).

We additionally test for self-preference bias, i.e. whether a judge scores extractions produced by the same underlying model family more favorably than it scores others' extractions. Gemini-3-Flash shows pronounced self-preference: its self-scored extractions average 4.80 vs. 3.52 for peer extractions ($+1.28$ raw difference), a difference-in-differences of $+0.44$ once the human-scored reference quality of those same extractions is accounted for, i.e., Gemini-3-Flash rates its own outputs far more favorably than either the human annotator or the other three judges do. Claude-Sonnet-4.6 and Qwen3.5-397B-A17B show negligible self-preference ($-0.12$ and $+0.003$ raw difference respectively), and DeepSeek-V3.2 is, if anything, harsher on its own outputs than on peers' ($-0.67$). This is consistent with Gemini-3-Flash's poor leave-one-out agreement above: removing its self-scored cells is necessary to obtain an unbiased estimate of its judging quality, and even so it remains the weakest-agreeing judge among the four.

\vspace{0.5em}
\paragraph{Name-matching methodology and validation}\label{app:sec:name-matcher-validation}

Materials mentioned by the human annotator and by an LLM extractor are frequently the same compound written differently (e.g. \texttt{"Bi2-xSbxTe3"} vs.\ \texttt{"Bi2Te3"}/\texttt{"Sb2Te3"}, or \texttt{"HA"} vs.\ \texttt{"Ca10(PO4)6(OH)2"} for hydroxyapatite), so linking a human-annotated material to its corresponding LLM extraction cannot rely on exact string equality alone. We first apply a string matcher (\texttt{SequenceMatcher} + word-Jaccard similarity, threshold 0.7; \texttt{eval\_utils.find\_best\_matches}), which alone links materials in only $\sim$15--17 of the 34 annotated papers per judge. As a fallback, an LLM judge (\texttt{DspyNameMatcherJudge}, which is the same aligner used for the thermocatalysis figure-linking evaluation, Section~\ref{app:sec:img-segmenter}) proposes matches for materials the string matcher leaves unmatched, each tagged with a confidence level (high/medium/low). Only high- and medium-confidence proposals are kept; low-confidence proposals were spot-checked as unreliable during development (e.g.\ a peptide sequence matched to an unrelated molecular formula) and are dropped entirely.

A domain expert on the team manually validated this procedure against the source papers, with results summarized in Table~\ref{table:name-matcher-validation}:

\begin{table}[t]
  \centering
  \caption{Manual validation of the LLM name-matcher, reviewed against source papers by a domain-expert annotator. The two precision rows are computed over all proposed matches at that confidence level, so ``Count'' is the number judged correct. The third row is a separate recall spot-check over candidate pairs the matcher did \emph{not} propose, so there ``Count'' is the number that were genuine misses; it is a miss rate, not a precision.}
  \resizebox{\linewidth}{!}{%
  \begin{tabular}{lccc}
    \toprule
    \textbf{Check} & \textbf{Count} & \textbf{Sampled} & \textbf{Rate} \\
    \midrule
    High-confidence proposals that were correct & 19 & 19 & 100\% precision \\
    Medium-confidence proposals that were correct & 15 & 19 & 79\% precision \\
    Non-proposed candidate pairs that \emph{should} have matched & 4 & 13 & 31\% missed \\
    \bottomrule
  \end{tabular}
  }
  \label{table:name-matcher-validation}
\end{table}

Four medium-confidence proposals were manually rejected. Two dropped a real structural component of the human-annotated name (\texttt{"LaAlO3/SrTiO3"} $\to$ \texttt{"LaAlO3"}; \texttt{"(La0.3Sr0.7)(Al0.65Ta0.35)O3/SrTiO3"} $\to$ \texttt{"(La0.3Sr0.7)(Al0.65Ta0.35)O3"}, both from arXiv:1605.04038); these are denylisted by name in the name-matching script's rejection list. The other two paired an unresolvably vague human annotation (\texttt{"Cyanine SMILES"}) with two different specific extractions, which is an instance of the naming-convention problem identified as a limitation of this project. Separately, and not fixable by any matching method: some human ground-truth names are themselves too vague to link to a specific extraction even in principle (e.g.\ \texttt{"Pf-AgNPs"}, \texttt{"Oxidized CNS"}, \texttt{"Cyanine SMILES"}), since the annotation guidelines did not enforce a standardized naming/formula convention. This caps the achievable $n$ regardless of matching method.

\vspace{0.5em}
\paragraph{Recall improvement: string-matcher-only vs.\ LLM-matched}\label{app:sec:string-vs-llm-match}

LLM-matching (above) substantially increases how many papers and materials can be linked between human and LLM annotations, which matters because agreement-metric uncertainty is driven by the number of independent \emph{papers} available, not materials (Section~\ref{app:sec:bootstrap-uncertainty}). \ref{table:string-vs-llm-match} compares the string-matcher-only baseline against the LLM-matched result used throughout this work (Table~\ref{table:human-llm-comparison}), both leave-one-out excluding self-scored cells.

\begin{table}[t]
  \centering
  \caption{String-matcher-only vs.\ LLM-matched material linking, leave-one-out excluding self-scored cells. $n$: number of linked (human, LLM) score pairs. Recall gain: additional linked pairs recovered by LLM-matching.}
  \resizebox{\linewidth}{!}{%
  \begin{tabular}{lcccccc}
    \toprule
    \textbf{Judge} & $n$\textbf{(string)} & $n$\textbf{(LLM-matched)} & \textbf{Recall gain} & \textbf{ICC(2,1) string} & \textbf{ICC(2,1) LLM-matched} \\
    \midrule
    Claude-Sonnet-4.6 & 39 & 75 & $+36$ & 0.378 & 0.276 \\
    DeepSeek-V3.2     & 47 & 87 & $+40$ & 0.176 & 0.166 \\
    Qwen3.5-397B-A17B & 48 & 93 & $+45$ & 0.072 & 0.126 \\
    Gemini-3-Flash    & 43 & 75 & $+32$ & 0.076 & 0.072 \\
    \bottomrule
  \end{tabular}
  }
  \label{table:string-vs-llm-match}
\end{table}

LLM-matching recovers 32--45 additional linked pairs per judge (roughly doubling $n$), meaningfully improving the evidence base for the agreement estimates above. The ranking on the string-matcher-only baseline is Claude-Sonnet-4.6 $>$ DeepSeek-V3.2 $>$ Gemini-3-Flash $>$ Qwen3.5-397B-A17B; LLM-matching gives Claude-Sonnet-4.6 $>$ DeepSeek-V3.2 $>$ Qwen3.5-397B-A17B $>$ Gemini-3-Flash. The top two judges (Claude-Sonnet-4.6, DeepSeek-V3.2) are unchanged either way; only the \#3/\#4 ranks of Qwen3.5-397B-A17B and Gemini-3-Flash swap, which does not affect the deployed extractor/judge pair (Qwen3.5-397B-A17B extractor, DeepSeek-V3.2 judge).

\vspace{0.5em}
\paragraph{Leave-one-out vs.\ all cells (self-preference sanity check)}

\texttt{loo\_no\_self} excludes each judge's own self-scored cells, so that a judge cannot inflate its apparent agreement by grading itself favorably (the primary metric used throughout, e.g.\ Table~\ref{table:human-llm-comparison}). \texttt{all} includes every cell, self-scored or not. Table~\ref{table:loo-vs-all} compares the two, LLM-matched.

\begin{table}[t]
  \centering
  \caption{Judge ranking with self-scored cells excluded (\texttt{loo\_no\_self}, primary) vs.\ included (\texttt{all}), LLM-matched, $n=110$ for \texttt{all}.}
  \begin{tabular}{lcccc}
    \toprule
    \textbf{Judge} & \textbf{ICC(2,1) (all)} & \textbf{ICC(3,1) (all)} & $\rho$\textbf{ (all)} & $\kappa$\textbf{ (all)} \\
    \midrule
    Claude-Sonnet-4.6 & 0.291 & 0.304 & 0.330 & 0.251 \\
    DeepSeek-V3.2     & 0.176 & 0.217 & 0.250 & 0.186 \\
    Qwen3.5-397B-A17B & 0.083 & 0.083 & 0.131 & 0.065 \\
    Gemini-3-Flash    & 0.070 & 0.117 & 0.318 & $-0.020$ \\
    \bottomrule
  \end{tabular}
  \label{table:loo-vs-all}
\end{table}

The ranking is identical either way (Claude-Sonnet-4.6 $>$ DeepSeek-V3.2 $>$ Qwen3.5-397B-A17B $>$ Gemini-3-Flash), so judge selection does not hinge on the leave-one-out correction; we nonetheless report \texttt{loo\_no\_self} as primary throughout, since it is the unbiased estimate and the correction matters most for Gemini-3-Flash, the judge with the largest self-preference effect (\textit{vide supra}).

\vspace{0.5em}
\paragraph{Uncertainty analysis: paper-level bootstrap vs.\ material-level}\label{app:sec:bootstrap-uncertainty}

The main text and \ref{table:human-llm-comparison} report point estimates of ICC(2,1) without error bars. At 19--20 independent papers per judge, uncertainty on a correlation-type statistic (ICC, $\rho$, $\kappa$) is genuinely wide, which is a direct consequence of sample size, not an artifact of which error-bar convention is used. We quantify this with a paper-level bootstrap (2,000 resamples, resampling whole papers with replacement, preserving all materials within a resampled paper together).

We resample whole papers because materials within the same paper share extraction context, chemistry difficulty, and the same annotator's reading of that paper. They are not independent observations. Resampling individual materials directly (ignoring paper boundaries) implicitly assumes any material could have come from any paper, which is false, and produces a tighter but overconfident error bar purely by over-counting correlated observations as independent (pseudo-replication). A toy simulation (paper-to-paper true differences only 3$\times$ the within-paper material noise, which is a conservative assumption, since real papers plausibly differ far more) showed material-level bootstrap standard error understating the true paper-level standard error by $\sim$20\% under those conditions; Table~\ref{table:bootstrap-paper-vs-material} confirms this same understatement on the real data.

\begin{table}[t]
  \centering
  \caption{Paper-level vs.\ material-level bootstrap uncertainty on ICC(2,1), LLM-matched, \texttt{loo\_no\_self}, 2,000 resamples. Material-level standard error is expected to be \emph{smaller}.}
  \begin{tabular}{lccc}
    \toprule
    \textbf{Judge} & \textbf{SE (paper)} & \textbf{95\% CI half-width (paper)} & \textbf{SE (material)} \\
    \midrule
    Claude-Sonnet-4.6 & 0.149 & 0.283 & 0.088 \\
    DeepSeek-V3.2     & 0.118 & 0.229 & 0.100 \\
    Qwen3.5-397B-A17B & 0.107 & 0.209 & 0.085 \\
    Gemini-3-Flash    & 0.058 & 0.110 & 0.047 \\
    \bottomrule
  \end{tabular}
  \label{table:bootstrap-paper-vs-material}
\end{table}

Material-level standard error understates the (correct) paper-level standard error by 15--41\% across judges in the real data (Claude-Sonnet-4.6 41\%, Qwen3.5-397B-A17B 21\%, Gemini-3-Flash 19\%, DeepSeek-V3.2 15\%), bracketing the 20\% toy-simulation estimate above and exceeding it for the judge whose agreement estimate matters most to judge selection. This confirms that resampling at the material level would overstate the precision of these agreement estimates. Showing wide paper-level CIs in the main figure would add visual clutter without changing the reported ranking (Section~\ref{app:sec:string-vs-llm-match} shows the top-2 ranking is stable under both matching methods and both bootstrap units); we therefore omit error bars from the main figure and report the full uncertainty analysis here.

\vspace{0.5em}
\paragraph{Extractor-choice robustness: ranking agreement across graders}\label{app:sec:judge-disagreement-examples}

A separate concern from judge selection is whether the choice of \emph{extractor} (Claude-Sonnet-4.6 identified as best by human judgment, Qwen3.5-397B-A17B deployed for cost reasons; main text) is itself an artifact of which judge is trusted to rank the extractors. Table~\ref{table:extractor-ranking-robustness} reports each judge's (and the human's) full ranking of the four extractor LLMs by mean \texttt{overall\_score}, and the Spearman rank correlation of that ranking against the human's.

\begin{table}[t]
  \centering
  \caption{Extractor ranking by each grader (mean \texttt{overall\_score} across all scored materials) and Spearman rank correlation against the human extractor ranking.}
  \begin{tabular}{llc}
    \toprule
    \textbf{Grader} & \textbf{Extractor ranking (best to worst)} & $\rho$\textbf{ vs.\ human} \\
    \midrule
    Human              & Claude $>$ Gemini $>$ Qwen $>$ DeepSeek & 1.00 \\
    Gemini-3-Flash     & Claude $>$ Qwen $>$ Gemini $>$ DeepSeek & 0.80 \\
    Claude-Sonnet-4.6  & Claude $>$ Qwen $>$ DeepSeek $>$ Gemini & 0.40 \\
    DeepSeek-V3.2      & Claude $>$ Qwen $>$ DeepSeek $>$ Gemini & 0.40 \\
    Qwen3.5-397B-A17B  & Claude $>$ Qwen $>$ DeepSeek $>$ Gemini & 0.40 \\
    \bottomrule
  \end{tabular}
  \label{table:extractor-ranking-robustness}
\end{table}

Every candidate judge ranks Claude-Sonnet-4.6 as the top extractor, matching the human ranking exactly at the top ($\rho=0.4$--$1.0$ across judges): the extractor-choice claim (Claude-Sonnet-4.6 best by human judgment, Qwen3.5-397B-A17B deployed as a cost trade-off) does not depend on which judge one trusts, even though the judges disagree on the ordering of the remaining three extractors.

\vspace{0.5em}
\paragraph{Example 1: Genuine extraction error, judges disagree on severity (Material: Bi\textsubscript{1.74}Pb\textsubscript{0.38}Sr\textsubscript{1.88}CuO\textsubscript{6+$\delta$})}

Source (arXiv:cond-mat/0602418): \textit{"High quality superstructure-free Bi\textsubscript{1.74}Pb\textsubscript{0.38}Sr\textsubscript{1.88}CuO\textsubscript{6+$\delta$} single crystal was grown by floating-zone technique."} Gemini-3-Flash extracted \texttt{synthesis\_method = "flux growth"}, which does not match the human-annotated ground truth (\texttt{"float zone \& Bridgman"}).

\begin{table}[t]
  \centering
  \caption{Evaluation scores for the Bi\textsubscript{1.74}Pb\textsubscript{0.38}Sr\textsubscript{1.88}CuO\textsubscript{6+$\delta$} extraction (Gemini-3-Flash as extractor) from all four candidate judges and the human annotator.}
  \begin{tabular}{lc}
    \toprule
    \textbf{Judge} & \textbf{Overall score} \\
    \midrule
    Human               & 4.50 \\
    Gemini-3-Flash      & 4.90 \\
    Claude-Sonnet-4.6    & 3.50 \\
    Qwen3.5-397B-A17B    & 3.50 \\
    DeepSeek-V3.2        & 2.50 \\
    \bottomrule
  \end{tabular}
\end{table}

Gemini-3-Flash (4.9) does not catch the method-name error at all, and instead praises the extraction for "not hallucinating" unstated parameters, which is consistent with its systematic permissiveness above. DeepSeek-V3.2 (2.5) catches the same error but goes further, characterizing the entire structured process and materials list as "hallucination," effectively penalizing the extractor for using the schema at all on a source paper that only mentions synthesis in passing (this is a characterization/ARPES paper, not a synthesis paper). Claude-Sonnet-4.6 and Qwen3.5-397B-A17B (both 3.5) land between the two extremes, identifying the method-name error as the central issue without over- or under-weighting it. On this particular example the human is unusually forgiving (4.5, close to Gemini-3-Flash's 4.9), so the two middle judges are the furthest from the human verdict, not the closest; the systematic pattern in the other direction emerges across the full set. What this example does show is the spread itself: a single, checkable method-name error produces a 2.4-point range across judges.

\vspace{0.5em}
\paragraph{Example 2: Correct extraction, judges still disagree by up to 2.1 points (Material: WFe\textsubscript{2}Ni-red)}

Source: \textit{"Addition of IPrNi(styrene)\textsubscript{2} [...] to WFe\textsubscript{2} in a THF:MeCN mixture results in conversion to a new $C_s$ species [...]"}, extracted by DeepSeek-V3.2. Here \texttt{synthesis\_method = "other"} matches the ground truth exactly, and every judge agrees the starting materials and single process step are correctly identified. Yet judges scored this extraction from 2.9 (DeepSeek-V3.2, scoring its own output) to 5.0 (Gemini-3-Flash), a 2.1-point spread on an objectively correct extraction.

\begin{table}[t]
  \centering
  \caption{Evaluation scores for the WFe\textsubscript{2}Ni-red extraction (DeepSeek-V3.2 as extractor) from all four candidate judges and the human annotator.}
  \begin{tabular}{lc}
    \toprule
    \textbf{Judge} & \textbf{Overall score} \\
    \midrule
    Human                & 4.75 \\
    Gemini-3-Flash       & 5.00 \\
    Claude-Sonnet-4.6    & 4.40 \\
    Qwen3.5-397B-A17B    & 4.50 \\
    DeepSeek-V3.2        & 2.90 \\
    \bottomrule
  \end{tabular}
\end{table}

DeepSeek-V3.2 penalizes the extraction for omitting quantitative detail (amounts, equipment) that the source text itself never specifies, treating textually-absent information as an extraction failure, the same over-strict pattern as in Example 1. Gemini-3-Flash rates it a perfect 5.0, again reflecting its general leniency. This illustrates that judge disagreement is not confined to cases with a genuine underlying error: even on fully correct extractions, judge calibration differences alone produce score gaps large enough to change which extractor model would be selected as "best," reinforcing why judge selection (choosing Claude-Sonnet-4.6, the best human-agreeing judge) matters independently of extractor quality. The two examples above and the four that follow are representative of the two qualitatively different sources of judge disagreement we observe: (1) disagreement about how heavily to penalize a genuine, checkable error, and (2) disagreement purely from scoring calibration on an extraction all judges agree is correct. Four further worked examples, all correct extractions, are given below to show this calibration gap is systematic rather than specific to WFe\textsubscript{2}Ni-red or to the main-text example.

\vspace{0.5em}
\paragraph{Example 3: Correct extraction, judges still disagree by up to 2.0 points (Material: GdCo\textsubscript{2})}

Source: \textit{"A GdCo\textsubscript{2} ingot was prepared by melting of the constituent metals on a water-cooled copper hearth in a high purity argon atmosphere."}, extracted by Gemini-3-Flash. The human annotator accepted \texttt{synthesis\_method = "arc melting \& induction melting"} as a match to ground truth, noting only that the field name itself contains "induction melting," a term the source text does not use (it only says "melting" on a water-cooled copper hearth, which implies arc melting).

\begin{table}[t]
  \centering
  \caption{Evaluation scores for the GdCo\textsubscript{2} extraction (Gemini-3-Flash as extractor) from all four candidate judges and the human annotator.}
  \begin{tabular}{lc}
    \toprule
    \textbf{Judge} & \textbf{Overall score} \\
    \midrule
    Human               & 4.50 \\
    Gemini-3-Flash      & 4.90 \\
    Claude-Sonnet-4.6    & 4.20 \\
    Qwen3.5-397B-A17B    & 4.10 \\
    DeepSeek-V3.2        & 3.20 \\
    \bottomrule
  \end{tabular}
\end{table}

DeepSeek-V3.2 (3.2) is the only judge to explicitly flag the same wording issue the human noted -- "it incorrectly classifies the synthesis method as `arc melting \& induction melting' when the text only states `melting'... on a water-cooled copper hearth" -- and additionally penalizes a hallucinated \texttt{"mix"} action label and the omission of the X-ray Debye-Scherrer diffraction characterization step as a separate process step. Gemini-3-Flash (4.9) never engages with the method-name wording at all, instead praising that "the synthesis section is brief but well-captured by the ontology." Claude-Sonnet-4.6 and Qwen3.5-397B-A17B (4.2 and 4.1) fall between the two, both naming the same "induction melting" issue as "partially inaccurate" or a "notable semantic error" without weighting it as heavily as DeepSeek-V3.2. As in Examples 1 and 2, the two extreme judges (Gemini-3-Flash highest, DeepSeek-V3.2 lowest) diverge the most from the human's 4.5, while the two middle judges track it most closely.

\vspace{0.5em}
\paragraph{Example 4: Correct extraction, judges still disagree by up to 1.7 points (Material: SrTiO\textsubscript{3} surface preparation)}

Source: \textit{"The samples were etched in 1:6 buffered oxide etch for 2 minutes to achieve TiO\textsubscript{2} termination."}, extracted by Gemini-3-Flash. \texttt{synthesis\_method = "other"} matches ground truth, and all three annealing/etching steps are captured.

\begin{table}[t]
  \centering
  \caption{Evaluation scores for the SrTiO\textsubscript{3} surface-preparation extraction (Gemini-3-Flash as extractor) from all four candidate judges and the human annotator.}
  \begin{tabular}{lc}
    \toprule
    \textbf{Judge} & \textbf{Overall score} \\
    \midrule
    Human               & 4.75 \\
    Gemini-3-Flash      & 4.90 \\
    Claude-Sonnet-4.6    & 3.60 \\
    Qwen3.5-397B-A17B    & 3.80 \\
    DeepSeek-V3.2        & 3.20 \\
    \bottomrule
  \end{tabular}
\end{table}

Gemini-3-Flash (4.9) again gives a near-perfect score, describing the extraction of temperature, duration, and atmosphere as "highly accurate" without qualification. DeepSeek-V3.2 (3.2) instead flags "several significant errors of commission": the buffered oxide etch is incorrectly listed as a starting material with a defined amount, and characterization equipment (AFM, detector) is included as if it were part of the synthesis. Claude-Sonnet-4.6 (3.6) and Qwen3.5-397B-A17B (3.8) land in between, both noting smaller completeness gaps (a missing beamline detail, a vendor hallucination for the detector) without the "over-extraction" framing DeepSeek-V3.2 uses.

\vspace{0.5em}
\paragraph{Example 5: Correct extraction, judges still disagree by up to 2.0 points (Material: (La\textsubscript{0.3}Sr\textsubscript{0.7})(Al\textsubscript{0.65}Ta\textsubscript{0.35})O\textsubscript{3}/SrTiO\textsubscript{3} interface).}

Source: \textit{"Deposit (La\textsubscript{0.3}Sr\textsubscript{0.7})(Al\textsubscript{0.65}Ta\textsubscript{0.35})O\textsubscript{3} layers onto TiO\textsubscript{2}-terminated STO (001) substrates"}, extracted by Gemini-3-Flash via pulsed laser deposition (PLD), matching ground truth.

\begin{table}[t]
  \centering
  \caption{Evaluation scores for the (La\textsubscript{0.3}Sr\textsubscript{0.7})(Al\textsubscript{0.65}Ta\textsubscript{0.35})O\textsubscript{3}/SrTiO\textsubscript{3} extraction (Gemini-3-Flash as extractor) from all four candidate judges and the human annotator.}
  \begin{tabular}{lc}
    \toprule
    \textbf{Judge} & \textbf{Overall score} \\
    \midrule
    Human               & 4.00 \\
    Gemini-3-Flash      & 4.90 \\
    Claude-Sonnet-4.6    & 3.90 \\
    Qwen3.5-397B-A17B    & 4.10 \\
    DeepSeek-V3.2        & 2.90 \\
    \bottomrule
  \end{tabular}
\end{table}

DeepSeek-V3.2 (2.9) again applies its strictest reading, calling out that "the extraction shows significant structural issues: it incorrectly populates the `starting\_materials' list... and conflates specific conditions" across the paper's two related interfaces (LAO/STO and LSAT/STO). Gemini-3-Flash (4.9) instead highlights only what was captured correctly -- laser fluence, repetition rate, growth temperature -- describing the duration calculation as showing "a high level of inference". Claude-Sonnet-4.6 (3.9) and Qwen3.5-397B-A17B (4.1) both note the same missing Hall-bar patterning step and incomplete oxygen-pressure range as DeepSeek-V3.2, but without its harsher framing.

\vspace{0.5em}
\paragraph{Example 6: Correct extraction, judges still disagree by up to 1.7 points (Material: 5-AGNR graphene nanoribbon)}

Source: \textit{"Annealing of 3,9(10)-dibromoperylene (DBP) monomer"}, extracted by Gemini-3-Flash via chemical vapor deposition (CVD), matching ground truth.

\begin{table}[t]
  \centering
  \caption{Evaluation scores for the 5-AGNR extraction (Gemini-3-Flash as extractor) from all four candidate judges and the human annotator.}
  \begin{tabular}{lc}
    \toprule
    \textbf{Judge} & \textbf{Overall score} \\
    \midrule
    Human               & 4.75 \\
    Gemini-3-Flash      & 5.00 \\
    Claude-Sonnet-4.6    & 3.60 \\
    Qwen3.5-397B-A17B    & 3.30 \\
    DeepSeek-V3.2        & 3.40 \\
    \bottomrule
  \end{tabular}
\end{table}

Gemini-3-Flash (5.0) calls the extraction's performance "exceptional," crediting it for correctly distinguishing the 5-AGNR variant's temperature (350\textdegree{}C) from other variants mentioned in the same source paragraph. DeepSeek-V3.2, Claude-Sonnet-4.6, and Qwen3.5-397B-A17B (3.4, 3.6, 3.3) all instead converge on the same completeness gap: the extraction omits the post-growth transfer procedure (PMMA coating, HF etching, gold removal) described later in the source text, which none of the three treats as disqualifying but all treat as a real gap worth a full point or more.

Across all six examples Gemini-3-Flash is the most permissive judge, and DeepSeek-V3.2 is the harshest in five of the six (the exception is Example 6, where Qwen3.5-397B-A17B scores 3.30 against DeepSeek-V3.2's 3.40, a gap well inside the scoring granularity). This gap is not confined to cases with a genuine underlying error -- it recurs even when every judge agrees the extraction is correct, indicating a systematic difference in judge calibration rather than noise from a single example.

\paragraph{Scaling LLM-as-a-judge across the dataset}\label{app:sec:llm-judge-scaling}

\ref{table:llm-syn-scores-synthesis-type} and Table~\ref{table:llm-syn-scores-material-type} show the performance of LLM-as-a-judge across the dataset.
Where we report Spearman rank correlations ($\rho$) between human and model scores, we compute their $p$-values using a permutation test (10{,}000 resamples, two-sided). This choice is motivated by the modest sample size and the bounded, quasi-ordinal nature of the scores, which induce many ties and can render asymptotic $p$-values anticonservative and unreliable. As the SciPy documentation recommends,\footnote{\url{https://docs.scipy.org/doc/scipy/reference/generated/scipy.stats.spearmanr.html}} “for small samples, consider performing a permutation test instead of relying on the asymptotic p-value,” especially when ties and discrete data violate large-sample assumptions. The permutation procedure generates the exact finite-sample null distribution of $\rho$ by permuting only one input (human scores) relative to the other while preserving marginal distributions. This approach provides valid inference under exchangeability, naturally handles ties, and ensures robust significance testing even with small, discrete datasets.

{
\scriptsize 
\setlength{\tabcolsep}{2.5pt} 
\renewcommand{\arraystretch}{1.1}

\begin{xltabular}{\textwidth}{l *{9}{>{\centering\arraybackslash}X}}
    \caption{Average LLM-judged extraction scores for the 35 synthesis methods in \texttt{LeMat-Synth} (full 81K-paper corpus, $N=58{,}345$ evaluated procedures). Scores are reported as mean $\pm$ standard deviation on a 1--5 scale. The Overall Score is the average of all seven evaluation criteria. \textbf{Note}: the ``other'' method row is dominated by hard extraction failures (e.g.\ schema-validation errors that empty out the ontology), which pulls its overall score (3.03) below all but one named method (liquid-phase epitaxy, 2.92); see the analogous note for material category below.}
    \label{table:llm-syn-scores-synthesis-type} \\
\toprule
\textbf{Synthesis} & \textbf{Structural} & \textbf{Material} & \textbf{Process} & \textbf{Equipment} & \textbf{Condition} & \textbf{Semantic} & \textbf{Format} & \textbf{Overall} & \textbf{Count} \\ 
\textbf{method} & \textbf{compl.} & \textbf{compl.} & \textbf{steps} & \textbf{extraction} & \textbf{extraction} & \textbf{accuracy} & \textbf{compliance} & \textbf{score} & \\
\midrule
\endfirsthead

\multicolumn{10}{c}{\textit{\tablename\ \thetable{} -- Continued from previous page}} \\[0.5em]
\toprule
\textbf{Synthesis} & \textbf{Structural} & \textbf{Material} & \textbf{Process} & \textbf{Equipment} & \textbf{Condition} & \textbf{Semantic} & \textbf{Format} & \textbf{Overall} & \textbf{Count} \\ 
\textbf{method} & \textbf{completeness} & \textbf{completeness} & \textbf{steps} & \textbf{extraction} & \textbf{extraction} & \textbf{accuracy} & \textbf{compliance} & \textbf{score} & \\
\midrule
\endhead

\bottomrule
\endlastfoot

other & 2.82$\pm$1.37 & 2.89$\pm$1.46 & 2.94$\pm$1.50 & 2.98$\pm$1.57 & 2.99$\pm$1.55 & 2.84$\pm$1.35 & 3.75$\pm$1.57 & 3.03$\pm$1.31 & 15,228 \\
solid-state & 3.52$\pm$0.89 & 3.43$\pm$1.13 & 3.61$\pm$0.99 & 3.59$\pm$1.12 & 3.84$\pm$1.01 & 3.44$\pm$1.11 & 4.76$\pm$0.34 & 3.74$\pm$0.75 & 7,304 \\
PVD & 3.00$\pm$0.89 & 2.85$\pm$0.93 & 2.90$\pm$1.05 & 3.50$\pm$0.88 & 3.29$\pm$1.08 & 2.77$\pm$1.04 & 4.58$\pm$0.35 & 3.27$\pm$0.69 & 6,271 \\
molecular beam epitaxy & 2.99$\pm$0.87 & 2.68$\pm$0.94 & 2.87$\pm$1.07 & 3.45$\pm$0.92 & 3.14$\pm$1.13 & 2.80$\pm$1.05 & 4.58$\pm$0.35 & 3.22$\pm$0.71 & 4,247 \\
flux growth & 3.81$\pm$0.91 & 3.64$\pm$1.26 & 3.73$\pm$1.12 & 4.05$\pm$1.12 & 4.04$\pm$1.14 & 3.73$\pm$1.11 & 4.80$\pm$0.32 & 3.97$\pm$0.80 & 3,678 \\
arc melting \& induction melting & 3.47$\pm$0.81 & 3.30$\pm$1.08 & 3.50$\pm$0.97 & 3.59$\pm$0.87 & 3.80$\pm$0.98 & 3.18$\pm$1.06 & 4.70$\pm$0.35 & 3.65$\pm$0.68 & 2,894 \\
CVD & 3.13$\pm$0.90 & 3.06$\pm$0.99 & 3.07$\pm$1.09 & 3.36$\pm$1.11 & 3.24$\pm$1.21 & 2.97$\pm$1.05 & 4.62$\pm$0.35 & 3.35$\pm$0.76 & 2,495 \\
precipitation & 3.39$\pm$0.93 & 3.53$\pm$1.01 & 3.78$\pm$0.95 & 3.30$\pm$1.18 & 3.48$\pm$0.99 & 3.36$\pm$1.09 & 4.74$\pm$0.32 & 3.65$\pm$0.73 & 1,936 \\
float zone \& Bridgman & 3.29$\pm$0.92 & 3.08$\pm$1.42 & 3.21$\pm$1.10 & 3.68$\pm$1.03 & 3.55$\pm$1.43 & 3.21$\pm$1.08 & 4.63$\pm$0.41 & 3.52$\pm$0.85 & 1,763 \\
pulsed laser deposition & 3.00$\pm$0.84 & 2.89$\pm$0.93 & 2.85$\pm$1.03 & 3.47$\pm$0.83 & 3.39$\pm$1.06 & 2.68$\pm$1.04 & 4.49$\pm$0.43 & 3.25$\pm$0.68 & 1,690 \\
solution-based & 3.35$\pm$0.84 & 3.31$\pm$1.04 & 3.51$\pm$0.97 & 3.37$\pm$1.17 & 3.49$\pm$1.03 & 3.31$\pm$1.05 & 4.70$\pm$0.34 & 3.58$\pm$0.70 & 1,311 \\
self-assembly & 3.28$\pm$0.79 & 3.31$\pm$0.90 & 3.50$\pm$0.92 & 3.28$\pm$1.21 & 3.24$\pm$1.02 & 3.22$\pm$0.97 & 4.69$\pm$0.33 & 3.50$\pm$0.67 & 1,053 \\
sol-gel & 3.39$\pm$0.85 & 3.27$\pm$1.08 & 3.65$\pm$0.92 & 3.32$\pm$1.09 & 3.65$\pm$0.89 & 3.29$\pm$1.03 & 4.71$\pm$0.34 & 3.61$\pm$0.69 & 994 \\
hydrothermal & 3.61$\pm$0.84 & 3.58$\pm$1.03 & 3.91$\pm$0.88 & 3.80$\pm$0.85 & 3.91$\pm$0.84 & 3.57$\pm$1.03 & 4.79$\pm$0.32 & 3.88$\pm$0.64 & 893 \\
mechanical mixing & 3.19$\pm$0.85 & 3.18$\pm$0.95 & 3.09$\pm$1.04 & 3.25$\pm$1.19 & 3.24$\pm$1.23 & 2.63$\pm$1.06 & 4.59$\pm$0.36 & 3.31$\pm$0.72 & 844 \\
solvothermal & 3.59$\pm$0.85 & 3.74$\pm$1.02 & 3.88$\pm$0.91 & 3.55$\pm$0.99 & 3.82$\pm$0.83 & 3.59$\pm$1.06 & 4.79$\pm$0.30 & 3.85$\pm$0.68 & 802 \\
electrochemical deposition & 3.19$\pm$0.78 & 3.17$\pm$0.93 & 3.16$\pm$1.00 & 3.38$\pm$0.96 & 3.10$\pm$1.09 & 2.95$\pm$1.05 & 4.60$\pm$0.33 & 3.36$\pm$0.66 & 774 \\
ball milling & 3.37$\pm$0.80 & 3.27$\pm$1.01 & 3.45$\pm$1.00 & 3.93$\pm$0.81 & 3.49$\pm$0.93 & 3.17$\pm$1.06 & 4.67$\pm$0.36 & 3.62$\pm$0.68 & 637 \\
atomic layer deposition & 2.59$\pm$0.94 & 2.80$\pm$0.92 & 2.76$\pm$1.03 & 3.36$\pm$0.96 & 3.11$\pm$1.16 & 2.72$\pm$1.04 & 4.57$\pm$0.33 & 3.13$\pm$0.71 & 560 \\
ion implantation & 3.47$\pm$0.69 & 3.08$\pm$0.80 & 3.45$\pm$0.93 & 3.66$\pm$0.83 & 3.53$\pm$0.93 & 3.15$\pm$0.97 & 4.63$\pm$0.35 & 3.57$\pm$0.58 & 392 \\
coprecipitation & 3.26$\pm$0.89 & 3.30$\pm$0.98 & 3.71$\pm$0.92 & 3.04$\pm$1.25 & 3.47$\pm$1.11 & 3.33$\pm$1.01 & 4.70$\pm$0.34 & 3.54$\pm$0.73 & 342 \\
microwave-assisted & 3.40$\pm$0.79 & 3.51$\pm$0.93 & 3.78$\pm$0.85 & 3.83$\pm$0.78 & 3.58$\pm$0.83 & 3.39$\pm$0.98 & 4.79$\pm$0.29 & 3.75$\pm$0.60 & 309 \\
template-directed & 3.47$\pm$0.75 & 3.34$\pm$0.87 & 3.76$\pm$0.75 & 3.40$\pm$0.96 & 3.46$\pm$0.91 & 3.49$\pm$0.92 & 4.73$\pm$0.30 & 3.67$\pm$0.60 & 289 \\
wet impregnation & 3.29$\pm$0.79 & 3.19$\pm$0.95 & 3.63$\pm$0.87 & 3.23$\pm$1.02 & 3.60$\pm$0.83 & 3.20$\pm$1.07 & 4.67$\pm$0.33 & 3.55$\pm$0.64 & 236 \\
combustion & 3.22$\pm$0.83 & 3.10$\pm$1.02 & 3.47$\pm$0.93 & 3.31$\pm$1.02 & 3.33$\pm$0.94 & 3.16$\pm$1.04 & 4.66$\pm$0.31 & 3.46$\pm$0.69 & 211 \\
liquid-phase epitaxy & 2.72$\pm$0.99 & 2.69$\pm$0.97 & 2.51$\pm$1.14 & 2.73$\pm$1.44 & 2.85$\pm$1.51 & 2.47$\pm$1.09 & 4.45$\pm$0.43 & 2.92$\pm$0.84 & 210 \\
spark plasma sintering & 3.39$\pm$0.73 & 3.01$\pm$0.92 & 3.40$\pm$0.96 & 3.87$\pm$0.70 & 3.76$\pm$0.94 & 3.01$\pm$1.01 & 4.70$\pm$0.36 & 3.59$\pm$0.58 & 194 \\
electrospinning & 3.43$\pm$0.73 & 3.33$\pm$0.88 & 3.76$\pm$0.79 & 3.80$\pm$0.76 & 3.55$\pm$1.00 & 3.36$\pm$0.96 & 4.72$\pm$0.32 & 3.71$\pm$0.56 & 135 \\
sonochemical & 3.64$\pm$0.69 & 3.60$\pm$0.95 & 3.99$\pm$0.69 & 3.90$\pm$0.80 & 3.67$\pm$0.74 & 3.49$\pm$0.90 & 4.78$\pm$0.30 & 3.87$\pm$0.54 & 127 \\
lithographic patterning & 3.47$\pm$0.77 & 3.27$\pm$0.77 & 3.41$\pm$0.94 & 3.64$\pm$0.94 & 3.26$\pm$1.38 & 3.04$\pm$1.03 & 4.70$\pm$0.38 & 3.54$\pm$0.66 & 122 \\
chemical bath deposition & 3.20$\pm$0.74 & 3.25$\pm$0.83 & 3.55$\pm$0.94 & 2.84$\pm$0.99 & 3.45$\pm$0.97 & 3.16$\pm$0.86 & 4.64$\pm$0.37 & 3.44$\pm$0.60 & 113 \\
spray pyrolysis & 3.24$\pm$0.79 & 2.99$\pm$0.99 & 3.40$\pm$0.99 & 3.58$\pm$0.77 & 3.27$\pm$0.92 & 3.02$\pm$1.07 & 4.60$\pm$0.35 & 3.44$\pm$0.68 & 103 \\
arc discharge & 3.34$\pm$0.70 & 3.02$\pm$0.86 & 3.17$\pm$0.95 & 3.52$\pm$0.80 & 2.99$\pm$1.13 & 3.16$\pm$0.89 & 4.63$\pm$0.31 & 3.40$\pm$0.63 & 85 \\
mechanochemical & 3.36$\pm$0.80 & 3.53$\pm$1.02 & 3.76$\pm$0.87 & 4.01$\pm$0.82 & 3.39$\pm$1.11 & 3.38$\pm$1.00 & 4.77$\pm$0.32 & 3.74$\pm$0.69 & 77 \\
incipient wetness impregnation & 2.92$\pm$0.58 & 3.04$\pm$0.84 & 3.58$\pm$0.84 & 2.44$\pm$1.18 & 3.44$\pm$0.85 & 3.27$\pm$1.00 & 4.67$\pm$0.37 & 3.33$\pm$0.54 & 26 \\
\end{xltabular}
}

\begin{table}[t]
    \caption{Average LLM-judged extraction scores for the 16 material categories in \texttt{LeMat-Synth} (full 81K-paper corpus, $N=58{,}345$ evaluated procedures). Scores are reported as mean $\pm$ standard deviation on a 1--5 scale. The Overall Score is the average of all seven evaluation criteria. \textbf{Note}: the ``other'' category (materials outside our 15 named classes) scores substantially lower across every criterion (overall $1.74\pm1.30$ vs.\ $3.4$--$3.8$ elsewhere) because it disproportionately contains hard extraction failures, e.g.,\ schema-validation errors where the model returns a target-compound-type outside the ontology's enumeration, emptying every downstream field. Its uniquely low format-compliance score ($1.46$, against $4.4$--$4.8$ for every other category) is the signature of this: these are records that failed schema validation, not records describing unusual materials.}
    
    \label{table:llm-syn-scores-material-type}
  \centering
  \renewcommand{\arraystretch}{1.15}
  \resizebox{\textwidth}{!}{
 \begin{tabular}{lccccccccc}
\toprule
\textbf{Material} & \textbf{Structural} & \textbf{Material} & \textbf{Process} & \textbf{Equipment} & \textbf{Condition} & \textbf{Semantic} & \textbf{Format} & \textbf{Overall} & \textbf{Count} \\ 
\textbf{category} & \textbf{completeness} & \textbf{completeness} & \textbf{steps} & \textbf{extraction} & \textbf{extraction} & \textbf{accuracy} & \textbf{compliance} & \textbf{score} & \textbf{} \\
\midrule
semiconductors \& electronic & 3.22$\pm$0.96 & 3.06$\pm$1.11 & 3.16$\pm$1.14 & 3.56$\pm$1.07 & 3.40$\pm$1.22 & 3.06$\pm$1.14 & 4.64$\pm$0.37 & 3.44$\pm$0.81 & 11,564 \\
ceramics \& glasses & 3.24$\pm$1.04 & 3.19$\pm$1.23 & 3.30$\pm$1.17 & 3.51$\pm$1.16 & 3.62$\pm$1.18 & 3.17$\pm$1.23 & 4.68$\pm$0.37 & 3.53$\pm$0.87 & 10,130 \\
metals \& alloys & 3.38$\pm$0.92 & 3.20$\pm$1.13 & 3.31$\pm$1.10 & 3.67$\pm$0.98 & 3.65$\pm$1.13 & 3.13$\pm$1.13 & 4.68$\pm$0.36 & 3.57$\pm$0.77 & 9,076 \\
nanomaterials & 3.32$\pm$0.85 & 3.25$\pm$0.98 & 3.44$\pm$1.02 & 3.49$\pm$1.06 & 3.47$\pm$1.06 & 3.21$\pm$1.06 & 4.68$\pm$0.34 & 3.55$\pm$0.70 & 8,210 \\
other & 1.69$\pm$1.40 & 1.76$\pm$1.53 & 1.76$\pm$1.54 & 1.78$\pm$1.55 & 1.77$\pm$1.55 & 1.93$\pm$1.19 & 1.46$\pm$0.92 & 1.74$\pm$1.30 & 4,426 \\
two-dimensional materials & 3.37$\pm$0.90 & 3.25$\pm$1.04 & 3.27$\pm$1.08 & 3.52$\pm$1.10 & 3.48$\pm$1.14 & 3.12$\pm$1.14 & 4.67$\pm$0.34 & 3.53$\pm$0.78 & 4,119 \\
hybrid \& organic-inorganic & 3.32$\pm$0.90 & 3.49$\pm$1.02 & 3.72$\pm$0.95 & 3.29$\pm$1.21 & 3.48$\pm$1.03 & 3.34$\pm$1.08 & 4.71$\pm$0.35 & 3.62$\pm$0.73 & 2,854 \\
polymers \& soft matter & 3.15$\pm$0.93 & 3.36$\pm$0.99 & 3.49$\pm$1.01 & 3.29$\pm$1.17 & 3.38$\pm$1.08 & 3.14$\pm$1.06 & 4.69$\pm$0.34 & 3.50$\pm$0.73 & 2,431 \\
functional materials \& catalysts & 3.47$\pm$0.92 & 3.46$\pm$1.14 & 3.60$\pm$1.06 & 3.55$\pm$1.24 & 3.67$\pm$1.15 & 3.43$\pm$1.10 & 4.71$\pm$0.37 & 3.70$\pm$0.79 & 1,477 \\
framework \& porous materials & 3.55$\pm$0.88 & 3.67$\pm$1.02 & 3.90$\pm$0.92 & 3.55$\pm$1.08 & 3.70$\pm$0.94 & 3.62$\pm$1.02 & 4.81$\pm$0.29 & 3.83$\pm$0.70 & 1,312 \\
composites & 3.42$\pm$0.75 & 3.22$\pm$0.90 & 3.60$\pm$0.89 & 3.51$\pm$0.97 & 3.56$\pm$0.97 & 3.29$\pm$0.97 & 4.67$\pm$0.35 & 3.61$\pm$0.63 & 1,006 \\
biomaterials \& biological & 3.25$\pm$0.86 & 3.32$\pm$0.93 & 3.41$\pm$0.96 & 3.48$\pm$1.13 & 3.28$\pm$1.04 & 3.02$\pm$1.05 & 4.67$\pm$0.36 & 3.49$\pm$0.71 & 664 \\
liquid materials & 3.17$\pm$0.97 & 3.44$\pm$1.03 & 3.32$\pm$1.12 & 3.33$\pm$1.23 & 3.19$\pm$1.18 & 2.93$\pm$1.17 & 4.63$\pm$0.40 & 3.43$\pm$0.79 & 610 \\
emerging \& quantum materials & 3.70$\pm$0.89 & 3.44$\pm$1.16 & 3.57$\pm$1.15 & 3.84$\pm$1.08 & 3.83$\pm$1.12 & 3.44$\pm$1.23 & 4.76$\pm$0.32 & 3.80$\pm$0.82 & 279 \\
energy \& sustainability & 3.53$\pm$0.86 & 3.48$\pm$1.01 & 3.50$\pm$1.14 & 3.75$\pm$0.98 & 3.67$\pm$1.03 & 3.27$\pm$1.21 & 4.71$\pm$0.32 & 3.70$\pm$0.76 & 170 \\
smart \& responsive materials & 3.35$\pm$0.77 & 3.38$\pm$0.86 & 3.18$\pm$0.85 & 3.85$\pm$1.01 & 3.35$\pm$1.07 & 2.88$\pm$0.91 & 4.74$\pm$0.36 & 3.53$\pm$0.59 & 17 \\
\bottomrule
\end{tabular}
}
\end{table}

\subsection{Figure extraction}
\label{app:sec:figure-extraction}

\paragraph{Segmenting large figures into subplots.} To extract individual subplots from figures in research papers, we employ Florence-2 \cite{xiao2024florence} fine-tuned with a LoRA adapter \cite{hu2021lora} on 695 annotated samples for joint subplot detection and classification. Given a figure, the model is prompted with the object-detection task token \texttt{<OD>} and outputs, in a single forward pass, both a bounding box and a label (\emph{quantitative plot} or \emph{qualitative plot}) for each detected subplot region. This unifies the two-stage detect-then-classify approach of earlier pipelines into one model call. Detected boxes covering more than 90\% of the figure area are discarded, as these correspond to the full figure. Remaining boxes are expanded by 5\% on each side to ensure complete coverage of axis labels and tick marks before cropping. Empirical results indicate that this approach reliably identifies and classifies subplots across a variety of figure types. \label{app:sec:img-segmenter}

\paragraph{Classifying plots with quantitative data.} The deployed pipeline uses the single-pass Florence-2 detect-and-classify model described above. Before settling on it, we evaluated a separate two-stage baseline in which classification is handled by its own model, reported here for completeness; it is not used to produce any result in this work. That baseline employs a ResNet-152 model \cite{he2016deep}, pretrained on ImageNet and fine-tuned on the DocFig dataset \cite{jobin2019docfigure}. The dataset is split into 19,000 samples for training and 13,000 samples for testing. The model is trained with default hyperparameters for 20 epochs using the Adam optimizer with a learning rate of 1e-3. Our classification task focuses exclusively on the plot types “line chart”, “bar plot” and “scatter plot” which are relevant for downstream information extraction; qualitative figures are excluded from further processing. The fine-tuned model achieves an F1-score of 88.03\% on the test set, indicating strong performance in accurately identifying quantitative plots for subsequent analysis.\label{app:sec:img-classifier}

\paragraph{Extracting data with a vision LLM.} To convert these numerical figures into a structured and interpretable format for further use, we explore the capabilities of advanced vision-language models to extract data from line plots, focusing on 2D coordinate retrieval. Inspired by \cite{polak2025leveraging}, where multimodal models were used to extract and regenerate plots, we use \texttt{Qwen3.5-397B-A17B} (selected per case study via the VLM benchmark in Table~\ref{tab:thermocat-vlm-extraction}; the extraction prompt is configurable per domain, see Section~\ref{app:subsec:prompts}) to extract 2D coordinates with their corresponding series names, as well as metadata fields like titles, axis labels, and units. The model is prompted to output a JSON object in a predefined schema, which is then parsed into a Pydantic object to ensure data consistency and structured integration into our data extraction pipeline.\label{app:sec:claude-extractor}

\paragraph{Figure extraction evaluation}\label{app:sec:fig-extr-eval}

Extracted 2D coordinates are evaluated against human-annotated ground truth using root mean square error (RMSE) and mean absolute error (MAE), computed on matched series only. Series-level detection is separately scored by precision, recall, and F1. Table~\ref{tab:thermocat-vlm-extraction} reports these metrics for the thermocatalysis line-chart extraction benchmark.

\subsection{Linking}

\paragraph{Thermocatalysis performance-synthesis linking quality (LLM-as-judge)}

For the thermocatalysis case study, each linked plot-series-to-material
mapping is scored by an LLM-as-judge (\ref{app:sec:img-segmenter},
\texttt{DspyNameMatcherJudge}'s companion linking-evaluation call) across
four criteria (material identity, performance-data correctness,
completeness, and format/structure), combined into a single
\texttt{overall\_score} (1--5 scale), and flagged against nine predefined
failure modes (name mismatch, one-to-many synthesis, many-to-one figure,
sample-code failure, precursor-vs-product confusion, characterization
confusion, dual-axis error, false negative, false positive). Per-paper
output is stored alongside each extracted material's record.

Of the four candidate extractor VLMs, Claude-Sonnet-4.6 has both the most
complete linking-evaluation coverage (26 of the 27 papers for which a
linking evaluation was attempted, the remaining paper's evaluation call
returned \texttt{null}) and the highest
scores (mean \texttt{overall\_score}$=4.54$, range $4.0$--$5.0$, 26/26
papers score $>3$, 25/26 score $>4$). Gemini-3-Flash and Qwen3.5-397B-A17B
have lower coverage (21/27 and 23/27 scored respectively) and lower, more
variable scores (mean $4.11$, range $1.5$--$4.8$, and mean $4.37$, range
$1.2$--$4.8$). DeepSeek-V3.2 produced linking-evaluation files for only 3
papers, none with a completed score, because its inference endpoint has no
image-input support and it could not process the figures at all
(\ref{tab:thermocat-vlm-extraction}); it is therefore not a candidate here
and no comparison should be drawn from those 3 files. Given its
completeness and score quality, Claude-Sonnet-4.6 is used for this
linking-quality analysis. Note that the main-text figure panels (Fig.~4b--d)
are rendered from the Qwen3.5-397B-A17B extraction, the VLM deployed for
this case study on the precision/recall benchmark
(\ref{tab:thermocat-vlm-extraction}); the two serve different purposes and
the linking-quality statistics below describe Claude-Sonnet-4.6's output.

Among Claude-Sonnet-4.6's 26 scored papers, the most common flagged failure
mode is \texttt{f8\_false\_negative} (23/26 papers have at least one
performance series the pipeline failed to link to a material), followed by
\texttt{f1\_name\_mismatch} (4/26) and \texttt{f2\_one\_to\_many\_synthesis}
/ \texttt{f3\_many\_to\_one\_figure} (2/26 each). No papers were flagged for
sample-code failure, precursor-vs-product confusion, characterization
confusion, dual-axis error, or false positives. Note that failure mode F8 is flagged because the pipeline links each paper's primary performance data but leaves secondary panels (stability tests, Arrhenius plots, literature-comparison curves) unlinked. This is enforced by the \texttt{PromptConfig} that compells to extract performance-temperature curves exclusively.

\paragraph{Failure cases}\label{app:sec:thermocat-failures}

The nine failure-mode flags are defined in
\ref{table:thermocat-failure-modes}, together with the number of
Claude-Sonnet-4.6-scored papers (of 26) in which each was triggered and a
concrete instance drawn from the judge's own reasoning trace for that
paper. Five of the nine modes were never observed in this corpus. We report
them for completeness since they remain part of the fixed evaluation
schema and may surface as the corpus grows.

\begingroup
\setlength{\LTcapwidth}{\textwidth}
\begin{longtable}{p{0.9cm}p{3.6cm}p{0.6cm}p{7.3cm}}
  \caption{The nine predefined linking-failure modes scored by the LLM-as-judge (Section~\ref{app:sec:thermocat-failures}), with the number of Claude-Sonnet-4.6-scored papers (of 26) in which each was flagged and a representative instance. F4--F7 and F9 were not observed in this corpus but remain part of the fixed evaluation schema.}
  \label{table:thermocat-failure-modes} \\
    \toprule
    \textbf{Code} & \textbf{Failure mode} & \textbf{N} & \textbf{Representative instance} \\
    \midrule
    \endfirsthead
    \toprule
    \textbf{Code} & \textbf{Failure mode} & \textbf{N} & \textbf{Representative instance} \\
    \midrule
    \endhead
    \bottomrule
    \endlastfoot
    F1 & Name mismatch -- text and figure use different names for the same material & 4 &
      Lorenzut et al. \cite{lorenzut_femo-based_2012}: the 10\%Fe-10\%Mo/8\%Y$_2$O$_3$-ZrO$_2$ catalyst's Figure 5 series was extracted as ``Fe2Mo8'' rather than ``Fe10Mo10'', so the correct synthesis record failed to match the plot series despite both referring to the same catalyst. \\
    F2 & One-to-many synthesis -- one synthesis produces multiple materials but only one was linked & 2 &
      Chen et al. \cite{chen_bimetallic_2020}: the NH$_3$-pretreated Ru-Fe/CNTs synthesis is present in the input, but only the H$_2$-pretreated variant's performance series (Figs.\ 6a, 7a, 7b) were linked -- the NH$_3$-pretreated material's own Figure 7b series was left with no associated performance data. \\
    F3 & Many-to-one figure -- multiple series in one figure not fully disambiguated & 2 &
      Duan et al. \cite{duan_mcm-412012}: Figure 6 plots four CoMo bimetallic ratios, but the extracted series names (e.g.\ ``Co\_9Mo\_3'', ``Co\_9Mo\_4'', ``Co\_3Mo\_3'') do not match any of the four true compositions (Co9Mo1, Co8Mo2, Co6Mo4, Co5Mo5), so none of the four were linked. \\
    F4 & Sample-code failure -- shorthand labels (e.g.\ ``S1'', ``NCA-3'') not resolved to their full material description & 0 & Not observed in this corpus. \\
    F5 & Precursor-vs-product confusion -- a precursor/intermediate synthesis linked to final-product performance data, or vice versa & 0 & Not observed in this corpus. \\
    F6 & Characterization confusion -- performance data confused with characterization data (XRD, TGA, BET, SEM, etc.) & 0 & Not observed in this corpus. \\
    F7 & Dual-axis error -- data series attributed to the wrong $y$-axis on a dual-axis plot & 0 & Not observed in this corpus. \\
    F8 & False negative -- a synthesis-performance pair clearly present in the paper was missed entirely & 23 &
      Wu et al.~\cite{wu_ammonia_2020}: the algorithm correctly linked the main NH$_3$-conversion activity curves, but missed the potassium-doped stability curve (Fig.\ 4d), the Arrhenius plots (Fig.\ 1c) for all three catalysts, and two of five bimetallic ratios in the primary activity plot (Fig.\ 1a). \\
    F9 & False positive -- a link produced for a material with no performance data, or no synthesis in the paper & 0 & Not observed in this corpus. \\
\end{longtable}
\endgroup

A concrete example of the F1 and F3 modes appears in Maleki et al.~\cite{maleki_coceo_2024} (Fig.~6, Co-Ce-Al oxide catalysts): one curve links correctly to its synthesis record; a second is a clean F1 case (one curve, one corrupted stoichiometry label -- ``Co$_{0.5}$Al$_{0.5}$O'' extracted instead of the true ``Co$_{0.9}$Al$_{0.1}$O'' -- a single fixable wrong answer); the remaining two curves are an F3 case, where the extracted names are close enough to each other and to the true compositions that no confident 1:1 assignment could be made, so the linker leaves the group unresolved. The same paper also exhibits F2 (one-to-many synthesis): the linked catalyst has unlinked performance data in three other figures (Figs.~7, 8, 12). Real mislinks (F1--F3) are rare and narrowly caused by naming or labeling issues, as in this example. No papers exhibit sample-code failures, precursor-product confusion, characterization confusion, dual-axis errors, or false positives. The more common flag, F8 (false negative: any secondary plot, such as a stability or kinetics curve, left unlinked while the paper's primary performance data links correctly), is technically not a linking error since the configuration specifically prompted to extract conversion-temperature curves only.

As with the superconductivity $T_c\geq150$~K flag
(\ref{app:sec:supercond-linking-fix}), the dominant failure mode here is a
missed link (F8): the judge's
\texttt{missing\_links} field shows the pipeline is conservative --
preferring to leave a series unlinked over guessing -- so recall on
secondary figures (stability tests, Arrhenius/kinetics panels, literature
comparison plots) is the main quality bottleneck.
Where mislinks do occur (F1--F3), the proximate cause is consistently a
numeric or naming discrepancy introduced upstream, either by the VLM's
plot-digitization step misreading a composition label (Lorentzut et al.~\cite{lorenzut_femo-based_2012},
Duan et al.~\cite{duan_mcm-412012}) or by the source figure itself using shorthand or
inconsistent stoichiometry notation across panels
(Chen et al.~\cite{chen_bimetallic_2020}).

\subsection{Benchmarks}\label{app:sec:benchmarking}

\subsubsection{Manual annotation}\label{app:sec:table-regen}
The human domain experts annotated 80 synthesis procedures as ground truth (held out from the main evaluation set). Four LLMs (Claude-Sonnet-4.6\cite{ClaudeSonnet46SystemCard2026}, DeepSeek-V3.2\cite{DeepSeekV32_2025}, Gemini-3-Flash\cite{Gemini3FlashModelCard2025}, Qwen3.5-397B-A17B\cite{Qwen3_5_2026}) were evaluated against this same set in both roles, as extractor and as judge. Because each procedure is scored once per extractor, and not every extractor produces a scoreable, name-matchable output for every procedure, the $4\times4$ extractor--judge matrix in Table~\ref{table:human-llm-comparison} and the main text is restricted to the $n=59$ (procedure, extractor) pairs for which a human score and all four judges' scores are jointly available.

\textbf{Selection criterion.} The judge model is selected as the one with the highest bias-corrected agreement with human scores, ICC(2,1), computed leave-one-out to exclude each judge's own self-scored cells (Table~\ref{table:human-llm-comparison}), with Spearman $\rho$ as a secondary, scale-invariant rank-consistency check. The extractor model is selected as the one with the highest mean \texttt{overall\_score} under the selected judge, cross-checked against the human-scored extractor ranking.


Claude-Sonnet-4.6 is simultaneously the best-agreeing judge (ICC(2,1)$_{\text{loo}}=0.276$, $\rho=0.406$, next-best DeepSeek-V3.2 at ICC(2,1)$=0.166$) and the top-ranked extractor by human judgment ($\mu=3.665$, see Table~\ref{table:human-llm-comparison} and the main text, Fig.~3b). Deploying Claude-Sonnet-4.6 for both extraction and judging across the full 80{,}940-paper corpus is nonetheless cost-prohibitive. We therefore deploy \textbf{Qwen3.5-397B-A17B as extractor and DeepSeek-V3.2 as judge} for the full-corpus run reported throughout this work. This is an explicit quality–cost trade-off rather than a claim that this pairing is optimal: DeepSeek-V3.2 is the second-best-agreeing judge in our benchmark (ICC(2,1)$_{\text{loo}}=0.166$, versus Claude-Sonnet-4.6's 0.276), and Qwen3.5-397B-A17B is the third-best extractor by human judgment ($\mu=3.251$, versus Claude-Sonnet-4.6's 3.665 and ahead of only DeepSeek-V3.2's 3.148). We report the benchmark above so readers can judge the quality this trade-off costs, rather than presenting the deployed pair as the best-performing one.

\subsubsection{Self-preference / cross-model bias analysis}

Self-preference is defined as a judge scoring its own model family's extractions systematically higher than it scores other models' extractions (diagonal vs. off-diagonal entries of the $4\times4$ extractor–judge matrix, corrected for the human-scored quality of the underlying extraction via difference-in-differences; see main text and Table~\ref{table:human-llm-comparison} discussion). Gemini-3-Flash is the only judge exhibiting substantial self-preference ($+1.28$ raw, DiD $+0.44$ vs. human); Claude-Sonnet-4.6 and Qwen3.5-397B-A17B show negligible self-preference, and DeepSeek-V3.2 is, if anything, harsher on its own outputs. Practical implication: outside of Gemini-3-Flash, self-preference is not a major confound in this framework, so judge selection can be made on agreement quality alone without needing to exclude same-family extractor–judge pairs as a separate precaution.

\subsection{Case study: Thermocatalysis}\label{app:casestudy-thermocat}
\begin{figure}
    \centering
    \includegraphics[width=\linewidth]{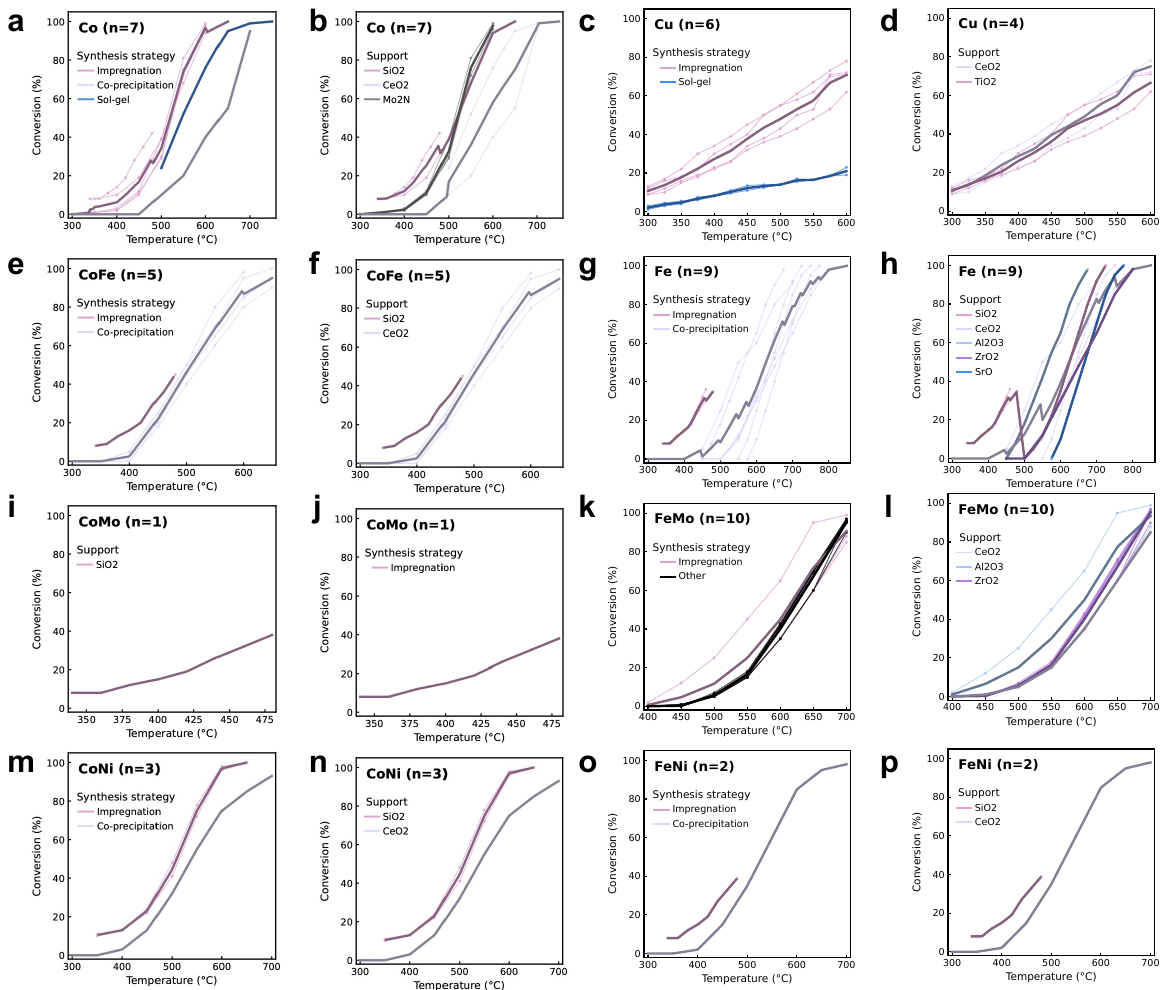}
    \caption{Extracted catalytic performance traces from the thermocatalysis corpus, one panel per metal system. Individual curves within each panel are colored by support or synthesis method.}
    \label{fig:si-tc-1-full}
\end{figure}

\begin{figure}
    \centering
    \includegraphics[width=\linewidth]{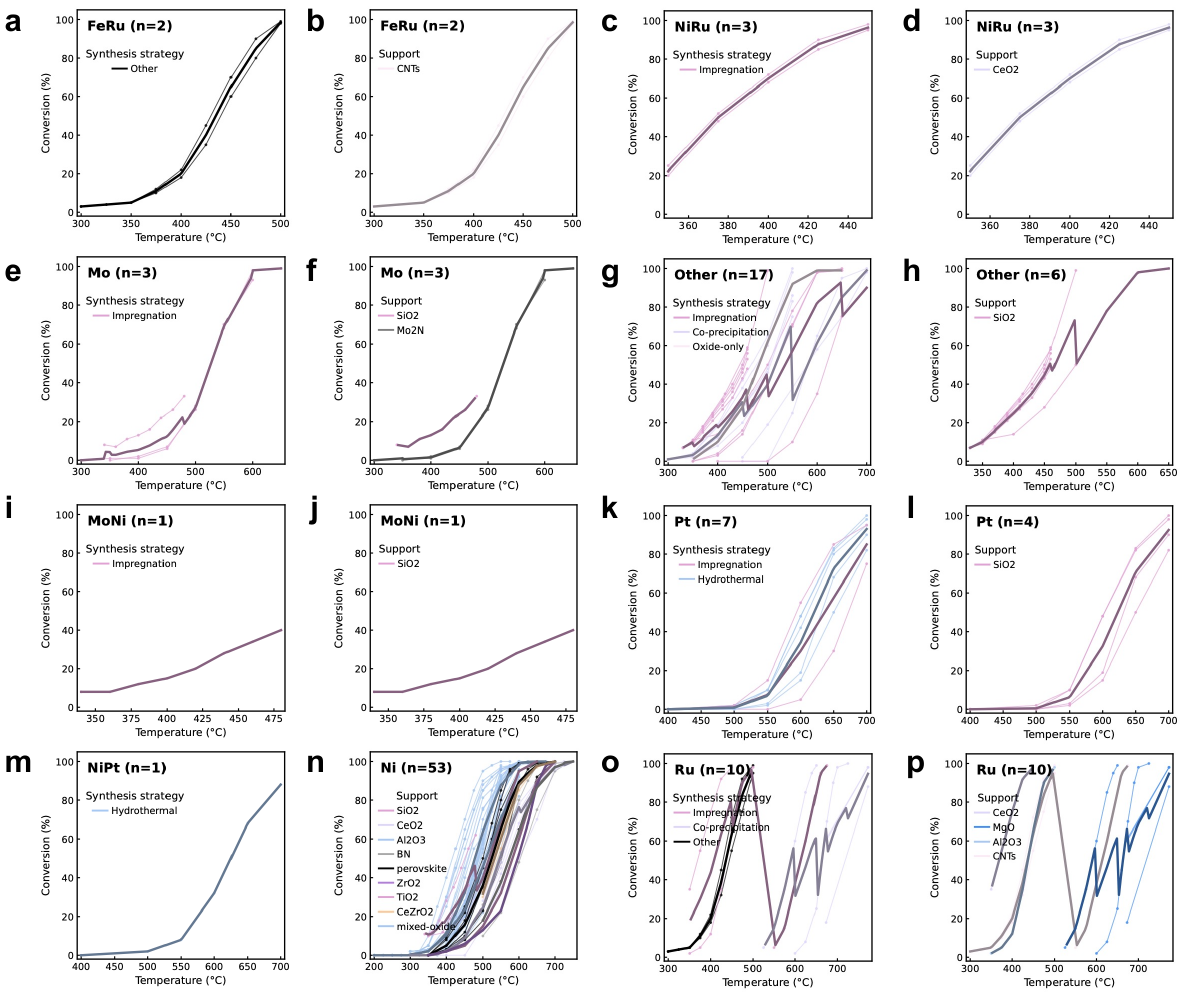}
    \caption{Extracted catalytic performance traces from the thermocatalysis corpus (continued), one panel per metal system. Individual curves within each panel are colored by support or synthesis method.}
    \label{fig:si-tc-2-full}
\end{figure}

The thermocatalysis corpus was assembled by domain experts working on thermocatalysis and comprises 28 largely closed-access papers, of which 23 contain at least one conversion-vs-temperature curve and supply the manually annotated ground truth and performance-curve data\cite{duan_mcm-412012,itoh_hydrogen_2002,noauthor_rare_2025,wu_ammonia_2020,chen_bimetallic_2020,do_turning_2024,fang_dispersed_2023,lucentini_catalytic_2019,maleki_coceo_2024,okura_ammonia_2018,podila_hydrogen_2016,teng_ru_2024,yi_plasma-assisted_2019,gao_plasma-assisted_2023,wu_engineering_2021,zhou_spatial_2021,dasireddy2021Fuel,meng_nh3_2024,podila_mgfe_2020,lorenzut_femo-based_2012,varisli_2012_cox,Srifa_2016_cox,chen_ru-based_2021}: 157 conversion-vs-temperature curves across 152 unique catalyst compositions, shown in full in Figures~\ref{fig:si-tc-1-full} and \ref{fig:si-tc-2-full}.

\begin{table}[h]
\centering
\caption{Figure extraction performance VLM benchmark on the thermocatalysis extraction task (23 papers; human ground truth: 152 catalyst compositions, 157 curves). Two matching strategies are compared: \textbf{fuzzy} (exact/substring normalization on material and series names, computed with no LLM judge) and \textbf{LLM-matched} (a DeepSeek-V3.2 name-matching judge, tolerant of paraphrase, unit, and notation differences). DeepSeek-V3.2 itself is excluded as a candidate VLM in this table. Its inference endpoint has no image-input support, so it could not be benchmarked on figure extraction, but is reused here purely as the name-matching judge for the LLM-matched columns. RMSE is computed on matched series only ($n_{\text{scored}}$); precision/recall/F1 are series-level detection metrics. Two denominator conventions differ between the two matching strategies and should be read with care. First, TP$+$FN is 95 in the fuzzy rows but 152 in the LLM-matched rows: the fuzzy strategy can only be scored against the subset of ground-truth series whose names normalize to a comparable form, whereas LLM matching is evaluated against all 152 annotated catalyst compositions. Recall is therefore not directly comparable across the two strategies, only within one. Second, $n_{\text{scored}}$ equals TP exactly in the fuzzy rows but is well below TP in the LLM-matched rows (e.g.\ 68 vs 89 for Qwen3.5-397B-A17B), because a series can be matched for detection purposes yet still lack the paired coordinate data needed to compute an RMSE (for instance when the ground-truth and predicted curves share no overlapping $x$-range); those series count as TP but are excluded from $n_{\text{scored}}$. Because the mean RMSE for each model is computed over a different-sized subset, RMSE comparisons across models carry a coverage confound: Gemini-3-Flash's lowest RMSE is computed over the fewest scored series.}
\label{tab:thermocat-vlm-extraction}
\resizebox{\linewidth}{!}{%
\begin{tabular}{llccccccc}
\toprule
VLM & Matching & Mean RMSE & $n_{\text{scored}}$ & TP & FP & FN & Precision & Recall / F1 \\
\midrule
\multirow{2}{*}{Qwen3.5-397B-A17B}
  & Fuzzy      & 0.084 & 42 & 42 & 16 & 53 & 0.724 & 0.442 / 0.549 \\
  & LLM-matched & 0.076 & 68 & 89 & 20 & 63 & 0.817 & 0.586 / \textbf{0.682} \\
\multirow{2}{*}{Claude-Sonnet-4.6}
  & Fuzzy      & 0.148 & 35 & 35 & 57 & 60 & 0.380 & 0.368 / 0.374 \\
  & LLM-matched & 0.140 & 65 & 107 & 53 & 45 & 0.669 & 0.704 / \textbf{0.686} \\
\multirow{2}{*}{Gemini-3-Flash}
  & Fuzzy      & 0.037 & 25 & 25 & 12 & 70 & 0.676 & 0.263 / 0.379 \\
  & LLM-matched & 0.048 & 41 & 57 & 18 & 95 & 0.760 & 0.375 / \textbf{0.502} \\
\bottomrule
\end{tabular}
}
\end{table}

\subsubsection{Linking module}
Plot series are matched to identified materials by a direct LLM prompt (\texttt{SeriesMaterialLinker}, no separate rule-based or NER normalization stage precedes it), configured with \texttt{gemini-3.1-pro-preview} as the linker model for this case study. The linker model is a per-domain configuration choice, not a fixed component: the superconductivity case study uses DeepSeek-V3.2 in the same slot (Section~\ref{app:sec:supercond}). Material-name normalization across synthesis text and figure legends is therefore delegated entirely to this LLM step.

\subsubsection{Extended statistics}
Figures are pre-filtered for catalysis-relevant performance data using \texttt{PlotFilterConfig.for\_catalysis()}: x-axis labels restricted to \{temperature, temp\} with units \{\textdegree C, K, \textdegree F, ...\}, and y-axis keywords restricted to \{conversion, yield, activity\} with percentage units, requiring a matching y-axis keyword when the axis is percentage-scaled.

Linked catalyst compositions are classified by active metal and support/oxide family (including spinel and mixed-oxide normalization) via a regex-based classification.

\subsection{Case study: Superconductors}\label{app:sec:supercond}
Superconductivity paper manual annotation\cite{Kang_2001,Rayaprol_2004,steiger2007pressurestudysuperconductivitymagnetism,Mudgel_2008,chen2008elementsubstitutioneffecttransition,Cheng_2008,Nakai_2009,wu2009superconductivity56ksamariumdoped,Matsuishi_2009,Liu_2009,Qi_2012,Lu_2014,Lou_2016,Moroni_2017,Liu_2019,Shang_2019,sigloch2022largeoutputvoltagemagnetic,Ou_2025}: 18 papers, hand-annotated
against 61 material rows
total, giving both a text-reported $T_c$ (\texttt{tc\_text\_human}, populated
for 40/61 rows -- the rest state no numeric text $T_c$) and a
plot-read $T_c$ (\texttt{tc\_human\_from\_plot}, populated for 59/61 rows).

The superconductor case study starts from the full 80,940-paper LeMat-Synth-Papers superset and narrows it by a category/keyword filter and a plot filter. Papers are first restricted to those tagged \texttt{Superconductor}; all three splits are then filtered to those whose abstract contains ``resistivity''. Each remaining paper's full text is passed to \texttt{gemini-2.5-flash} (temperature 0), asked whether the paper contains a resistivity/resistance-vs-temperature plot, explicitly excluding plots whose only varying parameter across curves is applied magnetic field or pressure (a common false positive, since many superconductor papers report $\rho(T)$ or $R(T)$ swept across $H$ or GPa). This two-stage filter yields 1,384 papers used for downstream $T_c$ extraction. Text and figures are extracted from these PDFs with Mistral OCR.

\paragraph{$T_c$ extraction pipelines}
For figure extraction, a per-figure VLM call (Qwen3.5-397B-A17B) reads three temperatures off the $\rho(T)$ or $R(T)$ curve: $T_\text{onset}$, where resistance first drops sharply; $T_\text{zero}$, where it first reaches approximately zero; and the midpoint $T_{c,\text{mid}} = (T_\text{onset}+T_\text{zero})/2$. The VLM is prompted to scan from high to low temperature and to prefer a zoomed inset over the main panel when both are present (Section~\ref{app:subsec:prompts}). The main challenge is distinguishing a genuine superconducting transition (a sharp, near-vertical drop over $\lesssim$3~K) from the gradual resistance decrease of ordinary metallic or Kondo-coherent behavior in heavy-fermion compounds, which the prompt explicitly warns against misreading as $T_c$.

Which VLM-extracted material a given plot series is linked to is a separate
step from reading $T_c$ off the curve, using the same plot-digitization and
linking components described in \ref{app:sec:claude-extractor}: a plot
digitizer (Claude-Sonnet-4.6, given the paper text and figure-caption
context) names each series, \texttt{PlotFilterConfig.for\_superconductivity()}
keeps only plots whose axes read as resistance/resistivity vs.\ temperature and discarding series with varying applied magnetic field and pressure,
and \texttt{SeriesMaterialLinker} (DeepSeek-V3.2) matches each surviving
series name to one of the paper's extracted materials. This structured
path is more accurate than reading $T_c$ and guessing the material in one
VLM call, but any one stage
failing (figure segmentation, axis-label digitization, or linking) drops
that material's $T_c$ to missing entirely. A paper-level fallback re-runs
the original one-shot VLM read, scoped only to materials the structured
path left without a value, recovering coverage the structured path alone
would lose. Claude-Sonnet-4.6 is used for this domain's plot digitizer rather than Qwen3.5-397B-A17B, unlike its selection for the thermocatalysis case study
(\ref{tab:thermocat-vlm-extraction}) because Qwen3.5-397B-A17B, accessed via OpenRouter, was substantially slower per call. We recommend users of \texttt{LeMat-Synth Parser}
test different VLM configurations on a small validation set for their own
domain (see Fig.~\ref{fig:app-pipeline}).

\begin{figure}[tp]
    \centering
    \includegraphics[width=0.45\linewidth]{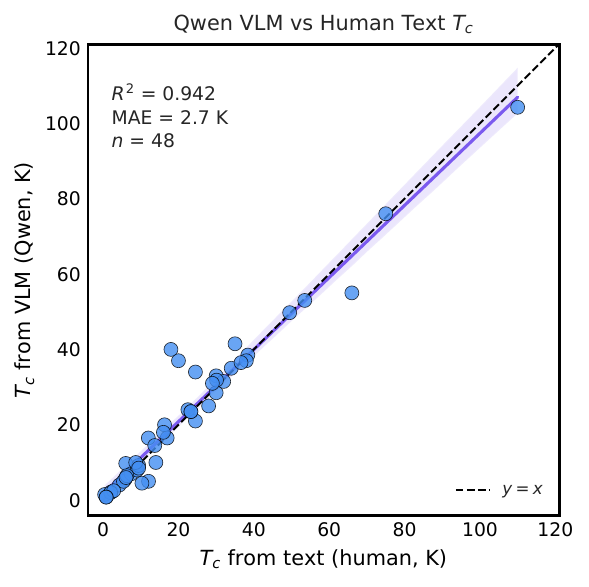}
    \caption{Validation of automated $T_c$ extraction: a vision-language model (VLM, Qwen3.5-397B-A17B) reading resistivity/resistance-vs-temperature plots directly is compared against $T_c$ values reported in the article text ($R^2=0.94$, MAE$=2.7$~K, $n=48$ matched pairs, Section~\ref{app:sec:supercond-linking-fix}). Good agreement confirms that the figure-extraction pipeline reads curves reliably where a text value exists to check against.}
    \label{fig:app-superc-calibration}
\end{figure}

\begin{figure}[tp]
    \centering
    \includegraphics[width=0.45\linewidth]{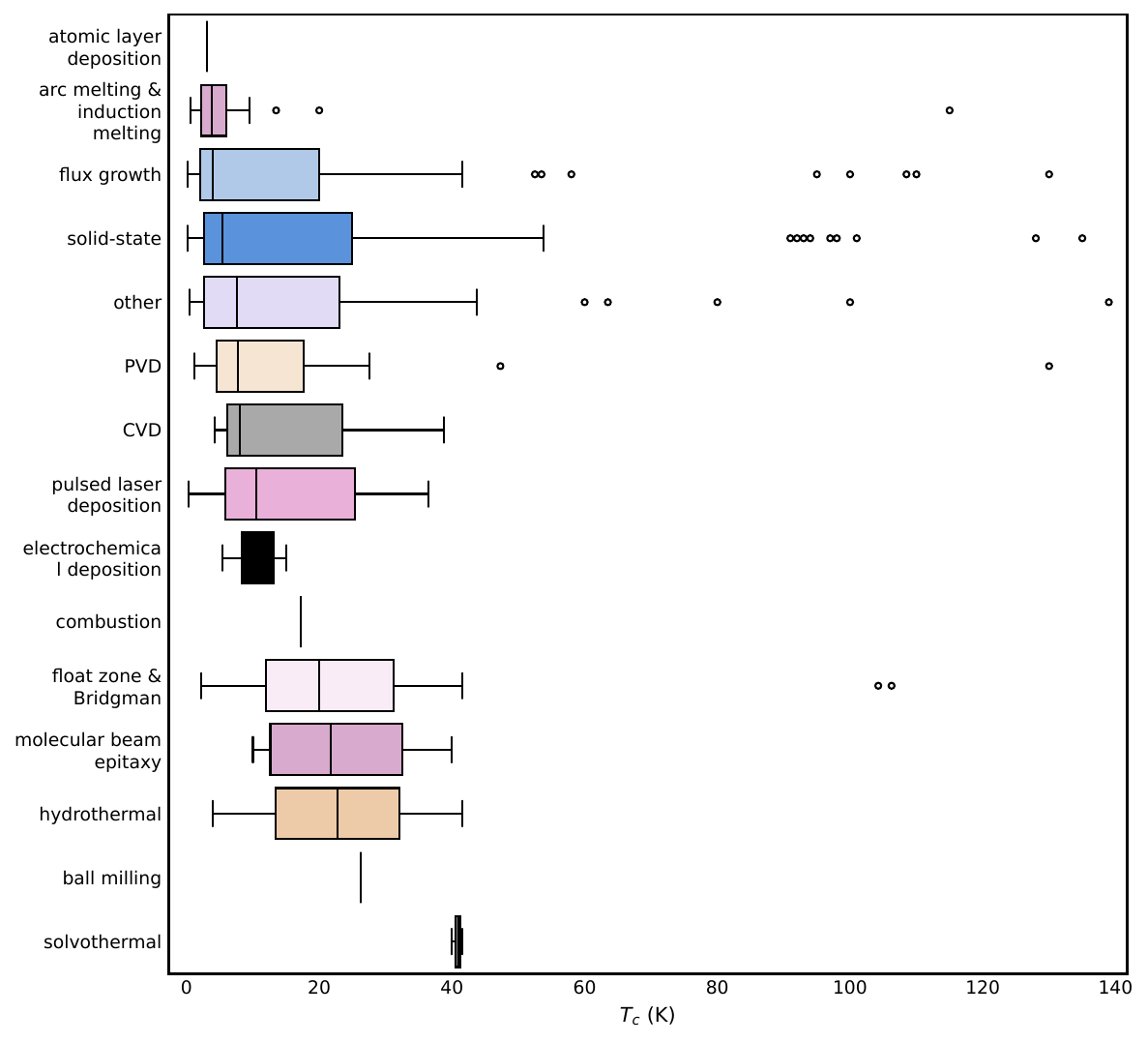}
    \caption{Extracted $T_c$ (\texttt{tc\_best}) by synthesis method, across the 497-paper corpus restricted to papers with a per-paper LLM-judge \texttt{overall\_score}$>4$ (Section~\ref{app:sec:supercond-failure-cases}) ($n=527$ rows spanning 15 distinct methods). The three most frequent methods are solid-state reaction ($n=183$), flux growth ($n=121$), and arc melting/induction melting ($n=40$). A residual ``other'' category not attributable to a specific named method accounts for $n=97$ rows.}
    \label{fig:app-superc-boxplot-synth-full}
\end{figure}

\begin{figure}[tp]
    \centering
    \includegraphics[width=0.45\linewidth]{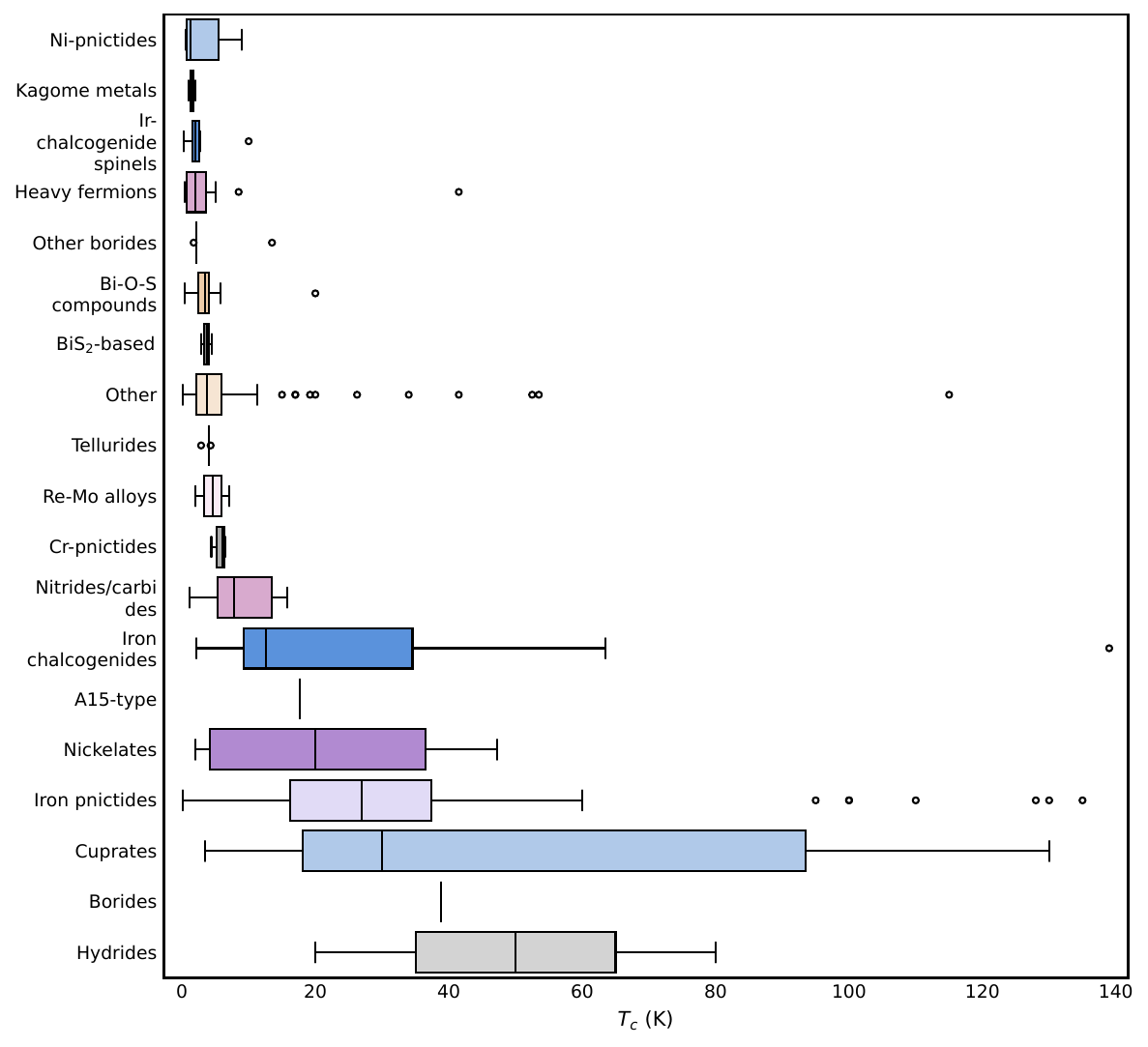}
    \caption{Extracted $T_c$ (\texttt{tc\_best}) by material family, on the same judge\_score$>4$-filtered corpus as Fig.~\ref{fig:app-superc-boxplot-synth-full}. Families are assigned by a rule-based classifier over each material's constituent elements, grouping compositions into categories such as cuprates, iron pnictides/chalcogenides, hydrides, borides (MgB$_2$-type), and heavy fermions, with an ``Other'' category for compositions matching none of the defined rules.}
    \label{fig:app-superc-boxplot-family-full}
\end{figure}

\begin{figure}[tp]
    \centering
    \includegraphics[width=\linewidth]{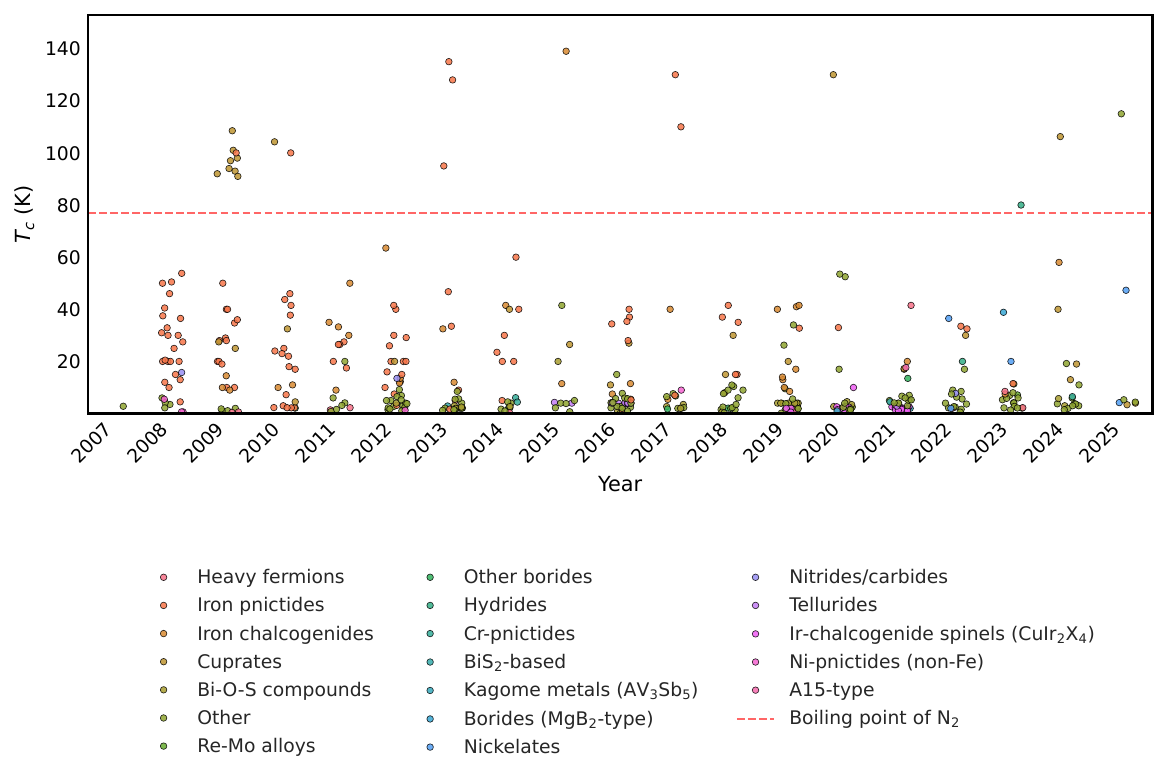}
    \caption{Extracted $T_c$ (\texttt{tc\_best}) vs.\ publication year, on the same judge\_score$>4$-filtered corpus as Fig.~\ref{fig:app-superc-boxplot-synth-full} ($n=465$ rows with both a numeric $T_c$ and year, spanning publication years 2007--2025). Note that the axis is publication year, not discovery year: the window opens well after the discovery of the cuprates (1986) and coincides with that of the iron pnictides (2008), so this figure shows the field's coverage of the major families over the past two decades. Within that window it adds per-composition detail within each series that individual papers report only for their best-performing composition.}
    \label{fig:app-superc-tc-by-year}
\end{figure}

\paragraph{Parity plot: text $T_c$ vs.\ figure $T_c$}

The manually-annotated ground-truth set was expanded from an initial 17
papers to 40 papers
(103 annotated rows; the
original 17-paper set is an exact subset). On the 40-paper set: $R^2=0.94$,
MAE$=2.7$~K, $n=48$ matched pairs (both a human text $T_c$ and a linked VLM
figure $T_c$ present). Materials are linked between the human annotation
and the VLM extraction using the same two-stage approach as the
judge-agreement analysis (\ref{app:sec:name-matcher-validation}):
exact/normalized string match first, then \texttt{DspyNameMatcherJudge} as
a fallback for stoichiometry-notation variants the string matcher misses
(e.g.\ \texttt{Ca1-xLaxFe2As2} vs.\ \texttt{Ca0.8La0.2Fe2As2}), keeping only
high- and medium-confidence LLM matches (42 string matches, 31 LLM
matches out of 103 rows).

\paragraph{Re-extraction of 10 papers affected by doping-series-label
collapse} An initial link against the original 40-paper VLM extraction run gave only
$n=30$: of the 62 rows with a numeric text $T_c$ (excluding the
pending-review paper below), 22 failed to link to any VLM-extracted
material and 10 more linked correctly but had no VLM $T_c$. Inspecting the
22 unlinked rows showed a systematic pattern: for 10 papers, the VLM
collapsed an entire doping series into a single generic parameterized
material label (e.g.\ \texttt{Bi2(Sr2-xLax)CuO6}$\delta$ instead of six distinct
compositions), reported zero materials for a paper, or explicitly stated
``No materials synthesized'' for a paper that clearly reports synthesized
compounds, consistent with the parameterized-legend-label limitation
already noted above (line~\ref{app:sec:supercond-linking-fix} context).
Re-running the identical, unmodified extraction pipeline
(no prompt or code changes) on
just these 10 papers recovered materials correctly in every case (e.g.\
13 distinct compositions for the paper that previously collapsed to one),
indicating the original failure was substantially attributable to LLM
extraction non-determinism. This
re-extraction is saved separately and does
\emph{not} modify the original 40-paper run or the separate 497-paper
\texttt{judge\_score}$>4$ corpus used for the family/synthesis/year figures
and the $T_c\geq150$~K failure analysis below, which were not re-run.
Re-linking with the 10 papers' corrected materials substituted in raised
$n$ from 30 to 48 (excluding the pending-review paper).

\label{app:sec:supercond-linking-fix}

\paragraph{Failure cases}
\label{app:sec:supercond-failure-cases}

The pipeline flags any extracted $T_{c,\text{mid}} \geq 150$~K as suspicious.
On the 497-paper \texttt{judge\_score}$>4$
corpus, 14 of 543 extracted $T_c$ rows (2.6\%) exceed 150~K, spanning 7 of
497 papers (1.4\%). We manually inspected all 7 papers against
their arXiv abstracts (\ref{table:high-tc-failures}).

\begin{table}[t]
  \centering
  \caption{Manual inspection of every $T_c \geq 150$~K extraction in the 497-paper \texttt{judge\_score}$>4$ corpus (14 rows (3\%), 7 papers(1\%)). Only one (H$_3$S, a genuine high-pressure hydride superconductor) is a correct extraction. The remaining six are VLM misreads, falling into two patterns: reading the wrong axis of a $T_c$-vs-doping phase diagram, or reading a non-superconducting resistivity anomaly (structural, charge/spin-density-wave) as if it were $T_c$.}
  \begin{tabular}{p{2cm}p{6.5cm}p{6cm}}
    \toprule
    \textbf{arXiv ID} & \textbf{Paper topic} & \textbf{Likely failure mode} \\
    \midrule
    1706.01744 & Pressure-induced structural transition in PbTaSe$_2$ (true $T_c\approx$3.8~K) & Misread high-$T$ structural/pressure feature as $T_c$ \\
    1812.09957 & Broad-range $R(T)$ of CrAs to fit phonon models (true $T_c=$4.2~K) & Misread normal-state background curve as $T_c$ \\
        2310.03609 & Si-doping study of Sc$_2$Ir$_{4-x}$Si$_x$, reports $T_c$-vs-doping phase diagram & Read composition axis of phase diagram as temperature axis \\
    2101.03473 & I-doping study of CuIr$_2$Te$_{4-x}$I$_x$ (true $T_c\approx$2.95~K), reports CDW+$T_c$-vs-doping diagram & Read composition axis of phase diagram as temperature axis \\
    2209.13299 & High-pressure H$_3$S hydride superconductor ($T_c=$197~K at 153~GPa) & Correct, genuine high-$T_c$, high-pressure physics \\
    1008.2029 & Magnetotransport of EuFe$_2$As$_2$ (parent compound: SDW/structural transition $\sim$190~K, not superconducting) & Misread SDW/structural anomaly (inset panel) as $T_c$ \\
    1504.00685 & Bi$_2$Rh$_3$S$_2$ vs.\ Bi$_2$Rh$_{3.5}$S$_2$: paper states Bi$_2$Rh$_3$S$_2$ is NOT superconducting (has $\sim$165~K structural transition instead) & Misread/mislinked structural transition of a non-superconducting compound as $T_c$ \\
    \bottomrule
  \end{tabular}
  \label{table:high-tc-failures}
\end{table}

Two failure patterns account for six of the seven papers. First,
\textbf{phase-diagram axis confusion} (2310.03609, 2101.03473): both
papers report $T_c$ as a function of doping/composition and not a
single $R(T)$ curve, and the VLM (or the digitization step feeding it)
appears to have read values off the composition axis as if it were a
temperature axis. The giveaway is that each paper yields several
distinct "$T_c$" values that track doping levels cleanly rather than
looking like independent temperature measurements. Second, a
\textbf{wrong-transition misread} (1706.01744, 1812.09957, 1008.2029,
1504.00685): the VLM latches onto a structural, charge-density-wave, or
spin-density-wave resistivity anomaly at high temperature and reports it
as a superconducting $T_c$. 1504.00685 is the clearest case, since the
source paper explicitly states the extracted compound is not
superconducting at all. The one true positive, 2209.13299 (H$_3$S), both
validates that the $\geq$150~K flag is not simply a hard rejection of
real high-pressure physics, and confirms the flag correctly separates a
genuine result from six spurious ones. All 7 flagged papers are excluded
from the material-family and $T_c$-vs-year visualizations in the main
text.


Linked superconductor compositions are classified into families
(cuprates, iron pnictides, hydrides, etc.) by a rule-based classifier
over their parsed element sets (e.g.\ Fe+As $\rightarrow$ iron pnictide,
Cu+O $\rightarrow$ cuprate), checked in a fixed priority order so that,
e.g., a hydride match is resolved before a more generic rule can claim
it. Compositions matching no rule fall into ``Other,'' which is a
genuine long tail of individually rare chemistries (skutterudites,
organic BEDT-TTF salts, pyrochlores, etc.) -- each appears too few times to justify a dedicated rule. 

\subsection{Prompts}
\label{app:subsec:prompts}

This section shows the system prompts employed and the full configurations used (incl. signatures and LLM configurations) to extract the data presented in this work.

\paragraph{Filtering papers}\mbox{}\\[0.5em]

\begin{tcolorbox}[enhanced,breakable,mytitle,myprompt,title=Prompt,top=0.6em,bottom=0.4em,left=0.6em,right=0.6em]
\begin{Verbatim}[breaklines,breakanywhere,fontsize=\small]
Analyze the following text and answer the questions in JSON format:
{chunk}
Questions:
1. Does it contain a material synthesis recipe?
    (Answer with true or false)
2. If yes, what is the material name?
    (Answer with the material name or "N/A" if no recipe)
3. If yes, which category of materials does it belong to?
    (Answer with the specific material type or "N/A" if no recipe)
List of material categories:
Metals, Ceramics, Semiconductors, Superconductors, Composites,
Biomaterials, Nanomaterials, Polymers, Magnetic, Textiles, Chemicals, Other
Format your response as a JSON object with the following structure:
{{
"contains_recipe": true/false,
"material_name": "material name or N/A",
"material_category": "material category or N/A"
}}
\end{Verbatim}
\end{tcolorbox}

\paragraph{Material extraction}\mbox{}\\[0.5em]

\begin{tcolorbox}[enhanced,breakable,mytitle,myprompt,title=Prompt,top=0.6em,bottom=0.4em,left=0.6em,right=0.6em]
\begin{Verbatim}[breaklines,breakanywhere,fontsize=\small]
You are a helpful assistant that extracts ONLY the final synthesized materials from scientific papers.

Your task is to identify ONLY the materials that are the final products of synthesis procedures described in the paper.

IMPORTANT GUIDELINES:
- ONLY include materials that are the final synthesized products
- DO NOT include starting materials, precursors, supports, gases, solvents, or other chemicals used in synthesis
- DO NOT include materials that are just mentioned or characterized but not synthesized
- Focus on the main target materials that are actually synthesized

EXAMPLES OF WHAT TO INCLUDE:
- "Ni/Al2O3" (if Ni is deposited on Al2O3)
- "Ir/SiO2" (if Ir is supported on SiO2)
- "LiFePO4 nanoparticles" (if LiFePO4 is synthesized)
- "Co-doped LiFePO4" (if this specific material is synthesized)

EXAMPLES OF WHAT TO EXCLUDE:
- "Ni", "Ir", "Ru" (if these are just precursors)
- "H-ZSM-5", "Al2O3", "SiO2" (if these are just supports)
- "Ammonia", "Argon", "Hydrogen" (gases)
- "Deionized water" (solvents)
- "Ammonium hydroxide" (reagents)

Return a simple comma-separated list of ONLY the final synthesized materials.

If no materials are synthesized in the paper, return No materials synthesized.

Keep the output simple and clean - just the final synthesized material names separated by commas.
\end{Verbatim}
\end{tcolorbox}

\begin{tcolorbox}[enhanced,breakable,mytitle,myyaml,title=Configuration (YAML),top=0.6em,bottom=0.4em,left=0.6em,right=0.6em]
\begin{Verbatim}[breaklines,breakanywhere,fontsize=\small]
architecture:
  _target_: llm_synthesis.transformers.material_extraction.dspy_extraction.DspyTextExtractor
  signature:
    _target_: llm_synthesis.transformers.material_extraction.dspy_extraction.make_dspy_text_extractor_signature
    signature_name: "TextToMaterials"
    instructions: "Extract ONLY the final synthesized materials from the publication text."
    input_description: "The publication text to extract the final synthesized materials from."
    output_name: "materials"
    output_description: "The final synthesized materials as a comma-separated list."
  lm:
    _target_: llm_synthesis.utils.dspy_utils.get_llm_from_name
    llm_name: "qwen3.5-397b-a17b"
    model_kwargs:
      temperature: 0.0
    system_prompt:
      _target_: llm_synthesis.utils.read_prompt_str_from_txt
      prompt_path: "examples/system_prompts/material_extraction/default.txt"
\end{Verbatim}
\end{tcolorbox}

\paragraph{Synthesis extraction}\mbox{}\\[0.5em]

\begin{tcolorbox}[enhanced,breakable,mytitle,myprompt,title=Prompt,top=0.6em,bottom=0.4em,left=0.6em,right=0.6em]
\begin{Verbatim}[breaklines,breakanywhere,fontsize=\small]
You are a helpful assistant that extracts the structured synthesis for a specific material from the paper text.

Focus ONLY on the synthesis procedure for the specified material. Search through the entire paper text to find the synthesis procedure that describes how this specific material is made.

IMPORTANT: You must output ONLY a valid JSON object with a "structured_synthesis" field. Do not include any reasoning, explanations, or markdown formatting.

If you cannot find a synthesis procedure for the specified material, return a minimal structure with the material name and an empty synthesis.

The JSON output must follow this exact structure:
{
  "structured_synthesis": {
    "target_compound": "string (required) - should match the specified material name",
    "target_compound_type": "string (required) - choose from: 'metals & alloys', 'ceramics & glasses', 'polymers & soft matter', 'composites', 'semiconductors & electronic', 'nanomaterials', 'two-dimensional materials', 'framework & porous materials', 'biomaterials & biological', 'liquid materials', 'hybrid & organic-inorganic', 'functional materials & catalysts', 'energy & sustainability', 'smart & responsive materials', 'emerging & quantum materials', 'other'",
    "synthesis_method": "string (required) - choose from: 'PVD', 'CVD', 'arc discharge', 'ball milling', 'spray pyrolysis', 'electrospinning', 'sol-gel', 'hydrothermal', 'solvothermal', 'precipitation', 'coprecipitation', 'combustion', 'microwave-assisted', 'sonochemical', 'template-directed', 'solid-state', 'flux growth', 'float zone & Bridgman', 'arc melting & induction melting', 'spark plasma sintering', 'electrochemical deposition', 'chemical bath deposition', 'liquid-phase epitaxy', 'self-assembly', 'atomic layer deposition', 'molecular beam epitaxy', 'pulsed laser deposition', 'ion implantation', 'lithographic patterning', 'wet impregnation', 'incipient wetness impregnation', 'mechanical mixing', 'solution-based', 'mechanochemical', 'other'",
    "starting_materials": [{"name": "string", "amount": "number or null", "unit": "string or null", "purity": "string or null", "vendor": "string or null"}],
    "steps": [{"step_number": "integer", "action": "string", "description": "string or null", "materials": [{"name": "string", "amount": "number or null", "unit": "string or null", "purity": "string or null", "vendor": "string or null"}], "equipment": [{"name": "string", "instrument_vendor": "string or null", "settings": "string or null"}], "conditions": {"temperature": "number or null", "temp_unit": "string or null", "duration": "number or null", "time_unit": "string or null", "pressure": "number or null", "pressure_unit": "string or null", "atmosphere": "string or null", "stirring": "boolean or null", "stirring_speed": "number or null", "ph": "number or null"}}],
    "equipment": [{"name": "string", "instrument_vendor": "string or null", "settings": "string or null"}],
    "notes": "string or null"
  }
}

Do not include any text before or after the JSON object. Output only the JSON. 
\end{Verbatim}
\end{tcolorbox}

\begin{tcolorbox}[enhanced,breakable,mytitle,myyaml,title=Configuration (YAML),top=0.6em,bottom=0.4em,left=0.6em,right=0.6em]
\begin{Verbatim}[breaklines,breakanywhere,fontsize=\small]
architecture:
  _target_: llm_synthesis.transformers.synthesis_extraction.dspy_synthesis_extraction.DspySynthesisExtractor
  signature:
    _target_: llm_synthesis.transformers.synthesis_extraction.dspy_synthesis_extraction.make_dspy_synthesis_extractor_signature
    signature_name: "SynthesisSignature"
    instructions: "Extract the structured synthesis for a specific material from the paper text."
    paper_text_description: "The complete paper text to search for the material's synthesis procedure."
    material_name_description: "The name of the specific material to extract synthesis for."
    output_name: "structured_synthesis"
    output_description: "The extracted structured synthesis for the specific material."
  lm:
    _target_: llm_synthesis.utils.dspy_utils.get_llm_from_name
    llm_name: "qwen3.5-397b-a17b"
    model_kwargs:
      temperature: 0.0
      max_tokens: 8000
      max_retries: 3
    system_prompt:
      _target_: llm_synthesis.utils.read_prompt_str_from_txt
      prompt_path: "examples/system_prompts/synthesis_extraction/default.txt"
\end{Verbatim}
\end{tcolorbox}

\paragraph{Figure extraction}\hfill\\
For figure extraction, we do not provide a separate DSPy configuration. 
Unlike material and synthesis extraction (which are wrapped with DSPy signatures and explicit 
input/output schemas), the figure extraction pipeline directly leverages the 
system prompt together with a \texttt{litellm}-backed multi-provider client (\texttt{LiteLLMPlotDataExtractor}), allowing the underlying VLM to be swapped per domain (Qwen3.5-397B-A17B for the thermocatalysis and superconductivity case studies) without code changes. In this setup, the model is
invoked with the raw prompt and image data, and the parsing into structured 
objects (\texttt{ExtractedLinePlotData}) happens entirely within the custom 
transformer implementation. Because no DSPy signature or schema mediation is 
involved, there is no corresponding YAML configuration block to display. Instead, 
the logic is captured in the prompt (shown below) and the Python implementation 
excerpted below. 

\begin{tcolorbox}[enhanced,breakable,mytitle,myprompt,
  title=Prompt,top=0.6em,bottom=0.4em,left=0.6em,right=0.6em]
\begin{Verbatim}[breaklines,breakanywhere,fontsize=\small]
LINE_CHART_PROMPT = """
You will be provided with a line chart. The chart may not be chunked very well, 
so you may need to read only the plot in the center of the image.
In the chart, there will be several lines representing different data series.

1. Identify the different lines by their colors and labels.
2. For each line, extract the coordinates of the points that make up the line. 
   Do not include any points that are not part of the line.
3. If the chart has metadata such as a title, x-axis label, y-axis labels, 
   or units, extract that information as well. 
   Keep the scientific terms in Markdown format.
4. Output the data in the specified format:

Name_of_Line_1: [[x1, y1], [x2, y2], ...]
title:
x_axis_label:
x_axis_unit:
y_left_axis_label:
y_left_axis_unit:

Do not output any other text, just the data in the format above.
"""
\end{Verbatim}
\end{tcolorbox}

\begin{tcolorbox}[enhanced,breakable,mytitle,myyaml,
  title=Implementation excerpt (Python), 
  top=0.6em,bottom=0.4em,left=0.6em,right=0.6em,
  colback=gray!5!white]
\begin{Verbatim}[breaklines,breakanywhere,fontsize=\scriptsize]
class ClaudeLinePlotDataExtractor(LinePlotDataExtractorInterface):
    def __init__(self, model_name: str,
                 prompt: str = resources.LINE_CHART_PROMPT,
                 max_tokens: int = 1024,
                 temperature: float = 0.0):
        super().__init__()
        self.claude_client = ClaudeAPIClient(model_name)
        self.prompt = prompt
        self.max_tokens = max_tokens
        self.temperature = temperature

    def forward(self, input: FigureInfoWithPaper) -> ExtractedLinePlotData:
        figure_base64 = input.base64_data
        self.claude_client.reset_cost()
        claude_response_obj = self.claude_client.vision_model_api_call(
            figure_base64=figure_base64,
            prompt=self.prompt,
            max_tokens=self.max_tokens,
            temperature=self.temperature,
        )
        return self._parse_into_pydantic(claude_response_obj)

    def _parse_into_pydantic(self, response: str) -> ExtractedLinePlotData:
        """Parse text into Pydantic object with regex pattern matching"""
        ...
\end{Verbatim}
\end{tcolorbox}

\paragraph{Synthesis evaluation}\hfill\\
In this case, the evaluation logic is fully captured within the DSPy configuration itself, 
so we do not provide a standalone prompt block. Both the task instructions and the 
system prompt are directly embedded inside the configuration file. The complete configuration is shown below:
\begin{tcolorbox}[enhanced,breakable,mytitle,myyaml,title=Configuration (YAML),top=0.6em,bottom=0.4em,left=0.6em,right=0.6em]
\begin{Verbatim}[breaklines,breakanywhere,fontsize=\small]
architecture:
  _target_: llm_synthesis.metrics.judge.general_synthesis_judge.DspyGeneralSynthesisJudge
  signature:
    _target_: llm_synthesis.metrics.judge.general_synthesis_judge.make_general_synthesis_judge_signature
    signature_name: "GeneralSynthesisJudgeSignature"
    instructions: >
      You are an expert materials scientist and data extraction specialist with extensive experience in:
        - Synthesis procedure analysis and documentation
        - Structured data extraction from scientific literature
        - Materials science ontology design and terminology standardization
        - Quality assessment of automated scientific information extraction systems

      Evaluate how well the GeneralSynthesisOntology extraction captures synthesis information from 
      the provided source text.

      IMPORTANT: Do NOT penalize the extraction system for failing to include information that is 
      not present in the original paper. Missing elements should only be considered errors if they 
      were clearly stated in the source but were not extracted. If an element is absent in both the 
      source and the extraction, and is correctly left blank or omitted, this should be considered 
      correct and scored highly.

      ASSESSMENT FOCUS:
        - Completeness: All synthesis components present in the source are captured
        - Accuracy: Correct values, units, and classifications based on the text
        - Structure: Proper organization and logical sequencing of elements
        - Semantic Preservation: Scientific meaning and intent faithfully maintained
        - Schema Compliance: Conforms to the expected ontology format and data types

      EVALUATION CRITERIA (Score 1-5 for each):
        1. Structural Completeness - Extraction of all relevant synthesis components from the source (materials, steps, equipment, conditions)
        2. Material Extraction - Correct names, quantities, units, purities as specified in the paper
        3. Process Steps - Accurate step order and correct action classification
        4. Equipment Extraction - Proper identification of all equipment explicitly mentioned
        5. Conditions Extraction - Accurate recording of parameters such as temperature, time, atmosphere, pressure, etc.
        6. Semantic Accuracy - Faithful preservation of scientific meaning without misinterpretation
        7. Format Compliance - Adherence to ontology schema, data types, and field structure

      For each criterion:
        - Assign a score between 1 and 5
        - Provide detailed technical reasoning for the assigned score
        - Offer specific, constructive recommendations for improvement, if applicable

  lm:
    _target_: llm_synthesis.utils.dspy_utils.get_llm_from_name
    llm_name: "deepseek-v3.2"
    model_kwargs:
      temperature: 0.1
      max_tokens: 4096
    system_prompt: >
      You are a senior materials scientist and data extraction expert with deep expertise in:
        - Inorganic and organic synthesis methodologies
        - Laboratory instrumentation and experimental workflows
        - Chemical nomenclature, stoichiometry, and unit conventions
        - Optimization of synthesis conditions and reaction parameters
        - Structured data modeling and materials science ontology design
        - Evaluation methodologies for automated information extraction systems

      Your assessments should reflect best practices in synthesis reporting and uphold the highest 
      standards of scientific accuracy, reproducibility, and structured data quality.

      When evaluating extracted synthesis data:
        - Rely on your domain expertise to assess technical correctness, semantic fidelity, and structural organization
        - Emphasize clarity, precision, and alignment with real-world experimental protocols
        - Consider the intended schema and use context to assess compliance and completeness
        - Do not penalize the extraction system for omitting elements that were not explicitly present in the source text

      Your evaluation should be technically rigorous, yet fair, grounded in both materials science principles and data extraction best practices.

  enable_reasoning_traces: true
  confidence_threshold: 0.7
\end{Verbatim}
\end{tcolorbox}

\paragraph{Thermocatalysis: performance–synthesis linking}\mbox{}\\[0.5em]

\texttt{SeriesMaterialLinker} is configured with \texttt{gemini-3-pro-preview} as the linker model (Section~\ref{app:casestudy-thermocat}).

\begin{tcolorbox}[enhanced,breakable,mytitle,myprompt,title=Prompt,top=0.6em,bottom=0.4em,left=0.6em,right=0.6em]
\begin{Verbatim}[breaklines,breakanywhere,fontsize=\small]
You are given materials studied in a scientific
paper and series/line names extracted from a plot in that paper. Match each
series name to the material it represents.

Materials: {materials}
Series names from plot: {series_names}
Figure context: {context}
Plot title: {plot_title}
X-axis: {x_axis_label} ({x_axis_unit})
Y-axis: {y_axis_label} ({y_axis_unit})

Return a JSON list of matches. Each match must have:
- "series_name": exactly one of the series names listed above
- "material_name": exactly one of the material names listed above
- "confidence": "high", "medium", or "low"
- "reasoning": brief explanation of why this match was made

Rules:
- Only match when there is a CLEAR connection between the series name and a
  material name (e.g., names are similar, or context makes the link obvious).
- NEVER guess or assign matches randomly. If you cannot identify a real
  connection, leave the series unmatched. An empty list [] is a valid response.
- If a series is a baseline, reference, substrate, equilibrium line, or
  pressure condition -- do NOT include it.
- series_name and material_name MUST be exactly from the lists above.
  Do not modify them.

Return ONLY a valid JSON list, no other text.
\end{Verbatim}
\end{tcolorbox}

\paragraph{Superconductors: Tc extraction from text}\mbox{}\\[0.5em]

Built inline with the same DSPy-based extractor architecture as the material/synthesis extraction prompts above. Shown here is the current multi-composition, multi-notation version, which superseded an earlier single-$T_c$-per-material version.

\begin{tcolorbox}[enhanced,breakable,mytitle,myyaml,title=Configuration (YAML),top=0.6em,bottom=0.4em,left=0.6em,right=0.6em]
\begin{Verbatim}[breaklines,breakanywhere,fontsize=\small]
architecture:
  _target_: llm_synthesis.transformers.material_extraction.dspy_extraction.DspyTextExtractor
  signature:
    _target_: llm_synthesis.transformers.material_extraction.dspy_extraction.make_dspy_text_extractor_signature
    signature_name: "TextToTcMulti"
    instructions: |
      Extract ALL superconducting critical temperature (Tc) values from this paper.

      RULES:
      1. NO HALLUCINATION: Only extract Tc values that appear as explicit numbers in
         this paper's own results. Never use general knowledge. Never extract values
         cited from other references. If no number is stated, report NR.

      2. Tc NOTATIONS -- look for all of these:
         onset: Tc_onset, Tconset, Tc,onset, Tc(onset), T_c^onset
         midpoint: Tc, Tc_mid, Tc,mid, T_c^mid
         zero-resistance: Tc_zero, Tc,zero, Tc(0), T_c^zero
         transition width: DeltaTc, delta_Tc -- report this as delta_Tc, NOT as a Tc value.

      3. WHERE TO FIND Tc -- check ALL of these locations in the paper:
         - Abstract (often states the main Tc result)
         - Results / Discussion sections (detailed Tc values per sample)
         - Tables (Tc columns, summary tables of properties)
         - Figure captions (e.g., 'Tc = 39 K as shown in Fig. 3')
         - Conclusions (often restates key Tc findings)
         Also look for INDIRECT phrasings:
         - 'superconductivity emerges below 39 K'
         - 'becomes superconducting at 23 K'
         - 'critical temperature of 92 K'
         - 'superconducting transition at 7.2 K'
         - 'zero resistance below 25 K'
         - 'onset of superconductivity near 30 K'

      4. MATERIAL NAME = fully resolved chemical formula.
         Substitute all variables with their numeric values.
         e.g. A(B1-xCx) with x=0.3 -> AB0.7C0.3
         Each distinct composition gets its own row.
         Use the chemical formula, not sample labels (S1, SC2, etc.).

      5. CONDITION = only external factors NOT encoded in the formula:
         pressure, magnetic field, sample form (single-crystal, thin-film, etc.).
         If none apply, use 'ambient'.
    input_description: "The full publication text from a superconductivity paper."
    output_name: "tc_values"
    output_description: |
      One line per (material, condition) pair in this pipe-delimited format:
      formula | condition: <value or ambient> | superconducting: YES/NO | T_onset: <number> K | Tc: <number> K | T_zero: <number> K | delta_Tc: <number> K

      Use NR for any value not explicitly reported in the text.

      Example lines:
      YBa2Cu3O7 | condition: ambient | superconducting: YES | T_onset: 93 K | Tc: 92 K | T_zero: 91 K | delta_Tc: NR
      Nb3Sn | condition: 12 GPa | superconducting: YES | T_onset: NR | Tc: 23 K | T_zero: NR | delta_Tc: NR
      SrTiO3 | condition: ambient | superconducting: NO | T_onset: NR | Tc: NR | T_zero: NR | delta_Tc: NR
  lm:
    _target_: llm_synthesis.utils.dspy_utils.get_llm_from_name
    llm_name: "deepseek-v3.2"
    model_kwargs:
      temperature: 0.0
      max_tokens: 16384
\end{Verbatim}
\end{tcolorbox}

\paragraph{Superconductors: Tc extraction from figures (geometric construction)}\mbox{}\\[0.5em]

A VLM prompt (\texttt{DIRECT\_TC\_PROMPT\_TEMPLATE}, same file) that reads $\rho(T)$/$R(T)$ plots directly, distinguishing genuine superconducting transitions (sharp, narrow, to zero resistance) from gradual metallic/Kondo behavior, and applying the standard onset/zero geometric construction for $T_c$.

\begin{tcolorbox}[enhanced,breakable,mytitle,myprompt,title=Prompt,top=0.6em,bottom=0.4em,left=0.6em,right=0.6em]
\begin{Verbatim}[breaklines,breakanywhere,fontsize=\small]
You are analyzing a Resistance (or Resistivity) vs Temperature plot from a
superconductivity paper. Your task is to determine the critical temperature
Tc for each series using the standard geometric construction.

CRITICAL DISTINCTION -- SUPERCONDUCTING TRANSITION vs NORMAL METALLIC BEHAVIOR:
Many materials (especially heavy-fermion compounds like CeCoIn5, CeRhIn5, etc.)
show a GRADUAL decrease in resistivity over a WIDE temperature range (e.g., from
50 K down to 5 K). This is NORMAL metallic behavior (Kondo coherence, phonon
scattering reduction, etc.) and is NOT a superconducting transition.

The superconducting transition has these characteristics:
  - It is a SHARP, near-vertical drop in resistance
  - It occurs over a NARROW temperature range (typically 0.1 to 3 K wide)
  - Resistance drops from a finite value all the way to ZERO
    (or very close to zero)
  - It looks like a cliff or step function, not a gradual slope

If you see a curve that gradually decreases over 10-50 K, that is NOT
superconductivity -- it is normal metallic/Kondo behavior.

STEP 0 -- EXAMINE THE FULL FIGURE (main plot + any insets/panels):
a) Identify ALL panels in the figure: the main plot and any insets, secondary
   panels, or embedded sub-plots. For each one, describe:
     - What quantity is on each axis (e.g., rho vs T, Tc vs x,
       phase diagram, etc.)
     - The axis ranges and tick marks
     - Whether it contains information relevant to determining Tc

b) CATEGORIZE each inset/panel into one of these types:
     (i)   ZOOMED R(T): A magnified view of the transition region of the same
           R(T) data as the main plot. Has the same axes (R vs T) but a narrower
           temperature range. -> Use this PREFERENTIALLY for geometric Tc
           construction (much better spatial resolution).
     (ii)  Tc SUMMARY: Shows Tc (or T_onset, T_zero) as a function
            of composition, pressure, doping, field, etc.
            (e.g., "Tc vs x" or a phase diagram with
           a Tc boundary). -> Read Tc values DIRECTLY from this panel for each
           series/composition. These are typically the most
            accurate values available.
     (iii) OTHER: Any panel not directly useful for Tc (e.g., Hall coefficient,
           magnetization, dR/dT, crystal structure). -> Note it but do not use it
           for Tc determination.

c) Read ALL numbered tick marks on the main plot axes. If a zoomed inset exists,
   read its tick marks separately.

d) CRITICAL -- SCALE AWARENESS: Before reading ANY value from the main plot,
   explicitly note the x-axis range. If the temperature axis spans a wide range
   (e.g., 0-300 K) and the transitions happen in a small fraction of that range,
   be extra careful with interpolation. Prefer reading from a zoomed inset or
   Tc-summary inset if available -- the main plot resolution may be too poor to
   determine precise transition temperatures.

STEP 0.5 -- EXTRACT Tc FROM SUMMARY INSETS (if any type-(ii) inset found):
If you identified a Tc-summary inset in Step 0b(ii), read the Tc values directly
from it for each series/composition. Report them as:
  inset_tc_<series_name>: <value> K

These values serve as a REFERENCE. In Step 4, if your geometric construction
from the R(T) curve gives a Tc that differs from the inset value by more than
20%, trust the inset value and report it instead -- noting "source: inset".

STEP 1 -- IDENTIFY SERIES:
{series_name_instruction}
List every distinct curve visible in the plot.

STEP 2 -- READ RESISTANCE VALUES AT LOWEST AND HIGHEST TEMPERATURE:
For EACH series, read two values CAREFULLY:
  a) R_at_lowest_T: the resistance/resistivity value at the LOWEST
     temperature shown in the plot. Look at the leftmost data point of
     this series -- what y-value does it have? Read the y-axis scale
     carefully. If the leftmost point is at y=0 on the scale, that is zero.
     If it is at y=5, y=10, y=20, etc. -- that is NOT zero.
  b) R_at_highest_T: the resistance/resistivity at the HIGHEST temperature
     (rightmost data point, i.e., normal-state value).

Report both values for every series. This is critical for determining
which materials are superconducting.

STEP 3 -- CONFIRM SUPERCONDUCTIVITY:
A series is superconducting ONLY if R_at_lowest_T is approximately zero
(within measurement noise of the zero line on the y-axis).

IMPORTANT: Be strict about what "approximately zero" means:
  - If the y-axis goes from 0 to 30 uOhm cm, then R_at_lowest_T must be
    below ~0.5 uOhm cm to count as zero (i.e., visually touching the x-axis)
  - A value like 5, 10, or 20 uOhm cm is NOT approximately zero
  - A curve that decreases gradually but never reaches the bottom of the
    plot is NOT superconducting

For non-superconducting series, report:
  superconducting: NO
  reason: <e.g., "R_at_lowest_T = 5 uOhm cm, clearly above zero">

STEP 4 -- GEOMETRIC Tc CONSTRUCTION (only for confirmed superconductors):
Use the INSET if available (better resolution), otherwise use the main plot.

For each superconducting series:

  a) NORMAL-STATE LEVEL: Read R_normal from the plateau IMMEDIATELY above
     the sharp superconducting drop. This is NOT the maximum resistance at
     high temperature -- it is the resistance just before the sharp drop begins.

  b) FIND T_onset: The temperature where the SHARP superconducting drop
     begins. This is where resistance SUDDENLY starts falling toward zero
     (not where it gradually decreases due to metallic behavior).

  c) FIND T_zero: Moving right from low T along R ~= 0, T_zero is the LAST
     point still at R ~= 0 before resistance starts rising. T_zero < T_onset.

  d) SANITY CHECK: T_onset - T_zero should typically be 0.1 to 5 K for
     most superconductors. If your T_onset - T_zero > 10 K, you are
     probably measuring a gradual metallic decline, NOT the SC transition.
     Re-examine the plot.

  e) Tc_mid = (T_onset + T_zero) / 2

  f) Delta_Tc = T_onset - T_zero

  g) CROSS-CHECK WITH INSET: If you extracted inset Tc values in Step 0.5,
     compare your geometric Tc_mid with the inset value. If they differ by
     more than 20%, the inset value is more reliable -- use it instead and
     set source to "inset".

STEP 5 -- RELATIVE ORDERING OF TRANSITIONS:
Even when transitions appear close together on the plot, they are almost
never at EXACTLY the same temperature. Look carefully:

  a) For each PAIR of superconducting series, compare which one starts
     dropping FIRST (at higher T). Look at the actual data points/markers --
     which series still has high resistance when the other has already
     started dropping?

  b) If series A starts dropping at a higher T than series B, then
     T_onset(A) > T_onset(B). Even if the difference is small (0.1-1 K),
     report it -- do NOT round both to the same value.

  c) NEVER report identical T_onset values for different series unless you
     are absolutely certain they are the same after careful comparison.

Output format:

inset_detected: <yes/no>
inset_type: <"zoomed_rt" | "tc_summary" | "other" | "none">
inset_description: <brief description of what the inset shows, or "N/A">
inset_axes: <tick marks of inset if present, otherwise "N/A">
inset_tc_values: <series: value K, series: value K, ... (if
    tc_summary type), otherwise "N/A">
main_x_axis_ticks: <list>
main_y_axis_ticks: <list>

Series: <name>
R_at_lowest_T: <value and unit>
R_at_highest_T: <value and unit>
superconducting: <YES/NO>
[If NO:]
reason: <why not superconducting>
[If YES:]
R_normal: <value and unit>
T_onset: <value> K
T_zero: <value> K
Tc_mid: <value> K
Delta_Tc: <value> K
source: <"inset" or "main plot" or "zoomed inset">

relative_ordering: <which series transitions first (highest T_onset),
second, etc. -- and by approximately how many K do they differ?>

Do not output any other text.
\end{Verbatim}
\end{tcolorbox}

\clearpage

\bibliographystyle{unsrt} 

\bibliography{LeMat-Synth}